\documentclass[aps,prb,nofootinbib,twocolumn,floatfix,superscriptaddress]{revtex4-2}
\usepackage{graphicx}
\usepackage{mathrsfs}
\usepackage{amssymb}
\usepackage{amsmath}
\usepackage{bm}
\usepackage{textcomp,color}
\usepackage[colorlinks=true,urlcolor=blue,citecolor=blue,linkcolor=blue,bookmarks=false, pdfstartview={FitH}]{hyperref}
\usepackage[capitalize]{cleveref} 
\usepackage{subcaption}
\usepackage{float}
\usepackage{ragged2e}
\usepackage{soul}

\newcommand{\beq}{\begin{equation}}
\newcommand{\eeq}{\end{equation}}
\newcommand{\bea}{\begin{eqnarray}}
\newcommand{\eea}{\end{eqnarray}}
\newcommand{\bse}{\begin{subequations}}
\newcommand{\ese}{\end{subequations}}
\newcommand{\nn}{\nonumber}
\newcommand{\bwt}{\begin{widetext}}
\newcommand{\ewt}{\end{widetext}}

\newcommand{\ve}{\varepsilon}
\newcommand{\e}{\epsilon}
\newcommand{\be}{{\bf\epsilon}}

\newcommand{\bk}{{\bf k}}

\newcommand{\bq}{{\bf q}}

\def\blue{\color{blue}}

\begin{document}

\title{Spin-orbit interaction enabled electronic Raman scattering from charge collective modes}
\author{Surajit Sarkar}
\affiliation{Department of Physics, Concordia University, Montreal, QC H4B 1R6, Canada}
\author{Alexander Lee}
\affiliation{Department of Physics and Astronomy, Rutgers University, Piscataway, NJ, 08854, USA}
\author{Girsh Blumberg}
\affiliation{Department of Physics and Astronomy, Rutgers University, Piscataway, NJ, 08854, USA}
\affiliation{National Institute of Chemical Physics and Biophysics, 12618 Tallinn, Estonia}
\author{Saurabh Maiti}
\affiliation{Department of Physics, Concordia University, Montreal, QC H4B 1R6, Canada}
\affiliation{Centre for Research in Molecular Modelling, Concordia University, Montreal, QC H4B 1R6, Canada}
\date{\today}

\begin{abstract}
Electronic Raman scattering couples to the charge excitations in the system, including the plasmons. However, the plasmon response has a spectral weight of $\sim q^2$, where $q$, the momentum transferred by light, is small. In this work, we show that in inversion-symmetry broken systems where Rashba type spin-orbit coupling affects the states at the Fermi energy (which is a known low energy effect) as well as the transition elements to other states (a high energy effect), there is an additional coupling of the plasmons to the Raman vertex, even at zero momentum transfer, that results in a spectral weight that is proportional to the spin-orbit coupling. The high energy effect is due to the breaking of SU(2) spin invariance in the spin-flip transitions to the intermediate state. We present a theory for this coupling near the resonant regime of Raman scattering and show that in giant Rashba systems it can dominate over the conventional $q^2$ weighted coupling. We also provide experimental support along with a symmetry based justification for this spin-mediated coupling by identifying a prominent $c$-axis plasmon peak in the fully symmetric $A_1$ channel of the resonant Raman spectrum of the giant Rashba material BiTeI. This new coupling could lead to novel ways of manipulating coherent charge excitations in inversion-broken systems. This process is also relevant for spectroscopic studies in ultrafast spectroscopies, certain driven Floquet systems and topologically non-trivial phases of matter where strong inversion-breaking spin-orbit coupling plays a role.
\end{abstract}

\maketitle

\section{Introduction}\label{Sec:Introduction}
\begin{figure}
\centering\captionsetup{justification=RaggedRight}
\begin{subfigure}{0.95\linewidth}
\includegraphics[width=\linewidth]{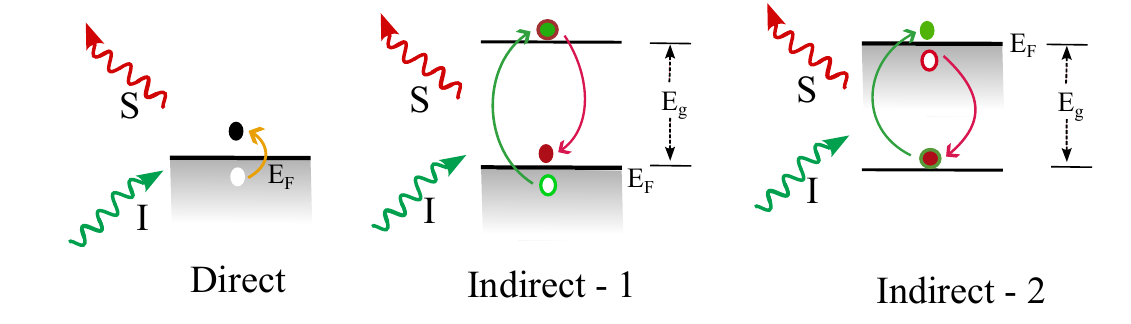}
    \caption{}
\end{subfigure}
\begin{subfigure}{0.95\linewidth}
    \includegraphics[width=\linewidth]{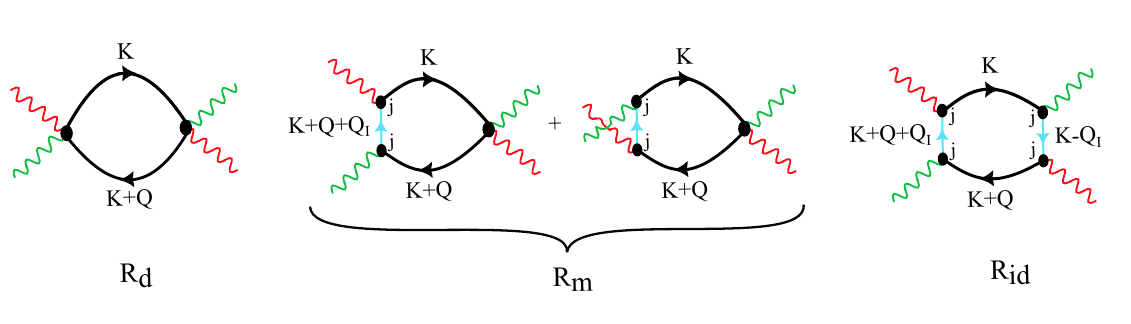}
    \caption{}
\end{subfigure}
\caption{(a) The direct and indirect Raman processes. (b) The response consists of contributions from the direct ($R_d$), mixed($R_m$), and indirect($R_{id}$) bubbles. The blue line indicates propagation in the intermediate state whereas the black lines are the particle and hole propagation near the Fermi level.}
\label{fig:RamanDiagrams}
\end{figure}

Raman scattering has long been used to study charge fluctuations in crystalline systems~\cite{Abstreiter1984,BookHayes2012,placzek1959rayleigh,Bairamova1993,Ivchenko2004}. The most commonly probed Raman excitations are optical phonon modes in solids and other vibrational modes of molecules. Electronic Raman Spectroscopy (eRS) that studies the charge dynamics of purely electronic degrees of freedom is relatively more challenging. In single band systems, the response is weighted by $q^2$, where $\bq$ is the momentum transferred by light scattering and usually corresponds to the long wavelength ($q\rightarrow 0$) regime in solids. In multi-band systems where the response could be viewed as a sum of independent contributions from each band, the same problem persists. In the literature this $q^2$ suppression has been attributed to charge conservation and/or to Coulomb screening. To overcome the smallness of the response due to $q^2$, one might turn towards \textit{resonant} eRS where the incoming laser light is made to resonate between states at the Fermi level and those at a higher or lower energy band (see Fig. \ref{fig:RamanDiagrams}). While this resonance boosts the overall signal for the electronic excitations near the Fermi surface\cite{Bairamova1993}, the strength is still difficult to detect. 

Updating the above description is one of the focuses of this work. What is known is that the eRS response from a spin-degenerate single band system consists of the incoherent excitations from the two-particle continuum of particle-hole (ph) excitations and coherent excitations from a collective mode, the plasmon, at energies \textit{above} the ph continuum. Both these excitations couple to eRS in the fully symmetric channel (which is captured by the parallel-polarization set-up of an eRS experiment) and are weighted by $q^2$. This channel couples to excitations that preserve all the crystal symmetries. Resonant eRS can amplify signals from both these contributions\cite{pinczuk1971resonant,pinczuk1989large,goni1991onedimensional}. In the presence of spin-splitting of the bands, which can be brought about due to the Zeeman effect in an external field or via the Rashba effect in inversion broken systems, the interband excitations also couple to eRS, but without the $q^2$ factor. However, these excitations involve spin-flips and are only accessible in the anti-symmetric channel (which is odd under swapping the incident and scattered polarizations and can be captured in the cross-polarization set-up of the experiment)\cite{lee2022chiral}. The response in the fully symmetric channel, however, is still weighted by $q^2$ even in the presence of spin splitting.

In this work we provide evidence for a novel coupling channel to plasmonic excitations in systems with broken inversion symmetry. Such systems have large spin-orbit coupling (SOC) of orbital origin which break the SU(2) spin invariance. In the presence of such a SOC, there are two effects that arise: (i) the usual spin splitting in the conduction band where the Fermi level lies; and (ii) the SOC induced spin-dependent modification of the interband transitions to and from the Fermi level to the intermediate states. Earlier studies on the effects of SOC on collective modes and their eRS signatures mostly focused on (i) which led to novel responses in the spin-sector in the form of new chiral-spin waves\cite{Perez2009,Baboux2013,Baboux2015,Perez2016, Kung2017,lee2022chiral}, but (ii) was not explored to our knowledge, which is what we do here. We substantiate our theory with support from the experimental observation of a prominent collective mode peak in the polar semiconductor, BiTeI, in the fully symmetric channel of the resonant eRS spectrum. The collective mode frequency and its carrier concentration dependence ascertain that this mode is the $c$-axis plasmon of the system. BiTeI is a giant Rashba material with the SOC parameter $\sim 4$ eV-\AA~and is about two orders of magnitude larger\cite{ishizaka2011giant,chen2013discovery,sakano2013strongly} than those in 2D electron gases in heterostructures where it is usually induced by gating and/or structural asymmetry of the interfaces.

In particular, we show that in resonant eRS, the orbital Rashba SOC between the conduction band states and the intermediate states (a high energy effect) generates an additional coupling to the plasmons, or any charge excitations in general, via a spin-charge susceptibility of the system which is also known to be induced (a low energy effect) due to Rashba SOC\cite{Maiti2015}. The spectral weight in this additional channel does not require $q$ to be finite but is controlled by a factor $E_{SOC}^2/E_F^2$, where $E_{SOC}$ is the SOC energy which is $\sim\lambda_{SOC}k_F$ (where $\lambda_{SOC}$ is the spin-orbit coupling constant with units of velocity and $k_F$ is the Fermi momentum), and $E_F$ is the Fermi energy. In usual semiconductors, this ratio is small ($\sim 10^{-2}$) [see e.g. Table I of Ref. \cite{Maiti2016}] and hence the effect was likely never observed. However, in giant Rashba systems including BiTeI, $E^2_{SOC}/E^2_F\sim 1$ and thus would a visible effect. This high energy effect is attributed to the breaking of SU(2) spin invariance in the spin-flip transitions between the Fermi level and intermediate bands. In fact, with regards to BiTeI, we argue that the direct observation of a $c$-axis plasmon in eRS is made possible due to a momentum-spin locking along the $c$-axis which would result in an out-of-plane canting of the spin states and a $c$-axis Drude-weight renormalization.

The discussion around coupling of charge fluctuations to eRS needs some consideration. For this reason, we organize the text as follows: in Section \ref{Sec:history} we summarize the current understanding of the eRS spectrum involving the charge and spin degrees of freedom, the $q^2$ suppression, and discuss the various effects of SOC that have been investigated using eRS. In Section \ref{Sec: Raman vertex}, for completeness, we review the formulation of the resonant Raman vertex and arrive at some known results. In Section \ref{Sec:SOC effect} we formulate, on general grounds, the origin of the non-$q^2$ coupling to the plasmon in SOC systems. In Section \ref{Sec:Graphene} we provide an explicit example (SOC in a Dirac system) for the high energy effect demonstrating the presence of the matrix elements that are responsible for the enhancement of the plasmon pole due to the spin-charge coupling. In Section \ref{Sec:BiTeI} we discuss the theoretical and experimental results for BiTeI, present some symmetry considerations, followed by some other predictions in regards to BiTeI. Finally, we conclude in Section \ref{Conclusion and outlook} where we contemplate the implications of this result to other systems. In the appendices we present details of the derivation of the resonant vertex, computation of the eRS for various polarizations, and the details of the toy models used in this work.

\section{eRS from charge and spin excitations}\label{Sec:history}
The electronic response to light can be modeled as a correlation between Raman vertices computed at the momentum $\bq$ and frequency $\Omega$ transferred to the system\cite{Devereaux2007}. The Raman vertices are density-like vertices dressed with factors arising from the direct (popularly called non-resonant) and indirect (which can be pre-resonant or resonant) Raman processes (see Fig.~\ref{fig:RamanDiagrams}). 
\paragraph{The $q^2$ factor:} In the non-resonant case, the entire Raman vertex (with both direct and indirect processes) is approximated by an effective-mass vertex which can be reduced to the usual density vertex but modulated with form factors consistent with the irreducible representations of the lattice. In this approximation, if there are poles in the associated correlation function, they manifest themselves as  $W_{q}^{2} / (\Omega^{2} - \Omega_{q}^{2})$, where $\Omega_{q}^{2} = v^2q^2$ for acoustic modes and $\Omega_{q}^{2} = \Omega_{0}^{2} + O(q^2)$ for optical modes, with $v$ being the mode velocity, $\Omega_0$ being the mode mass, and $W_{q}$ representing the spectral weight. In the non-interacting case, $W_q\propto\Omega_q$ and thus, the acoustic modes have a spectral weight $\propto q^2$. This is why in optical experiments, where $aq\ll1$ ($a$ being the lattice constant of the material), acoustic modes don't have noticeable spectral weights. 

In the presence of interactions, the pole at $\Omega_{q}$ is renormalized to $\Omega_{q}^{*}$, while $W_{q}$ could be renormalized differently. For acoustic modes in charge neutral systems, short range interactions renormalize $v$ (which is usually related to the Fermi velocity), thereby changing the correlation function to $q^2/(\Omega^2-v^{*2}q^2)$. This is ultimately what is responsible for zero sound in liquid He-3\cite{LL_StatPhys2,Aldrich1976}. In charged systems, however, long range interactions in the form of the unscreened part of the Coulomb interaction are also present, and are typically accounted for within the random-phase-approximation (RPA)\cite{Bohm1951_3}. The Coulomb renormalization due to RPA has two important consequences for eRS: it affects the Raman response only in the fully symmetric channel; and it shifts the acoustic poles to either optical bulk plasmons in 3D or to `super-acoustic' ($\Omega_{q} \sim q^{1/2}$) ones in 2D, without affecting $W_q^2$ factor which remains $\propto q^2$. Thus, the bulk plasmon in 3D systems, although technically an optical mode, still couples to the Raman process in the acoustic sense. In some sense, the plasmon is an acoustic mode that is singularly renormalized due to the Coulomb interaction. This renormalization is similar to the Anderson's mechanism in superconductors where the Goldstone mode expected from the spontaneously broken symmetry gets lifted to the bulk plasmon\cite{Anderson1963}.

It is sometimes presented that the $q^2$ factor arises from charge conservation. This association is derived from the fact that for the fully symmetric channel in the non-resonant limit the Raman vertex reduces to the density vertex. In this case, the Raman correlation functions capture the density fluctuations. In the limit of $\bq\rightarrow 0$ and finite $\Omega$ one is then probing the changes in total density, which is conserved. Hence, the correlation function and the corresponding response is zero. This argument also holds in the presence of interactions as charge conservation is still preserved. However, this argument cannot be directly extended to the resonant eRS as the Raman vertex does not reduce to the density vertex. Nevertheless, the response can still be proportional to $q^2$ as will be shown below.

\paragraph{Finite charge response in eRS.} In single band, spin-degenerate systems, only acoustic modes (which includes plasmons) are possible and hence $\{\Omega_q,W_q\}\propto q^2$. However, if there could be two bands or if the spin degeneracy could be lifted, the interband transitions would allow for modes of the form $\Omega_q^2=\Omega_0^2+\mathcal{O}(q^2)$ (along with the two-particle continuum of incoherent excitations). The eRS spectrum in such cases would not suffer from the $q^2$ suppression discussed above. Historically, the eRS observable interband effects could only be observed in either semiconductors in magnetic field (transition between Landau levels due to the orbital effect of the field\cite{Pinczuk1992} and between the spin-split bands due to the Zeeman-effect\cite{Perez2009}), or in superconductors\cite{Klein1981,Klein2010}(excitations in the particle-hole symmetric Bogoliubov bands). Interband excitations in most other materials are usually of the order of eV which is beyond the typical energy scales for the Raman shifts, and hence not observed. Note that the \textit{total} charge is still conserved in multiband systems, but we get a response that is not suppressed by $q^2$ and electronic correlations often play an important role in causing the spectral weight to be observable. This is already known and understood in multiband superconductors as the Leggett response\cite{Blumberg2007,Klein2010,Cea2016,Maiti2017b}. Thus, the association of charge conservation and $q^2$-suppressed eRS is only of significance in single band systems (or systems modelled as copies of single band systems) for the fully symmetric Raman channel.

In this context, it is relevant to discuss the result of Ref. \cite{Sarma1999}, where there was no consideration of SOC and it showed that accounting for the resonant processes boosted the signal from the ph continuum of excitations in addition to the plasmon. This is relevant as it seemed to explain the results of experiments (in 1D systems) with two comparable peaks, where one corresponded to the plasmon and the other was attributed to the ph excitation\cite{Goni1991}. The conventional theory involving RPA failed to explain the comparable weights of the ph excitations and the plasmon. While Ref. \cite{Sarma1999} provided an explanation for the comparable weights of those excitations, it is important to note that one needed finite $q$ to get the response itself as the overall result still has to be proportional to $q^2$. As we will demonstrate, our results for the eRS in an inversion-symmetry broken system remains finite even in the $q\rightarrow 0$ limit, and hence it really provides a novel coupling channel to the plasmon. It is also relevant to reiterate here that we are only interested in purely electronic mechanisms to couple to plasmons. See Ref. \cite{lee2023resonant} that describes other ``direct" means of plasmon observation, which still require aide of phonons or some incoherent superposition of many $q$ excitations.

\paragraph{Finite spin response in eRS.} When one accounts for the indirect-terms in the Raman vertex, one is able to also couple to the spin-flip excitations in the cross-polarization set-up of the eRS, creating an effective spin vertex (e.g. in III-V semiconductors\cite{Abstreiter1984,Bairamova1993,Ivchenko2004} and perovskite semiconductors\cite{Rodina2022}). This effective spin-vertex was later reformulated for a square-lattice Hubbard model wherein it was pointed out that the indirect Raman processes could be used to extract the excitations in the antisymmetric channel\cite{Shastry1990,Shastry1991} (which only contained information about the spin-flip excitations in the system). This property is what allowed the study of spin collective modes in strongly correlated systems (magnons)\cite{Devereaux2007}, in semiconductors (Silin mode, at the Larmour frequency, and other spin-flip excitations)\cite{Perez2009,Baboux2013,Baboux2015,Perez2016}, in topological insulators (chiral-spin modes)\cite{Kung2017}, and in the heavy fermion superconductor URu$_2$Si$_2$ (chiral density wave)~\cite{Buhot2014,kung2015chirality,Kung2016}. 

\paragraph{Previous works on the effect of SOC on resonant eRS.} Of the two effects of SOC on eRS discussed in the introduction, the theory for effect (i) which leads to new collective mode behavior is well understood\cite{Shekhter2005,Ashrafi2012,Maiti2015,Kung2017,raghu2010collective,baboux2012giant}. The effect on resonant eRS has also been investigated theoretically but only in the context of the usual spin coupling of the Raman vertex. That is, the models were such that the coupling of light to the spin vertex was not modified but the spin-fluctuations in the system were\cite{Vitlina2012,Maiti2017}. In such treatments, the resonant processes only trivially enhanced the spin susceptibility via the factor $\sim\frac{1}{(\Omega-\Omega_I)^2+\Gamma^2}$ which was derived for a Kane model~\cite{Kane1957,Ivchenko2004} for III-V semiconductors in cubic systems. Other works that accounted for indirect processes in the eRS were formulated for Graphene (without SOC) theoretically~\cite{Kashuba2009,Heller2016} and experimentally \cite{Riccardi2016}. But none of the works explored the effect (ii) mentioned in the introduction. 

\section{The Resonant Raman scattering cross-section.}\label{Sec: Raman vertex}
We shall model the eRS response in equilibrium where the cross-section per unit solid angle $d\mathcal{O}$ per unit scattered energy $d\Omega_S$ is given by\cite{cardona1983lightscattering,Shastry1990,Shastry1991,Devereaux2007}
\bea\label{eq:cross section}
\frac{d^2\sigma}{d\mathcal{O} d\Omega_S}&=&r_0^2\frac{\Omega_S}{\Omega_I}\left[1+n_B(\Omega)\right] R(\Omega,\bq),
\eea
where $r_0$ is the classical radius of the electron, $\Omega_{I,S}$ are the frequencies of incident and scattered light with the Raman shift $\Omega=\Omega_I-\Omega_S$, $n_B(\Omega)$ is the Bose-Einstein distribution function, and $R(\Omega,\bq)$, which has dimensions of density of states, is computed from the branch cuts of the analytic continuation of\cite{Shvaika2005}
\beq\label{eq:corr}
\chi(Q)=\int_0^\beta d\tau e^{i\Omega_m\tau}\langle T_\tau\hat\gamma_{\bq}(\tau)\hat\gamma_{-\bq}(0)\rangle,
\eeq
where $Q\equiv (i\Omega_m,\bq)$ with $\Omega_m$ being the bosonic Matsubara frequency, $\beta=1/k_BT$, and $\hat\gamma_\bq(\tau)$ is the Heisenberg evolution of $\hat\gamma_\bq$ in imaginary time. Specifically, $R(\Omega,\bq)={\rm Im}\chi(\Omega+i\eta,\bq)$. Further, $\hat\gamma_{\bq}=\sum_\bk\gamma^{nn'}_\bk\hat c^\dag_{n,\bk+\bq/2}\hat c_{n',\bk-\bq/2}$, where (repeated indices are summed)
\bea\label{eq:Raman vertex}
    &&\gamma^{nn'}_\bk=\epsilon^I_\alpha\gamma^{nn'}_{\alpha\beta}(\bk)\epsilon^S_\beta,\nn\\
    &&\gamma^{nn'}_{\alpha\beta}(\bk)=\underbrace{\delta_{\alpha\beta}\delta_{nn'}}_{\rm direct(d)}\nn\\
    &&+m_e\underbrace{\sum_m\left(\frac{[j^S_\beta]_{nm}[j^I_\alpha]_{mn'}}{\varepsilon_n-\varepsilon_m+\Omega_I-i\Gamma}+\frac{[j^I_\alpha]_{nm}[j^S_\beta]_{mn'}}{\varepsilon_n-\varepsilon_m-\Omega_S-i\Gamma}\right)}_{\rm indirect(id)}.\nn\\
\eea
The $\hat c^\dag_{n,\bk}, \hat c_{n,\bk}$ correspond to the creation and annihilation operators for states in band $n$ with momentum $\bk$. The direct and indirect labels refer to the processes shown in Fig. \ref{fig:RamanDiagrams}, $\alpha,\beta\in\{x,y,z\}$ (the spatial components of the polarization vectors $\be^I$ and $\be^S$). $\ve_n$ are the eigenstates of the system, and the labels $n,n',m\in$ Hilbert space of the system in the eigenbasis. $\Gamma$ encapsulates the lifetime effects of the intermediate states. The operator $\hat j_{\alpha}\equiv \frac1e\frac{\partial\hat H}{\partial A_\alpha}|_{\mathbf A=0}$, where $\mathbf A$ is the vector potential\footnote{Note that this derivative is carried out not in the eigenbasis, but in the basis where the $\mathbf A$ is covariantly introduced to the system. At the level of the tight-binding model, this is usually done via Peierl's substitution. It can also be done by $\bk\rightarrow \bk+e\mathbf A$ in effective Hamiltonians. It is not guaranteed that the different prescriptions would agree.}, and $m_e$ is the effective electron mass for Hamiltonians with parabolic dispersion and the bare electron mass for those with linear dispersion (in which case there would also be no direct term). The two terms in Eq. (\ref{eq:Raman vertex}) lead to three types of terms in $R(\Omega,\bq)$\cite{Shvaika2005}:
\beq\label{eq:ramanterms}
R(\Omega,\bq)=R_d(\Omega,\bq)+R_m(\Omega,\bq)+R_{id}(\Omega,\bq),
\eeq
where $R_{d}(\Omega,\bq)$ is the contribution from the correlations between the direct processes ($\sim\langle\gamma_{d}\gamma_d\rangle$), $R_{m}(\Omega,\bq)$ is the mixed contribution $\sim\langle\gamma_d\gamma_{id}\rangle$, and $R_{id}(\Omega,\bq)$ is from $\langle\gamma_{id}\gamma_{id}\rangle$. 

\paragraph*{Note:} The non-resonant contribution arises from \textit{both} direct and indirect processes while the resonant contribution arises only from indirect processes. In a field theory sense, the contributions from indirect processes can be further classified into on-shell and off-shell contributions. The former refers to transitions between the real energy states of the system, while the latter are virtual processes. These virtual processes could be interband or intraband in nature. In the absence of any on-shell contributions, the effects of off-shell contributions become relevant (see e.g \cite{Shvaika2005,Kashuba2009,Heller2016} where such terms were considered in non-semiconducting systems). In resonant eRS one usually focuses on the on-shell contributions. We refer the reader to Appendix \ref{app:1} for a more thorough discussion.

In this work, we focus on the on-shell contribution from the $R_{id}$ term. This is formally achieved by approximating $\gamma^{nn'}_{\alpha\beta}\approx m_e\sum_m\frac{[j^S_\beta]_{nm}[j^I_\alpha]_{mn'}}{\varepsilon_n-\varepsilon_m+\Omega_I-i\Gamma}+\frac{[j^I_\alpha]_{nm}[j^S_\beta]_{mn'}}{\varepsilon_n-\varepsilon_m-\Omega_S-i\Gamma}$. The electronic structure is usually such that resonance is only possible in either the term containing $\Omega_I$ (resonance with bands above Fermi level) or the one containing $\Omega_S$ (resonance with bands below). Without loss of generality, we retain the former (with $\Omega_I$). Further, if the gap, $E_g$, between the Fermi surface states and those in the intermediate band is large, the dispersion of the intermediate states around the $k_F$-wavevectors could be ignored. We can then write $\gamma^{nn'}_{\alpha\beta}\approx m_e\sum_m\frac{[j^S_\beta]_{nm}[j^I_\alpha]_{mn'}}{\Omega_I-E_g-i\Gamma}$. In this approximation, 
\bea\label{eq:Raman susceptibilty}
\chi(Q)&\approx&-\Pi_{\alpha\beta;\gamma\delta}(Q)\e^I_\alpha\e^{S*}_\beta\e^{I*}_\gamma\e^{S}_\delta,\nn\\
\Pi_{\alpha\beta;\gamma\delta}(Q)&\equiv&\int_K{\rm Tr}\Big[\hat \gamma_{\alpha\beta}({\bk})\hat G_K\hat \gamma_{\gamma\delta}({\bk})\hat G_{K+Q}\Big],
\eea
where $K\equiv (i\omega_n,\bk)$ with $\omega_n$ being the fermionic Matsubara frequency, and $\int_K\equiv T\sum_n\int_{\bk}$. The hat represents a matrix structure in the subspace containing only the Fermi surface states, and $\hat G_K$ represents the Green's function in the same subspace. Note that the transition operators $\hat j_\alpha$ are defined in the Hilbert space made of both the Fermi surface states and the intermediate states. However, the resonance condition allows us to pick the dominant contributions which effectively factor out the Hilbert subspace associated with the intermediate states from those at the Fermi level. The steps demonstrating this are detailed in Appendix \ref{app:2}.

\begin{figure}
\centering\captionsetup{justification=RaggedRight}
\includegraphics[width=1.0\linewidth]{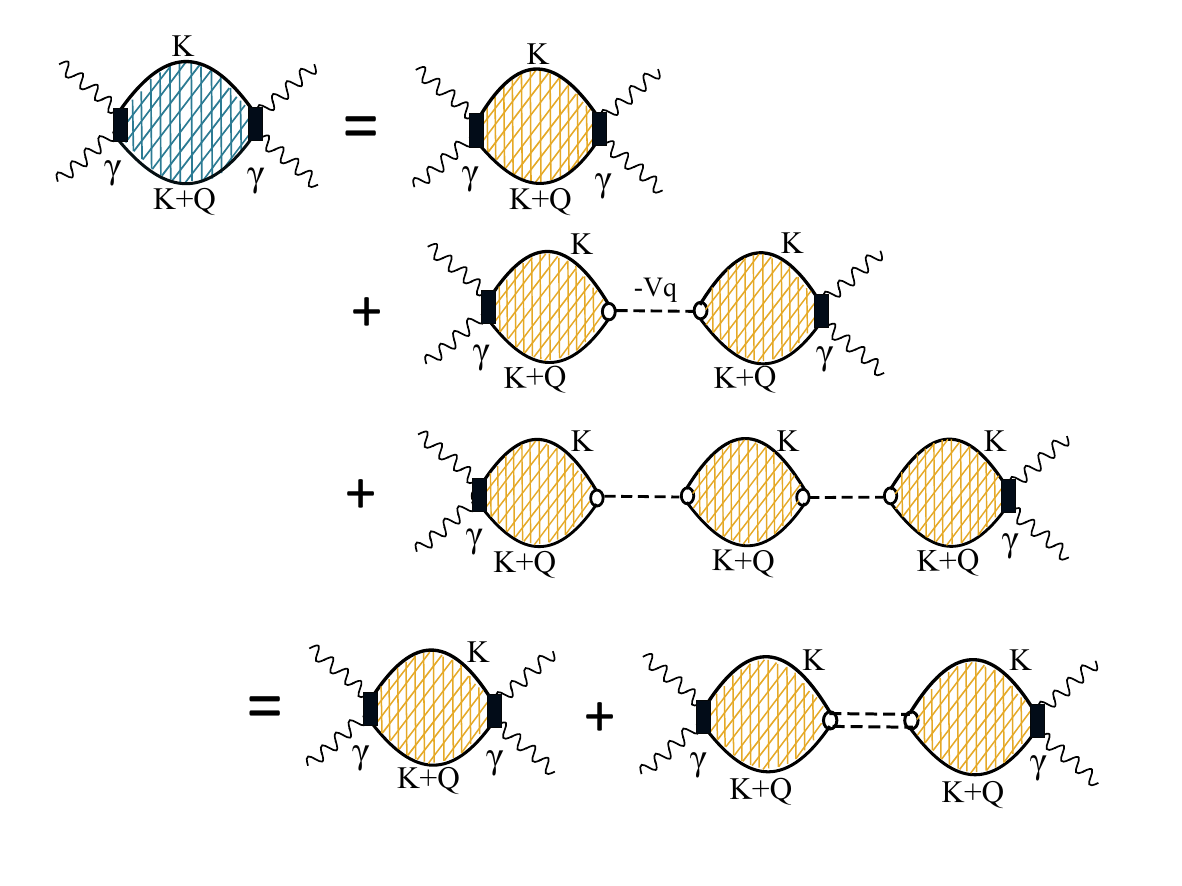}
\caption{Renormalization from the unscreened Coulomb interaction $V_{\bq}$ in the RPA scheme. The shaded yellow bubbles indicate vertex corrections which are ignored in this work. The doubled line is the RPA-renormalized Coulomb interaction.}
\label{fig:CoulombRenorm}
\end{figure}

\paragraph{Interaction renormalization.} In general, interactions renormalize the three terms $R_{d,m,id}$ differently, and the renormalization also varies with the choice of polarizations of the incoming and scattered light. Furthermore, the short-range part of the interaction predominantly leads to vertex corrections (which are captured in the ladder approximation in a diagrammatic approach), while the long-range part due to the unscreened part of Coulomb interaction leads to additional contributions which are captured via the random-phase-approximation (RPA) (see Fig. \ref{fig:CoulombRenorm}). This leads to
\bea\label{eq:dressed Raman susceptibilty}
\Pi_{\alpha\beta;\gamma\delta}(Q)&=& \Pi^V_{\alpha\beta;\gamma\delta}(Q)+\frac{\Pi^V_{\alpha\beta;0}(Q)V_\bq\Pi^V_{0;\gamma\delta}(Q)}{1-V_\bq\Pi^V_{00}(Q)},
\eea
where $\Pi^V$ denotes the vertex corrected form of the bare $\Pi$ (which are described below), and $V_{\bq}=e^2/\e_0q^2$ (the Fourier transform of the Coulomb interaction in 3D). $\Pi^V$ leads to either spin collective modes\cite{Maiti2015} or other acoustic modes. If we wish to focus on the plasmons, we can drop the vertex corrections in all the terms in Eq. (\ref{eq:dressed Raman susceptibilty}), and can identify the various $\Pi$'s as:
\bea\label{eq:RPA Raman susceptibility}
    \Pi_{\alpha\beta;0}(Q)&=&\int_K{\rm Tr}[\hat \gamma_{\alpha\beta}\hat G(K)\hat\sigma_0\hat G(K+Q)],\nn\\
    \Pi_{0,\alpha\beta}(Q)&=&\int_K{\rm Tr}[\hat \sigma_0\hat G(K)\hat \gamma_{\alpha\beta}\hat G(K+Q)],\nn\\
    \Pi_{ij}(Q)&=&\int_K{\rm Tr}[\hat \sigma_i\hat G(K)\hat\sigma_j\hat G(K+Q)],
\eea
where $\hat\sigma_{1,2,3}$ are the spin vertices and $\hat \sigma_0$ is the Coulomb vertex in the reduced Hilbert space of states at the Fermi-surface. $\Pi_{ij}(Q)$ is usually referred to as the generalized susceptibility of a system. The denominator in Eq. (\ref{eq:dressed Raman susceptibilty}) is the same one that appears in the renormalization of the charge susceptibility and contains the plasmon pole of the system. Note that the plasmon pole couples to the Raman response via $\Pi^V_{\alpha\beta;0}$ and $\Pi^V_{0;\gamma\delta}$, which are sensitive to the choice of polarization vectors of the eRS experiment. Since we wish to work with frequencies close to the plasmon pole, we could further write
\bea\label{eq:dressed Raman susceptibilty2}
\Pi_{\alpha\beta;\gamma\delta}(Q)&\approx& \frac{\Pi_{\alpha\beta;0}(Q)V_\bq\Pi_{0;\gamma\delta}(Q)}{1-V_\bq\Pi_{00}(Q)},\nn\\
&=&\frac{e^2}{\e_0q^2}\frac{\Pi_{\alpha\beta;0}(Q)\Pi_{0;\gamma\delta}(Q)}{1+\frac{\Omega^2_{\rm pl}}{\Omega_m^2}},
\eea
where the plasma frequency is given by $\Omega_{\rm pl}^2=e^2\nu_Fv_F^2/3\e_0$. In semiconductors the $\ve_0$ should be replaced by $\ve_\infty$ to account for the background charges. Here we have used the fact that $\Pi_{00}(Q)=-\nu_Fv_F^2q^2/3\Omega_m^2$, where $\nu_F$ is the density of states at the Fermi level and $v_F$ is the fermi velocity (which are to be seen as Fermi surface averages in anisotropic systems).

\begin{figure}
\centering\captionsetup{justification=RaggedRight}
\includegraphics[width=\linewidth]{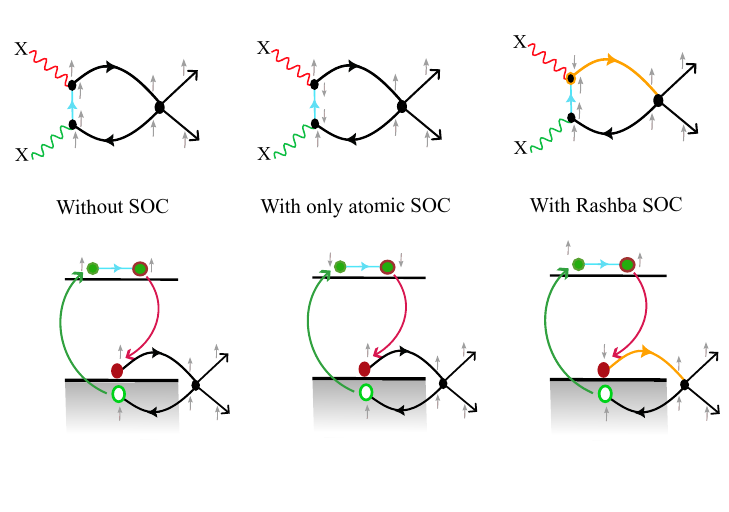}
\caption{Schematic view of the polarization-charge bubble which represents the final state renormalizations by particle-hole propagation near the Fermi level in the fully symmetric channel of eRS. In the cases without SOC and with atomic SOC the final state is dressed by the usual charge-like bubble (black propagators) with a weight of $q^2$. Broken-inversion contributes additional spin-flip processes (as yellow vertices and propagators) that provide a stronger coupling to the plasmons in the fully symmetric channel.}
\label{fig:Schematic}
\end{figure}

\paragraph{Known results.} For the plasmon pole to show up in the response, one needs $\Pi_{\alpha\beta;0}$ and $\Pi_{0;\gamma\delta}$ to be non-zero. Diagrammatically, this object is a bubble with the $\alpha\beta$-vertex on one side and the charged vertex on the other. We may call this the \textit{polarization-charge} bubble. This is a central object in our calculations and is composed of (see Fig. \ref{fig:Schematic}) three propagators: the first one reflects propagation in the intermediate state after the photon excitation, the other two reflect the propagation of the created particle-hole pair at the Fermi level after photon emission. This bubble is unique to the contribution in the symmetric channel of eRS. We summarize the computation of this bubble in Appendix \ref{app:3} but the end result is that 
\begin{eqnarray}\label{eq:Pi_gamma_final}
\Pi_{\alpha\beta;0}(Q)&\approx&\frac{m_e\mathcal{F}^{\beta\alpha}_i\Pi_{i0}(Q)}{\Omega_I-E_g-i\Gamma},
\end{eqnarray}
where $i\in\{0,1,2,3\}$, and $\mathcal F^{\beta\alpha}_i$ is a \textit{polarization-factor} that couples the charge-charge($i=0$) and spin-charge($i=1,2,3$) susceptibilities to the Raman vertex and has units of $v_F^2$. It is easy to show that $\Pi_{0;\gamma\delta}(Q)$, within the same approximations, has $i\Gamma\rightarrow-i\Gamma$. The computation of the polarization factor is sensitive to the choice of polarizations of light and is central to our problem. In the parallel polarization set-up, $\alpha\beta=XX$ (or $YY$), and $\mathcal F^{XX}_0\neq0$, while $\mathcal F^{XX}_i=0$ for $i\in\{1,2,3\}$. Thus, the charge-charge susceptibility [$\Pi_{00}(Q)$], which is renormalized by the unscreened Coulomb interaction, is picked up in the parallel polarization set-up. On the other hand, for a cross-polarization set up where $\alpha\beta=XY$, we have $\mathcal F^{XY}_0=0$, while some of $\mathcal F^{XY}_i\neq0$ for $i\in\{1,2,3\}$. Thus, the cross-polarization set-up couples to the spin-charge susceptibility of the system. The $XY$ setup can also couple to the spin-spin susceptibility in the non-RPA channel via vertex corrections, but those diagrams are presently not in consideration and have been a subject of previous studies\cite{Vitlina2012,Maiti2017}. Further, the spin-charge susceptibility itself is zero when there is spin-charge separation. This is why we can usually only couple to the charge-charge susceptibility $[\Pi_{00}(Q)]$ in the $XX$ set-up leading to the conventional result:
\bea \label{eq:dressed Raman susceptibilty3}
\Pi_{\rm \alpha\alpha;\alpha\alpha}(Q)&=& \frac{v_F^2q^2}{\Omega_m^2+\Omega_{\rm pl}^2}\frac{\nu_F(m_e\mathcal F_0^{\alpha\alpha})^2}{(\Omega_I-E_g)^2+\Gamma^2},
\eea
where $\alpha$ could be X or Y. The circular polarizations of eRS can be constructed out of polarization vectors $\e^X\pm i\e^Y$. Despite the enhancement from the interband resonance, the plasmon pole in parallel polarization setup is still weighted by $q^2$. 

If we were to account for spin-splitting of the Fermi surface states either due to the Zeeman effect or due to the Rashba effect, the charge susceptibility would still be $\sim q^2$\cite{Maiti2015} and thus would not really affect the result in Eq. (\ref{eq:dressed Raman susceptibilty3}). The presence of Rashba SOC, however, modifies the spin-charge susceptibility and $\mathcal F$ in a characteristic manner. Formulating this is the subject of the next section. This modification will not be present when the spin splitting is induced by the Zeeman effect.

\section{Novel effect due to SOC}\label{Sec:SOC effect} 
It is clear from Eq. (\ref{eq:dressed Raman susceptibilty2}) that the plasmon coupling to the Raman vertex is controlled by the polarization-charge bubble $\Pi_{\alpha\beta;0}(Q)$, which near resonance can be approximated by Eq. (\ref{eq:Pi_gamma_final}). Thus, the computation of the polarization factors and the generalized susceptibilities is crucial. From Appendix \ref{app:3} we learn that the term $j_\beta j_\alpha$ in the Raman vertex [Eq. (\ref{eq:Raman vertex})] is only evaluated between states where the Fermi level lies (say the $v$ states) and intermediate states (say the $c$ states), and can be expressed, on general grounds, as
\beq\label{eq:deff2}\Big[j_{\beta,vc}j_{\alpha,cv}\Big]=\mathcal{F}^{\beta\alpha}_i(\bk)\hat\sigma_i.\eeq
This is where the polarization factors are introduced. It will be more instructive to actually arrive at this form by modelling the matrix elements of the transition elements $j_{\alpha,cv}$ themselves. Without loss of generality, we can say
\begin{equation}\label{eq:current}
\hat j_{\alpha,cv}=p^\alpha_0\hat\sigma_0+p^\alpha_1\hat\sigma_1+p^\alpha_2\hat\sigma_2+p^\alpha_3\hat\sigma_3,
\end{equation}
where $p^\alpha_0$ corresponds to the average of the two spin non-flip transitions, $p^\alpha_3$ corresponds to the differences in the two spin non-flip transitions, $p^\alpha_1$ and $p^\alpha_2$ correspond to the average and difference of the spin-flip transitions. This leads to 
\begin{eqnarray}\label{eq:f}
&&\hat j_{\beta,vc}\hat j_{\alpha,cv}\nn\\
&=&(\underbrace{p^\beta_0p^\alpha_0+\vec p^\beta\cdot\vec p^\alpha}_{\mathcal F_0})\hat\sigma_0+(\underbrace{[p^\beta_0\vec p^\alpha+p^\alpha_0\vec p^\beta]+i\vec p^\beta\times\vec p^\alpha}_{\vec{\mathcal F}})\cdot\hat{\vec \sigma}\nn\\
     &=&\mathcal F^{\beta\alpha}_0\hat\sigma_0+\mathcal F^{\beta\alpha}_1\hat\sigma_1+\mathcal F^{\beta\alpha}_2\hat\sigma_2+\mathcal F^{\beta\alpha}_3\hat\sigma_3\nonumber\\
     &\equiv&\mathcal F^{\beta\alpha}_i\hat\sigma_i.
\end{eqnarray} 

In systems with inversion and time reversal symmetry, parity is conserved and the SU(2) invariance of the spins is maintained which leads to the transition elements being such that either only $p^\alpha_0$ is present (when there is no atomic SOC splitting), or only $p^\alpha_{1,2,3}$ are present and equal (in the presence of atomic SOC splitting).  This means that $\mathcal{F}_0^{\beta\alpha}=p^\beta_0p^\alpha_0$ or $\vec p^\beta\cdot \vec p^\alpha$, and $\mathcal{F}_i^{\beta\alpha}=i\vec p^\beta\times \vec p^\alpha$. Observe that if $\alpha=\beta$ only $\mathcal F_0$ component contributes and couples the Raman vertex to the charge-charge susceptibility of the system near resonance. Whereas if $\alpha\neq\beta$, then $\mathcal{F}_i$ components are non-zero and couples the Raman vertex to the spin-charge susceptibilities of the system. However, in systems with inversion, these susceptibilities are zero and the plasmon pole can only couple to the Raman vertex through the $\mathcal{F}_0$ factor and hence the charge-charge susceptibility, leading to Eq. (\ref{eq:dressed Raman susceptibilty3}). 

In inversion-broken systems, parity is broken which results in the following changes: (a) SU(2) invariance is broken which no longer constraints $p^\alpha_0$ and $p^\alpha_{i}$ as it did in the conventional case; (b) it allows for the coupling of spin and charge degrees of freedom which makes the spin-charge susceptibilities non-zero. The effect (a) leads to the important result that $\mathcal F_i^{\beta\alpha}$ now is no longer just the cross product of the $\vec p$ factors but also contains the $p_0\vec p$ terms as in Eq. (\ref{eq:f}). This means that if $\alpha=\beta$ and the cross product vanishes, there is still a contribution from the mixing of the $p_0$ and $p_i$ terms which was prevented when SU(2) invariance was preserved. This is an effect that happens at large energy scales. The effect (b) is a low energy effect that was explicitly shown in a 2D metal with Rashba type SOC\cite{Maiti2015} and is expected to hold in general. In fact, for a system with a spin-orbit velocity of $\lambda_{SOC}$, the form is expected to be $\Pi_{i0}(Q)\approx\nu_F\lambda_{SOC}q/\Omega_m$ for $i\in\{1,2,3\}$ (the form for $\Pi_{00}$ is still $q^2/\Omega_m^2$). These two effects lead to the fact that an effective spin vertex is created in the \textit{parallel} polarization geometry, which couples to the spin-charge susceptibility of the system, which is also rendered non-zero due to SOC. 

Plugging the form for $\Pi_{i0}$ we just discussed into Eq. (\ref{eq:dressed Raman susceptibilty2}) we get
\begin{eqnarray}\label{eq:fullresponse}
\Pi_{\alpha\beta;\gamma\delta}(Q)&\approx&\frac{e^2}{\epsilon_0q^2}\frac{m_e^2}{1+\frac{\Omega_{pl}^2}{\Omega_m^2}}\frac{\mathcal{F}^{\beta\alpha}_i\Pi_{i0}(Q)\mathcal F^{\delta\gamma}_j\Pi_{j0}(Q)}{(\Omega_I-E_g)^2+\Gamma^2}\nn\\
&\xrightarrow{XX}&\underbrace{\frac{\lambda^2_{\rm SOC}}{v_F^2}}_{\rm SOC~effect}\underbrace{\frac{\Omega_{pl}^2}{\Omega^2_m+\Omega_{pl}^2}}_{\rm plasmon~pole}\underbrace{\frac{\nu_F(m_e\mathcal F_0^{XX})^2}{(E_g-\Omega_I)^2+\Gamma^2}}_{\rm Res.~enhancement}\nn\\
&&+\mathcal{O}\left(\frac{v_F^2q^2}{\Omega_m^2+\Omega_{pl}^2}\right).
\end{eqnarray}
The regular $q^2$ weighted plasmon pole that arises from $\Pi_{00}\sim q^2$ contribution is still present, but appears as a correction to the SOC induced term. Comparing Eqs. (\ref{eq:dressed Raman susceptibilty3}) and (\ref{eq:fullresponse}) we see the that the ratio of the SOC induced term to the regular $q^2$-weighted term is $\sim[\lambda_{SOC}\Omega_{pl}/v_F^2q]^2\equiv r^2$. This is the main result of this article. Thus, if we want the SOC induced response to dominate we would need to increase $r$. In usual semiconductors (used in heterostructure quantum wells) we have $v_F\sim 10^4-10^5$ m/s, $\lambda_{SOC}\sim 10^3$ m/s for the Rashba parameters of about $10$ meV-\AA~and $\Omega_{pl}\sim 100$ meV \cite{Bercioux2015}. These numbers lead to $\lambda_{SOC}/v_F\sim 10^{-2}$, and $v_Fq\sim0.1-1$ meV (where $q$ is the wavenumber of light). This leads to $r\sim 1$-$10$. However, for giant Rashba systems $\lambda_{SOC}/v_F\sim 1$ leading to $r\sim 100$. Since the response is $\propto r^2$ one can have up to 4 orders of magnitude increase in the effect. It should be asserted that the coupling to the spin-charge susceptibility is a general result that is a consequence of inversion breaking SOC, but the particular form of the eRS response in the above form is valid only near the plasmon pole.


Note that to get this effect, it is important to have a spin-dependent correction to the transition elements $j_{cv}$. It is not sufficient to have spin-splitting in the intermediate states and the Fermi surface states. The SOC based modification of $j_{cv}$ is expected in the presence of inversion breaking SOC of the Rashba type that couples spin and momentum. This spin-momentum coupling is essential because the transition element is proportional to the interband component of the current operator. A spin structure to the current operator can only be introduced in the presence of Rashba type SOC. Zeeman like spin splitting will not achieve this. In fact, in the next section, we demonstrate this precisely for a simple toy model. Finally, we note that a Rashba coupling can also be externally introduced by gating 2D systems. But this effect usually manifests itself in the Rashba coupling of the quasiparticles at the Fermi surface. The intermediate states are unlikely to get coupled to the Fermi surface states. However, in the orbital Rashba systems, the Rashba coupling emerges from the inversion breaking in the unit cell and thus couples multiple bands.

\section{Toy model demonstrating the high energy effect: spin-sensitivity of the transition element}\label{Sec:Graphene} 
We now demonstrate a case where the proposed structure of $\mathcal F^{\alpha\beta}_i$ is induced due to Rashba type SOC. Consider the Dirac system: a hypothetical model of doped Graphene on a transition metal dichalcogenide such that the chemical potential is in one of the valence bands. The Hamiltonian is given by
\bea\label{hamil_gra}
\hat{\mathcal{H}}&=&v_F\left(\tau_z\hat s_0\hat \sigma_xk_x+\hat s_0\hat \sigma_y k_y\right)+\Delta\hat s_0\hat \sigma_z+\nn\\
&&\frac{\Delta_{{\text{R}}}}{2}\left(\tau_z\hat s_y\hat\sigma_x-\hat s_x\hat\sigma_y\right)+\frac{\Delta_{{\text{Z}}}}{2}\tau_z\hat s_z\hat\sigma_0,
\eea
where $\Delta$ is the charge gap, $\Delta_{\rm R}$ is the Rashba SOC energy and $\Delta_{\rm Z}$ is the Valley-Zeeman SOC energy~\cite{Wang2015,Cummings2017,Garcia2018}, $\hat s,\hat\sigma$ are the spin and sublattice matrices and $\tau_z=\pm1$ marks the valley index. This Hamiltonian is written in the sublattice basis $\{{a}_\uparrow,{a}_\downarrow,{b}_\uparrow,{b}_\downarrow\}$, where $a,b$ represent the two atoms of the unit cell. We refer to this as hypothetical because $\Delta$, in real graphene, is too small to resonate with visible light. However, we can artificially let $\Delta\sim$ frequency of visible light. We choose this model for it's analytical tractability and the fact that it allows for dipole-active transitions between the two bands with intraband and interband spin-splitting. Finally, since we will be interested in the resonant terms, the analysis will not be affected by the lack of the direct process in Dirac systems.

Let us first ignore the Valley-Zeeman coupling ($\Delta_Z=0$). In the resulting Hamiltonian we can perform $\bk\rightarrow\bk+e\mathbf A$, which leads to the definition of current to be $\hat{\mathbf j}=v_F\tau_z\hat s_0\hat{\boldsymbol{\sigma}}$ in the $a$-$b$ basis. To proceed with our calculations, we need to move into the eigenbasis denoted by $\{{c}_\uparrow, {c}_\downarrow,{v}_\uparrow,{v}_\downarrow\}$. However, instead of solving the problem exactly, it is sufficient to tackle the problem perturbatively in $\Delta_R$. Upon transforming $\mathbf{j}$ to the eigenbasis, we show in Appendix \ref{app:Graphene} that $\hat{\mathbf{j}}$ acquires a spin-dependence in the interband sector.

To get an intuitive understanding of how this dependence arises, consider an intermediate basis where there is no SOC. This basis was used in Ref. \cite{Kumar2021}. In this basis the Hamiltonian (with SOC) takes the form
\begin{eqnarray}
H=\left(\begin{array}{cc}
H^{cc}&H^{cv}\\
H^{vc}&H^{vv}\end{array}\right),
\end{eqnarray}
where 
\bea\label{eq:Hdetails}
H^{cc}&=& \hat{s}_0\xi_k+\frac{\Delta_Rv_F}{2\xi_k}(\bk\times\hat{\mathbf{s}})\cdot\hat{z},\nn\\
H^{vv}&=&-\hat{s}_0\xi_k-\frac{\Delta_Rv_F}{2\xi_k}(\bk\times\hat{\mathbf{s}})\cdot\hat{z},\nn\\
H^{cv}&=&\frac{\Delta_R}{2}\left(\frac{i\xi_kk_x-\Delta k_y}{k\xi_k}\hat{s}_x+\frac{i\xi_kk_y+\Delta k_x}{k\xi_k}\hat{s}_y\right),\nn\\
H^{vc}&=&-\frac{\Delta_R}{2}\left(\frac{i\xi_kk_x+\Delta k_y}{k\xi_k}\hat{s}_x+\frac{i\xi_kk_y-\Delta k_x}{k\xi_k}\hat{s}_y\right).\nn\\
\eea
Here $\xi_k=\sqrt{v_F^2k^2+\Delta^2}$. In the absence of $\Delta_R$, $H$ is block diagonal. $\Delta_R$ induces a correction to the diagonal terms which results in the usual intraband Rashba SOC (which leads to the modification of the low energy behavior) and also induces block-off-diagonal terms $(H^{cv},H^{vc})$ which leads to an interband Rashba SOC (this is what we call the high energy effect). In this model, both interband and intraband Rashba couplings are the same $\Delta_R$. However, when we allow for renormalizations from interactions and other higher order interband processes, the interband $\Delta_R$ will be renormalized differently from the intraband $\Delta_R$. To reflect this, we shall use $\Delta_R$ to denote the intraband Rashba coupling and $\delta_R$ to denote the interband coupling. This changes the prefactors of $H_{cv}$ and $H_{vc}$ to $\delta_R$. Because of $H_{cv,vc}$, which is $\bk$- and spin-dependent, the resulting dipole transition element (the interband current operator) also acquires spin-dependence. This coupling does not arise for a Zeeman effect and hence Rashba type SOC is essential to get this effect.

Returning back to the eigenbasis, the current operator can be computed to leading order in SOC (both $\Delta_R$ and $\delta_R$) as
$$\hat{\mathbf j} =\begin{pmatrix}\hat{\mathscr J}_{cc}&\hat{\mathscr J}_{cv}\\\hat{\mathscr J}_{vc}&\hat{\mathscr J}_{vv}\end{pmatrix}.$$ The explicit forms of the $\mathscr J$'s are listed in Appendix \ref{app:Graphene}. We can then use Eq. (\ref{eq:f}) to infer that (to linear order in $\{\Delta_R,\delta_R\}$)
\bea\label{eq:FABa}
&&\frac{\mathcal{F}^{XX}_0}{v_F^2}=\frac{\mathcal{F}^{YY}_0}{v_F^2}=\frac{v_F^2k^2+2\Delta^2}{2\xi_\bk^2},~\frac{\mathcal{F}^{XY}_0}{v_F^2}=\frac{i\tau_z\Delta}{\xi_\bk};\nn\\
&&\frac{\mathcal{F}^{XX}_3}{v_F^2}=\frac{\mathcal{F}^{YY}_3}{v_F^2}=\frac{-v_Fk\Delta^2\delta_R}{2\xi_\bk^4},~\frac{\mathcal{F}^{XY}_3}{v_F^2}=\frac{-i\tau_zv_Fk\Delta\delta_R}{2\xi_\bk^3},\nn\\
\eea
and $k$ will ultimately be restricted to the Fermi surface. Notice that if $\delta_R\rightarrow 0$, then only $\mathcal F_0$ survives which couples the Raman vertex to the charge-charge susceptibility leading to the conventional result. However, in the presence of $\delta_R$, only the contribution (upon summing over valleys) from $\mathcal F^{XX}_3$ and $\mathcal F^{YY}_3$ survive but not from $\mathcal F^{XY}_3$, indicating that this SOC induced contribution is only present in the XX/YY set up, but not in the $XY$ set-up. To estimate the size of this high energy effect, note that $k\rightarrow k_F$ and that the SOC induced $\mathcal F_i$'s are $\sim E_F\Delta\delta_R/[{\rm max}(E_F,\Delta)]^3$. In semiconductors, $E_F<\Delta$, and hence $\mathcal F_i\sim E_F\delta_R/\Delta^2$.

Next, if we now only consider the Valley-Zeeman term ($\Delta_R=0$, $\Delta_Z\neq0$), we note that $H_{cv,vc}=0$ (Appendix \ref{app:Graphene}). This leads to $\mathscr J_{cv,vc}=0$. Since there is no spin-dependent modification of the $\mathbf j$ operators in the interband sector, we trivially get that $\mathcal F^{\beta\alpha}_0$ are unchanged (as expected), and $\mathcal F^{\beta\alpha}_i=0$ for $i\in\{1,2,3\}$, indicating that the plasmon pole would not couple to the Raman vertex. These explicit calculations of the spectrum in the two scenarios with $\Delta_R=0$ and $\Delta_Z=0$ validate the assertions made in Sec. \ref{Sec:SOC effect}. While we have explicitly demonstrated that Rashba SOC leads to a spin-flip vertex in parallel polarization (and hence in the fully symmetric channel of the eRS), in this specific example of the 2D Dirac system, $\mathcal F^{\beta\alpha}_3$ was found to be non-zero. This means that the appropriate susceptibility that couples to the resonant Raman vertex is $\Pi_{30}$. In a strict 2D Dirac system, this component of the spin-charge susceptibility is zero unless a second layer (or multiple layers) is included. The physics of inducing a spin-flip vertex does not change, but the spin-charge susceptibility along the $c$-axis can now be probed. This motivates the study of a 3D system with large Rashba SOC. This is precisely what we do next.

\section{Resonant eRS in ${\rm BiTeI}$}\label{Sec:BiTeI} 
We begin with a theoretical analysis for BiTeI which has one of the largest known bulk Rashba coupling factors of any metallic system with the Rashba parameter $\alpha_{R}$ estimated to be in the range $\approx3.7$eV-\AA~to $4.0$eV-\AA~\cite{ishizaka2011giant,Ishizakaprb12,VanGennep14,Maiti2015, cai_22}, satisfying the requirements of the first factor in Eq.~(\ref{eq:fullresponse}). To investigate the plasmon coupling, we need to evaluate $\Pi_{\alpha\beta;\gamma\delta}(Q)$. As noted in the previous section, the form of $\mathscr J_{cv}$ is essential for the effect of interest. To compute the transition elements in BiTeI, one needs knowledge of the wavefunctions of the intermediate states. Although there isn't a model that provides information about the higher energy states, there exists an effective low energy model for the conduction band electrons. We can proceed to model the intermediate states phenomenologically.

Consider the following low-energy continuum model for BiTeI
\begin{eqnarray}
\hat{H}&=&\left(\frac{k_1^2+k_2^2}{2m_1}+\frac{k_3^2}{2m_3}\right)\hat \sigma_0+\alpha_{R}(\vec \sigma\times\vec k)_3+\lambda\hat \sigma_3 k_3,\nn\\
\end{eqnarray}
where $m_1$ and $m_3$ are the in-plane and out-of-plane masses, $\alpha_{R}$ is the in-plane Rashba SOC in BiTeI, and $\lambda$ is a SOC for the out-of-plane direction. 
Here $1,2$ refer to the $x,y$ axes, and $3$ refers to the $z$-axis. This model, without the $\lambda$-term, was previously used to study possible spin-collective modes~\cite{Maiti2015}, however, this resulted in the expectation value of the spin for its eigenstates being strictly in-plane. The $\lambda$-term is added\footnote{The correct form of this term is of the type $\bar \lambda k_3f(k_1^2,k_2^2)$ such that along the $\Gamma$-A line of the Brillouin zone there is no SOC splitting. However, due to the chemical potential lying below the Dirac point, the Fermi surface has a donut topology, so that we will always be dealing with states away from the $\Gamma$-A line. We absorb all such dependency into a constant $\lambda$. What is essential is that it needs to be odd in $k_3$ and even in other momenta.} to account for the canting of the spins in the $z$-direction as computed in Ref.~\cite{Maas2016}. The Matsubara Green's function for the non-interacting system is then given by
\begin{eqnarray}\label{GF_beta}
\hat G_K(i\omega_m)&=&\sum_{s=\pm}\frac{1}{i\omega_m-\varepsilon^{s}_{\bk}+\mu}\hat\Omega^s_{\bk},\nn\\
\hat\Omega^s_{\bk}&\equiv&\frac12\Big[\hat \sigma_0+s\left(\hat \sigma_1\frac{\alpha_{R} k_2}{D_k}-\hat \sigma_2\frac{\alpha_{R} k_1}{D_k}+\hat \sigma_3\frac{\lambda k_3}{D_k}\right)\Big]\nn\\
\end{eqnarray}
where $D_k=\sqrt{\alpha_{R}^2k_{||}^2+\lambda^2k^2_3}$, $k_{||}^2=k_1^2+k_2^2$, and $\varepsilon^{\pm}_{\bk}=\frac{k_{||}^2}{2m_1}+\frac{k_3^2}{2m_3}\pm D_k$. 

Using $\hat{G}_{K}(i\omega_m)$, we can compute $\Pi_{i0}(Q)$ using Eq. (\ref{eq:RPA Raman susceptibility}) as (we have carried out the Matsubara sums)
\bea\label{eq:BiTeIPi}
\Pi_{i0}(Q)&=&\sum_{s,s'=\pm}\int_{\bk}\mathcal{N}^i_{s,s'} \left[\frac{n_F(\varepsilon^s_{\bk})-n_F(\varepsilon^{s'}_{\bk+\bq})}{i\Omega_m+\varepsilon^s_{\bk}-\varepsilon^{s'}_{\bk+\bq}}\right],\nn\\
\mathcal{N}^i_{s,s'}&\equiv&{\text{Tr}}[\sigma_i\,\hat\Omega^{s}_{\bk}\sigma_0\hat\Omega^{s'}_{\bk+\bq}],
\eea
where $n_F(\ve)$ is the Fermi distribution function. For BiTeI, we cannot compute $\mathcal{F}_i^{\beta\alpha}$ as we don't have the information about $\hat{\mathbf j}$. However, from Sections \ref{Sec:SOC effect} and \ref{Sec:Graphene} we know that in the presence of orbital Rashba SOC, a spin-dependence arises in the interband transition elements such that $\mathcal{F}^{XX}_i\neq0$. This leads to the coupling to the susceptibilities $\Pi_{i0}$. In line with known experiments in BiTeI\cite{lee2023resonant}, we restrict the chemical potential ($\mu$) to lie in the lower band $\ve^-_{\bk}$ and the momentum transfer from the light to the $z$-direction. Next, we need $\Pi_{i0}$'s to leading order in $q_3$. Evaluating this we get (see Appendix \ref{app:BiTeI})
\bea\label{eq:BiPi}
\Pi_{10}(Q)&=&\Pi_{20}(Q)=0\nn\\
\Pi_{30}(Q)&=&-\lambda q_3\int_{\bk}\frac{[-\partial_\ve n_F(\ve^-)]}{D_k}\frac{k_3\partial_{k_3}\ve^-}{i\Omega_m}\nn\\
&&+\alpha^2\lambda q_3\int_{\bk}\frac{[n_F(\ve^+)-n_F(\ve^-)]}{D_k^3}\frac{i\Omega_m k^2_\parallel}{\Omega_m^2+4D_k^2}.\nn\\
\eea
As ascertained in Sec. \ref{Sec:SOC effect}, the spin-charge susceptibility $\Pi_{30}\propto \lambda q_3$. These can now be plugged into Eq. (\ref{eq:Pi_gamma_final}) to get
\bea\label{eq:BiRaman}
\Pi_{\alpha\beta;0}(Q)&\approx&\frac{\sqrt{m_\alpha m_\beta}\mathcal{F}^{\beta\alpha}_i\Pi_{i0}(Q)}{\Omega_I-E_g-i\Gamma}\nn\\
&=&\frac{\sqrt{m_\alpha m_\beta}\mathcal{F}^{\beta\alpha}_3\Pi_{30}(Q)}{\Omega_I-E_g-i\Gamma}.
\eea
Here $m_\alpha$ refers to the mass in the $\alpha$-direction. Since our polarizations will be in plane, $m_\alpha=m_\beta=m_1$. This leads to
\bea\label{eq:ReamanBite}
\Pi_{\alpha\beta;\beta\alpha}(Q)&\approx&\frac{e^2}{\e_\infty q^2}\frac{\Pi^2_{30}(Q)}{(\Omega_I-E_g)^2+\Gamma^2}\frac{(m_1\mathcal F^{\beta\alpha}_3)^2}{1-V_\bq\Pi_{00}(Q)},\nn\\
&\xrightarrow{\bq=q_3\hat z}&\frac{e^2}{\e_\infty q^2_3}\frac{\Pi^2_{30}(Q)}{(\Omega_I-E_g)^2+\Gamma^2}\frac{(m_1\mathcal F^{\beta\alpha}_3)^2}{1-V_\bq\Pi_{00}(Q)},\nn\\
\eea
where $\Pi_{00}(Q)$ is calculated as (see Appendix \ref{app:BiTeI})
\bea\label{eq:PPP}
\Pi_{00}(Q)&=&q_3^2\int_{\bk}[-\partial_\ve n_F(\ve^-)]\frac{(\partial_{k_3}\ve^-)^2}{-\Omega_m^2}\nn\\
&&+\lambda^2q_3^2\int_{\bk}\frac{[n_F(\ve^+)-n_F(\ve^-)]}{D_k^3}\frac{\alpha^2k^2_\parallel}{\Omega_m^2+4D_k^2}.\nn\\
\eea
As expected, it is $\propto q_3^2$ which counters the $q_3^2$ from $V_\bq$. Further, the $q_3^2$ in the denominator of Eq. (\ref{eq:ReamanBite}) is also countered by the $q_3^2$ in $\Pi_{30}^2$, leading to a finite response even if $q_3\rightarrow 0$. Observe that, and this is expected from the results discussed in Sec. \ref{Sec:Graphene}, the result is proportional to the SOC component $\lambda$ which accounts for canting of the spins out-of-plane. This is necessary to couple the spin degree of freedom to the $q_3$ dispersion. Had we chosen to work with an in-plane momentum ($q_1$ or $q_2$) then we would not need this canting effect (see Appendix \ref{app:BiTeI}). However, we will be interested in the $q_3$ component as this is the scenario that will correspond to the eRS setup we will use to study BiTeI where the momentum transferred by light will be along the $c$-axis. 

To calculate the location of the plasmon pole itself we would need the solution to the equation $V_{\bq}\Pi_{00}(Q)=1$, which is the condition for zero of the denominator of Eq. (\ref{eq:dressed Raman susceptibilty2}). This is computed in Appendix \ref{app:omega_pl} and the result in real frequencies is:
\bea\label{eq:plasmafreq}
\Omega_{\rm pl}^2=\Omega_0^2\frac{m^*_1}{\sqrt{m^*_3}}f_{pl}(\tilde\alpha, \tilde\lambda),
\eea
where, $\Omega_0^2\equiv n_0e^2/m_0\epsilon_\infty$ with $n_0\equiv \frac{(2m_0|\mu|)^{3/2}}{6\pi^2}$ being the number density of single spin parabolic electrons with a chemical potential $|\mu|$; $m_1^*\equiv m_1/m_0$, $m_3^*\equiv m_3/m_0$ (with $m_0$ being the bare electron mass); and $f_{pl}$ is a dimensionless function of dimensionless variables $\tilde \alpha\equiv\alpha_R \sqrt{2m_1/|\mu|}$ and $\tilde \lambda\equiv\lambda \sqrt{2m_3/|\mu|}$, such that in the limit $\tilde\alpha\rightarrow 0,~\tilde\lambda\rightarrow 0$, we get $f_{pl}\rightarrow 1$. 

The function $f_{pl}$ only weakly depends on $\Omega$ and hence can be treated as $\Omega$-independent for the estimation of the plasma frequency. The full numerical calculation that we use shortly will find the true pole in the eRS response function. However, as is clear from Eq. (\ref{eq:PPP}), the function $f_{pl}$ consists of interband \textit{spin-flip} transitions between the Rashba subbands with weight $\propto\lambda^2q_3^2$. Consequently, this leads to a continuum of spin-flip excitations with the same weight. This means that the plasmon that would result in the $q_3$ direction would likely be damped by this continuum of spin-flip excitations. This is certainly not new. Even in 2DEG, the 2D plasmon is damped by the spin-flip continuum\cite{Pletyukhov2006}. But this was the plasmon that dispersed in the plane. The same remained true in the 3D-in-plane plasmon that was considered in Ref \cite{Maiti2015} (also a model for BiTeI). In that work, however, the canting effect was not modelled and this led to the fact that the $c$-axis plasmon was neither renormalized by SOC, nor affected by the spin-flip continuum. What is new here is that in general for a system belonging to a polar group, the $3$-component of spin can couple to the $3$-component of momentum leading to the the $c$-axis plasmon being renormalized by SOC and also be damped by the continuum. This is what is accounted for by the $\lambda$ term in the model considered above. 

\begin{figure}[t]
\centering\captionsetup{justification=RaggedRight}
\includegraphics[width=0.95\linewidth]{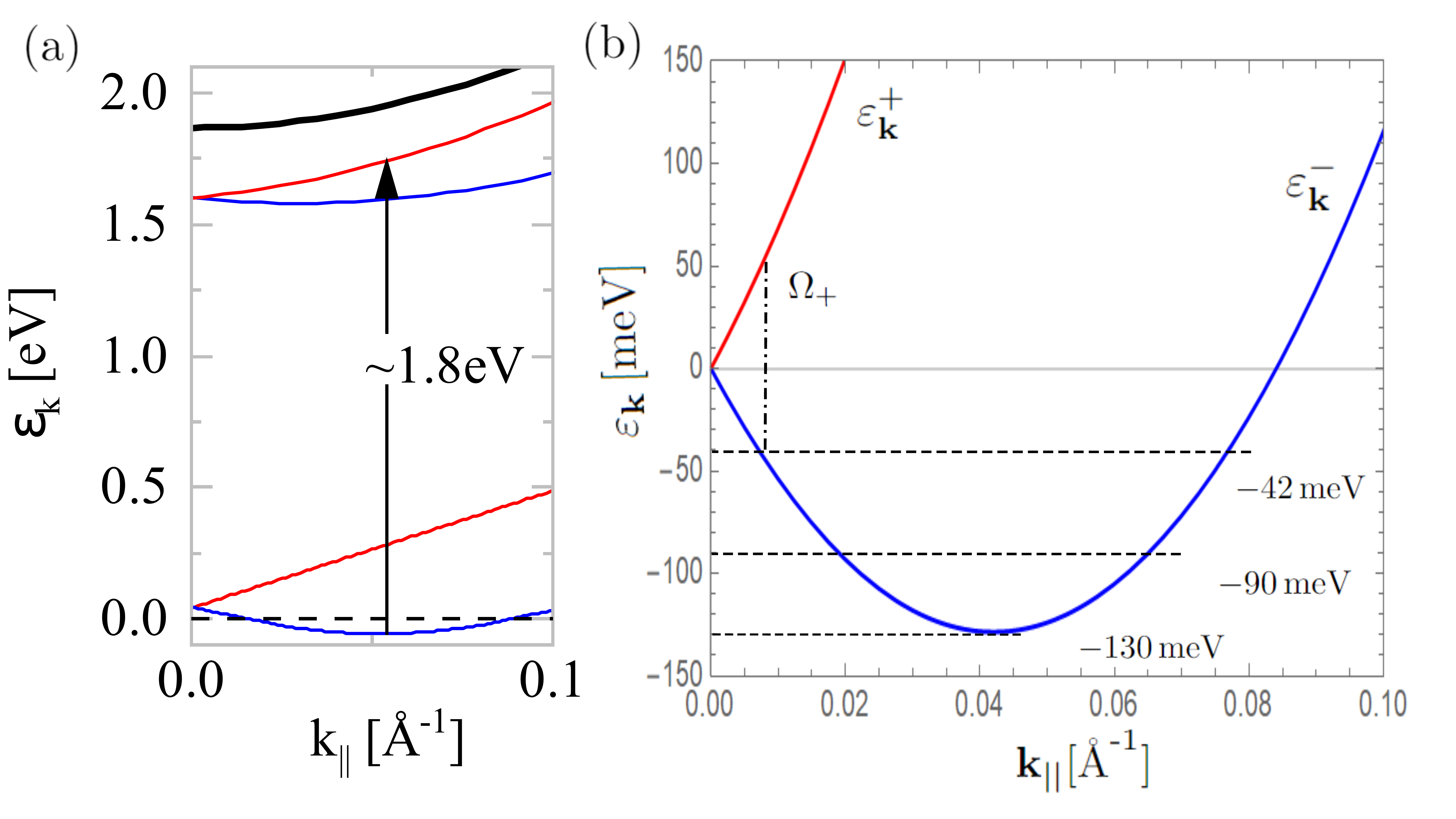}
    \caption{(a) Low energy (where the chemical potential - dashed line - lies) and high energy bands of BiTeI plotted against $k_{\parallel}$ at $k_{z}=\pi/c$. The colors represent the spin-split bands and the black band at $\approx 2$ eV represents a spin-degenerate band. (b) A cartoon of the band structure of BiTeI with the zero fixed at the Dirac point. The chemical potential is varied between $-90$ meV to $-42$ meV in the samples we investigated\cite{lee2023resonant}.}
\label{fig:bandstructure}
\end{figure}

\subsection{Resonant eRS experiment}
To test our theoretical analysis of BiTeI we took Raman measurements of an electronic collective mode in the fully symmetric channel under the following conditions. The incident light enters the samples parallel to the z-axis of the crystal with light polarized along the optically equivalent x- and y-axes. The single crystals of BiTeI that were grown using the vertical Bridgman technique had concentrations of elemental I sufficient to ensure the chemical potential lay below the Dirac point (for characterization details, see~\cite{lee2023resonant}). 
The Raman spectra were taken in multiple polarization geometries to perform algebraic decomposition of the spectra into the different symmetry channels. 
Calculations of the Raman spectra used a model band structure of BiTeI shown in Fig.~\ref{fig:bandstructure}. 

\begin{figure}[t]
\centering\captionsetup{justification=RaggedRight}
\includegraphics[width=\linewidth]{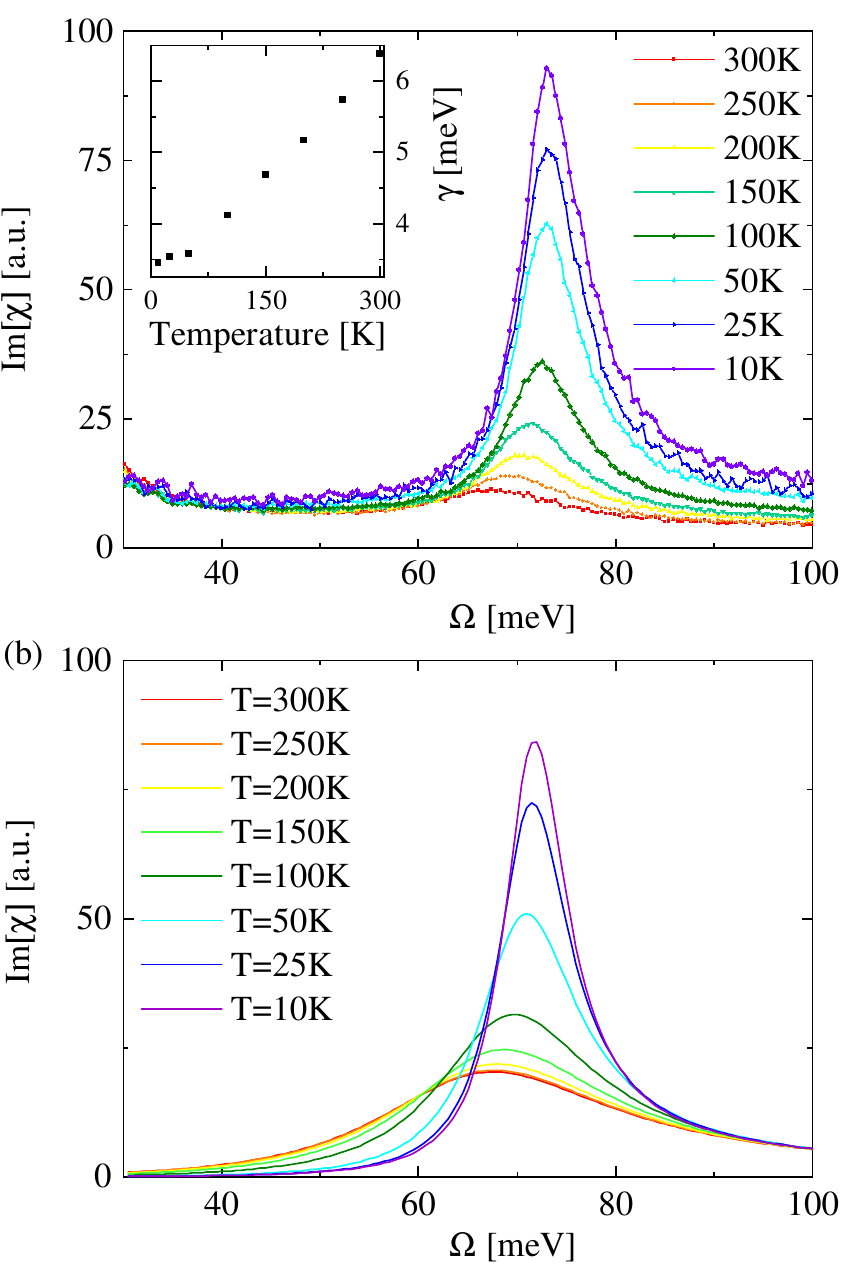}
    \caption{(a) The temperature dependence of the Resonant eRS spectrum of the collective mode in the fully symmetric A$_1$ channel of BiTeI. The inset shows the corresponding half width at half maxima. (b) The theoretical calculation of eRS for the parameters mentioned in the text.}
\label{fig:temperatureprofile}
\end{figure}

\begin{figure}[t]
\centering\captionsetup{justification=RaggedRight}
\includegraphics[width=\linewidth]{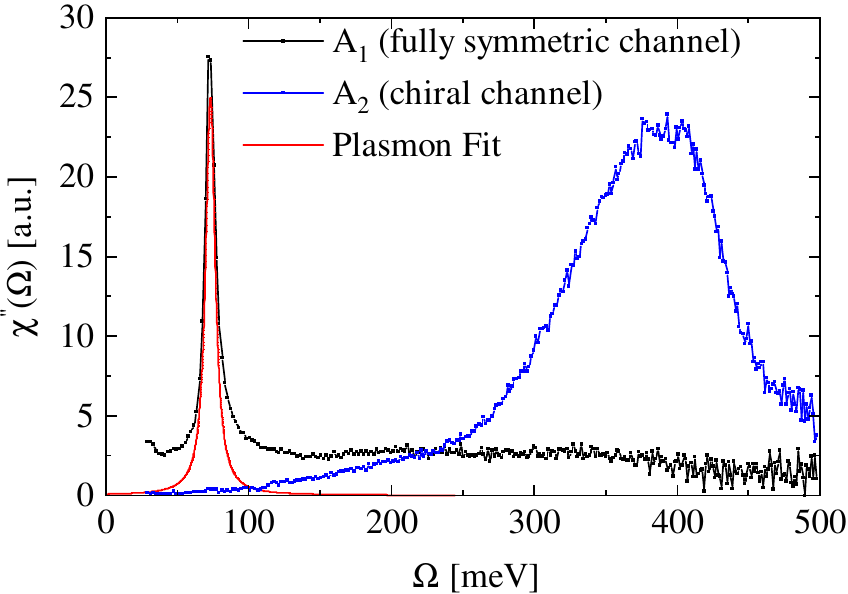}
    \caption{The response at 10 K over extended Raman shifts in the $A_{1}$ (fully symmetric) and $A_{2}$ (chiral) irrep decompositions of the collected eRS spectrum. The red line is the calculation from Fig. {\ref{fig:temperatureprofile}(b). The $A_2$ response shows a spin-flip continuum starting from $\approx 100$ meV, while the $A_1$ response shows a much weaker feature in that frequency range.}
    }
\label{fig:symmetrydecomposition}
\end{figure}

\begin{figure}[t]
\centering\captionsetup{justification=RaggedRight}
\includegraphics[width=\linewidth]{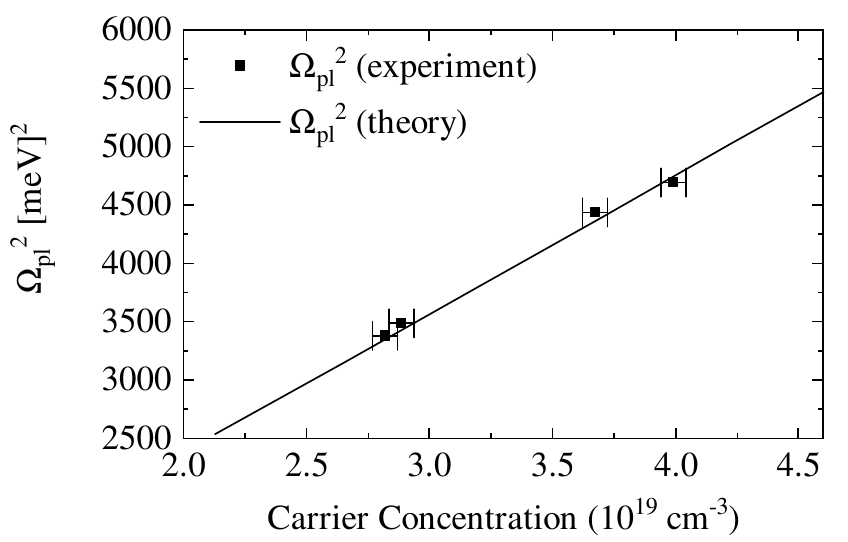}
    \caption{
    The collective mode peak position ($\Omega_{pl}$) vs carrier concentration deduced from Hall-measurements for various samples~\cite{lee2023resonant} (data points) and the theoretical plot of the same for the parameters given in the text. The theoretical number density was estimated from the Fermi surface volume. The linear scaling of $\Omega^2_{pl}$ with the carrier density suggests that it is a plasmon. 
    }
\label{fig:carrierconcentrationprofile}
\end{figure}

\begin{figure}[t]
\centering\captionsetup{justification=RaggedRight}
\includegraphics[width=0.9\linewidth]{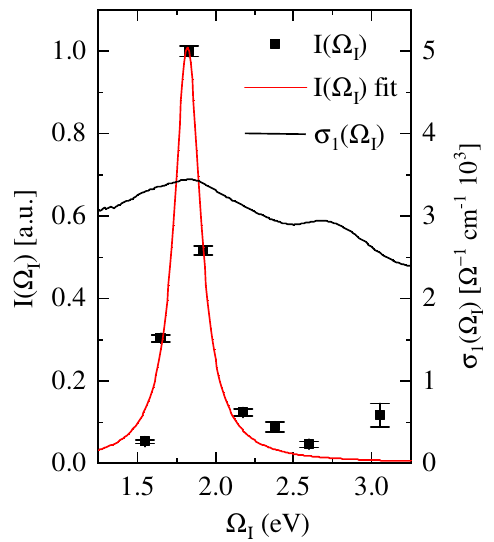}
    \caption{Resonant Raman excitation profile of the collective mode [$I(\Omega_I)$] as a function of the incident laser energy showing sharp feature at $1.83$ eV. The absorption spectrum $\sigma_{1}(\Omega)$ is also plotted to demonstrate that there are indeed dipole active transitions at that excitation energy. 
    }
\label{fig:excitationprofile}
\end{figure}

In Fig.~\ref{fig:temperatureprofile}(a), we show the collected eRS spectrum in the symmetric channel for temperatures ranging from $10$\,K-$300$\,K around Raman shifts near the collective mode energy of $\approx 74$ meV. In Fig.~\ref{fig:temperatureprofile}(b), we plot the calculated temperature dependence of the eRS for the XX polarization for the following choice of parameters that are within the ranges reported in the literature: $\alpha_R=3.82\,$eV\AA~\cite{ishizaka2011giant,Ishizakaprb12,VanGennep14,Maiti2015, cai_22}, $\lambda=0.35\,$eV\AA,  $m_1=0.14\,m_0$~\cite{VanGennep14,Maiti2015}, $m_3=0.91\,m_0$~\cite{lee_prl11}, $\epsilon_\infty=11.9\,\epsilon_0$~\cite{schwalbe_prb16}.
This leaves us with two additional parameters for the above plot. One is the chemical potential, $\mu$, which has been chosen to be $\approx -42$  meV  [the minimum of band is at $\varepsilon^-_{\text{min.}}\approx -130$ meV, see Fig.~\ref{fig:bandstructure}(b)]. All energy measurements are relative to the Dirac point in the spin-split spectrum. This choice of $\mu$ leads to an onset of spin flip transitions at $\Omega_+\approx 90$ meV consistent with the result in Ref. \cite{lee2022chiral}. 
The other parameter is the damping rate which includes contributions from iodine vacancies, the nearby Rashba spin-flip continuum of excitations, finite temperature, and electron-electron interactions such as in Refs. \cite{Gavoret,Sharma2021,Goyal2023}. To model these, we simply introduce a scattering rate that reproduced the mode width of 7 meV at 10K and the plots at other temperatures evolved accordingly.

We saw in Eq. (\ref{eq:PPP}) that due to the $\lambda$ parameter, we expect to also pick up contribution from the continuum of spin-flip excitations, albeit weakly. In Fig.\,\ref{fig:symmetrydecomposition}, we show the eRS in the symmetric channel (which is the $A_1$ irreducible representation (irrep) of the point group C$_{\rm 3v}$ relevant for BiTeI) over extended range of frequencies. The continuum is evident by a weak feature in the black line in Fig. \ref{fig:symmetrydecomposition}. For comparison, we also show the continuum obtained in the $A_2$ irrep of C$_{\rm 3v}$ which couples directly to spin-flip excitations and hence shows a much stronger response. The onset of the spin-flip continuum in the channels belonging to different irreps is at the same value, but the spectral weight of the continuum in the response in $A_2$ is controlled predominantly by $\alpha_R$, whereas that in $A_1$ is controlled by $\lambda$ and hence is smaller. From the $A_2$ response we deduce that the onset of the spin-flip continuum is $\approx 100$ meV.
Having presented the theory we now understand that this coupling of charge fluctuations to the spin ones is due to Rashba SOC. However, since the plasmon is a fully symmetric excitation of the system, it does not show up in the $A_2$ channel. We also present the calculated plasmon response in the XX polarization in red [taken from Fig. \ref{fig:temperatureprofile}(b)], where we see that the spectral weight of the plasmon completely overshadows the already weak continuum. We had explicitly demonstrated in Sec. \ref{Sec:Graphene} that this coupling is not present in the cross-polarization setup, i.e. $\mathcal F^{XY}=0$, consistent with the data in the $A_2$ channel.

To verify that we are indeed observing the plasmon peak, we tracked the collective mode frequency for different samples grown from the same batch with light variations in the chemical potential induced by changes in iodine deficiencies. In Fig.~\ref{fig:carrierconcentrationprofile}, we plot the square of the energy of the electronic collective mode as a function of carrier concentration (as determined from Hall measurements\cite{lee2023resonant}) and compare it against the calculation of the renormalized plasma frequency obtained from Eq. (\ref{eq:plasmafreq}) for number densities determined as below (we restrict ourselves to the case where the chemical potential only lies in the lower Rashba sub-band) :
\bea\label{eq:nodens}
n_{\rm BiTeI}&=&\int_{-\infty}^\mu dE\int_{\bk}\delta(E-\ve_{\bk}^-)\nn\\
&=&n_0m^*_1\sqrt{m^*_3}~f_n(\tilde \alpha,\tilde\lambda),
\eea
with
\bea\label{eq:Irashba}
f_n(\tilde \alpha,\tilde\lambda)&\equiv&\frac{3}{2}\int_{-\infty}^{\tilde\mu}d\tilde E\int_{0}^\infty d\tilde k_3\int_0^\infty dy~\delta(\tilde E-\tilde\varepsilon^-_{\sqrt{y}, \tilde k_3}),\qquad
\eea
where $\tilde \mu\equiv \mu/|\mu|$, $\tilde E\equiv E/|\mu|$, $\tilde k_i\equiv \hbar k_i/\sqrt{2m_i|\mu|}$ and $y$ enters after setting $k^2_\parallel=y$. The function $f_n(\tilde \alpha,\tilde\lambda)$ is such that $f_n(0,0)=1$. To remain below the Dirac point of BiTeI, we need to set $\tilde\mu=-1$. From Eqs. (\ref{eq:plasmafreq}) and (\ref{eq:nodens}) the slope of $\Omega_{pl}^2$ vs $n_{\rm BiTeI}$ is $$s\equiv 
\frac{e^2}{m_3\epsilon_\infty}\frac{f_{pl}(\tilde\alpha,\tilde\lambda)}{f_n(\tilde\alpha,\tilde\lambda)}.$$
With the above choice of parameters, the theoretical estimate aligns well with the experiment. It is interesting to note that experimentally we deduce the number density from the Hall measurements whereas theoretically, we calculate the number density from the Fermi-surface volume. The observed agreement suggests that the effective 1-band Hall-effect theory still seems to apply insofar as to determine the carrier concentration. This justifies the Hall analysis done in Ref. \cite{lee2023resonant}.

To validate the requirement of resonance for the spin-mediated coupling to plasmons, we changed the incident photon energy from 1.55-2.60\,eV to scan above and below the resonance with the spin-split bands shown in Fig. \ref{fig:bandstructure}. We plot the resonant Raman excitation profile in Fig.~\ref{fig:excitationprofile} against the optical absorption data. The latter captures all the dipole-allowed transitions indicating the presence of interband transitions. We see that there is a broad absorption peak around 1.7-2.2 eV and then another around 2.7-3 eV. The first one corresponds to the gap shown in Fig. \ref{fig:bandstructure} which has spin-split bands at $\approx 1.8$ eV and a spin-degenerate band $\approx 2.1$ eV. Observe that the excitation profile only resonates with the bands around 1.8 eV. Since the excitation profile is obtained by integrating across the collective mode spectral weight, we can infer that the collective mode is only excited while the resonance is between the spin-split bands and not the spin-degenerate ones. This is relevant because, as stated earlier, we need SOC in the transition elements for our effect to materialize. And such a coupling ensures SOC splitting in \textit{both} bands. The lack of such a splitting for the band at $\approx 2.1$ eV [in Fig.~\ref{fig:bandstructure}(a)] suggests a lack of such coupling and hence lack of weight in the excitation profile. This in turn suggests that the polarization factor $\mathcal F_0^{XX}$ for the band at $\Omega_I=1.83$ eV is likely finite, while that at $\Omega_{I}=2.1$ eV is negligible.

\subsection{Symmetry considerations of the observed $c$-axis plasmon in BiTeI}
Being a polar self-doped metal, the planar properties of BiTeI are dictated by the C$_{\rm 3v}$ point group. A unique property of C$_{\rm nv}$ groups is that the dipole field $z$ belongs to the $A_1$ representation, along with the usual quadrupolar fields $x^2+y^2,z^2$. The quadrupolar form is another way of seeing why any collective excitation in the $A_1$ channel only ends up coupling quadratically in $q$. The presence of the $z$-dipole term in $A_1$ allows for the $q_3$ to be present. Since plasmon is a longitudinal mode, this would imply that the excitation would correspond to the $c$-axis plasmon. \footnote{We inform the reader of our decision to simultaneously use $3$-, $z$, and $c$. The first one is useful for theoretical modeling, the second is the convention reported in the character tables and the third is prevalent in crystallography. We choose them according to the relevant context, as is standard practice.} 

While the possibility of the $c$-axis plasmon showing up is evident, it still remains to be explained why SOC is necessary. The way this plays out is this: Without SOC, the only dipolar field ($z$) in the system could arise from finite $q_3$. For the $A_1$ excitation, the plasmon mode appears as $q_3^2$ as we need the result to be comprised of a $z^2$ contribution. Of course, this physics was reflected in the calculations that took the form of $q^2.q^2/q^2$, where the numerators are from the polarization-charge bubbles and the denominator is from the Coulomb interaction renormalization. In the presence of SOC of the type $\lambda$, the dipole field ($z$) is provided by $\lambda$ itself resulting in a $\lambda^2$ contribution in the $A_1$ response. In the calculations this appeared as $\lambda q.\lambda q/q^2$ leading to the $\lambda^2$ coupling. Note that in both cases, you need to break inversion for the plasmon to couple to the Raman vertex. In the former case the inversion breaking was due to finite $q_3$, but in the presence of SOC, this was intrinsically present due to $\lambda$.

\subsection{SOC induced Drude-weight renormalization along the $c$-axis}\label{subsec:SOCmass}
\begin{figure}
\centering\captionsetup{justification=RaggedRight}
\begin{subfigure}{0.85\linewidth}
\includegraphics[width=\linewidth]{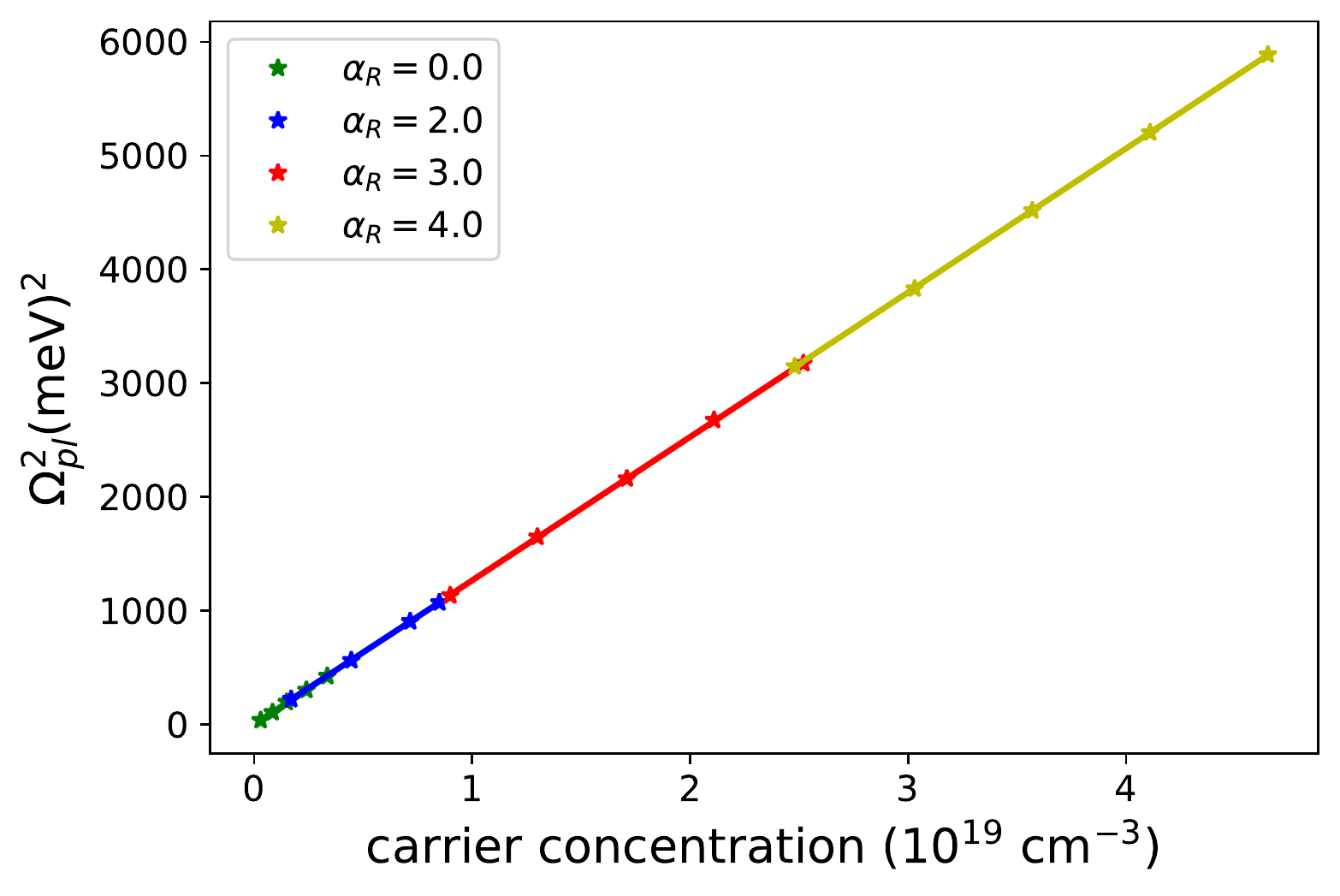}
\caption{}\end{subfigure}
\begin{subfigure}{0.85\linewidth}
\includegraphics[width=\linewidth]{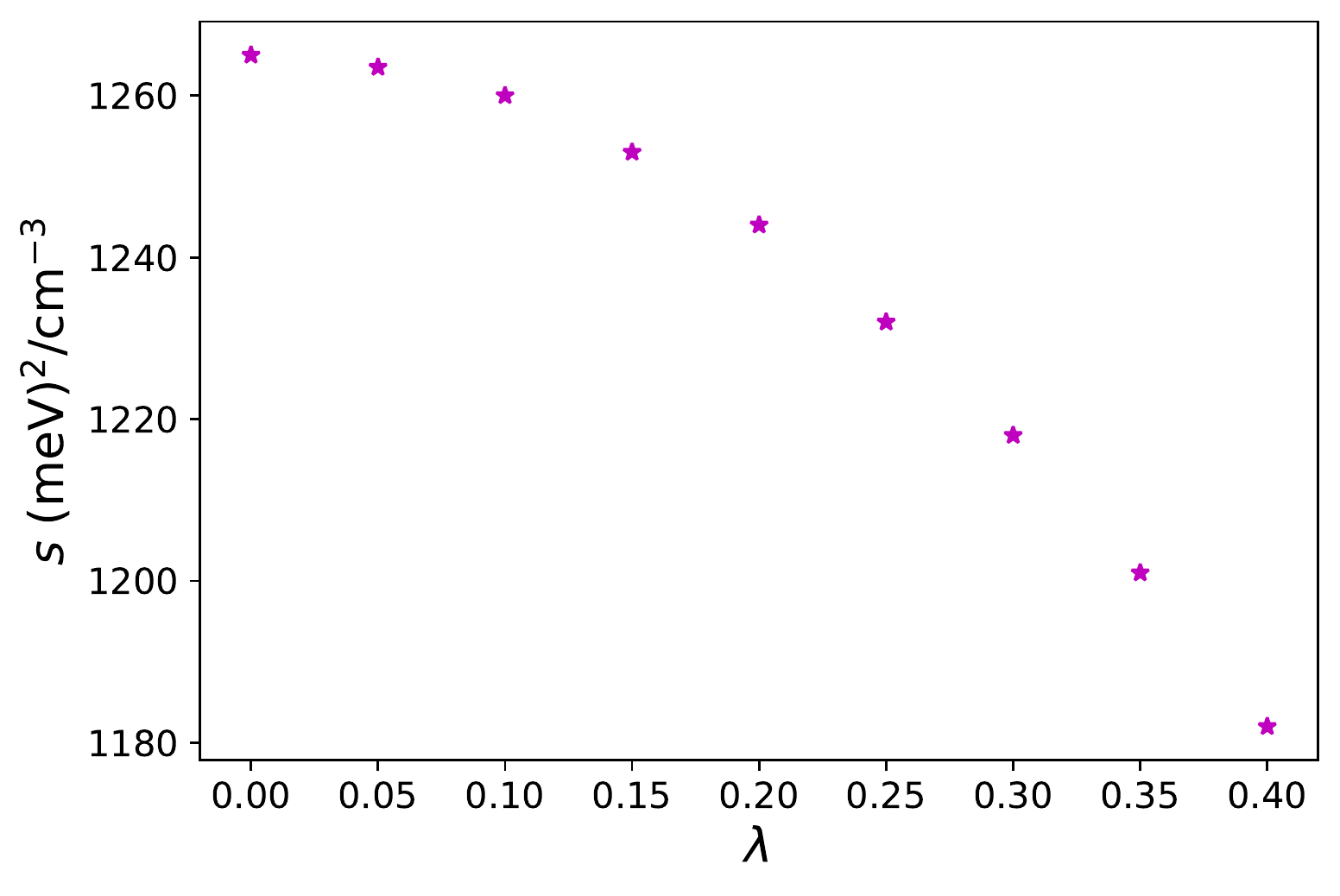}
\caption{}\end{subfigure}
\caption{(a) The square of the plasmon energy $\Omega^2_{\rm pl}$ is plotted against $n_{\rm BiTeI}$ for different values of $\alpha_R$ and $\lambda=0$. The fixed slope for different $\alpha_R$ demonstrates a lack of Drude-weight renormalization. (b) Sensitivity of the slope $s$ to $\lambda$ (in units of 
 eV-\AA). The reduction of the slope indicates downward renormalization of the Drude weight.}
\label{fig:DrudeWeight}
\end{figure}

It is instructive to theoretically investigate the behavior of the slope $s$ introduced above. Observe that in the absence of SOC, since $f_{pl}=1=f_n$, $s$ measures $1/m_3\epsilon_{\infty}$ irrespective of the carrier concentration and the in-plane masses. Further, if $\lambda=0$, $s$ remains unaffected even when we change $\alpha_R$. This is demonstrated in Fig. \ref{fig:DrudeWeight}(a). This does not mean that the $\Omega_{pl}$ and $n$ are unaffected, it simply means that they change in a proportional manner as shown by the constant slope. However, in the presence of $\lambda$, the slope evolves and this is plotted in Fig. \ref{fig:DrudeWeight}(b). 

The meaning behind the sensitivity of $s$ to a SOC parameter can be understood from the following. Consider a spinless free electron gas (with possibly different masses along the $ab$ axis and $c$ axis). Its conductivity in the $\alpha^{\rm th}$ direction is given by $\sigma_{\alpha\alpha}(\Omega)=\frac{n_0e^2}{m_\alpha}\frac{\tau}{1-i\Omega\tau}$, where $\tau$ is the scattering lifetime and $m_\alpha$ is the mass in the $\alpha^{\rm th}$ direction. Let us now define two quantities: (i) $\mathcal D\equiv\lim_{\Omega\tau\rightarrow\infty}\Omega{\rm Im}[\sigma(\Omega)]$ which is $n_0e^2/m_\alpha$; and (ii) $W\equiv\frac2\pi\int_0^\infty d\Omega{\rm Re}[\sigma(\Omega)]$, which also evaluates to $n_0e^2/m_\alpha$. The latter is nothing but the optical sum rule. The relation between the integrated optical weight, $W$, and $n_0,m_\alpha$ remains the same irrespective of the details of the system (even if we include spin and even SOC)
The quantity $\mathcal D$, however, represents the spectral weight carried by the free carriers in the system and is referred to as the Drude weight. In systems without SOC, $\mathcal D=W$ and is {\blue thus} protected by the optical sum rule value preventing any renormalization. This protection remains valid even in the presence of spin splitting induced by a Zeeman field. However, in the presence of spin-splitting due to SOC, while the relation involving $W$ still holds (irrespective of how $n$ is renormalized), the relation involving $\mathcal D$ no longer holds and in fact is renormalized downward, meaning that the Drude weight decreases. The lost spectral weight is recovered at higher energies pertaining to the interband spin-flip transitions, thereby restoring the sum rule. This is reminiscent of the spectral weight rearrangement in a BCS superconductor from the $\delta$-peak in optical conductivity to the finite energy bump at twice the gap energy, as we increase disorder\cite{Chen1993} and has been reported as the ``color change" effect\cite{Klein1999,basov1999}.

Since the slope $s$ is derived from the plasma frequency which is a response from the free carriers, $s$ actually tracks the Drude weight $\mathcal D$. Further, since $s\sim \mathcal D/n$ and $W\sim n$, $s$ could also be seen as a measure of $\mathcal D/W$. This should be constant without SOC and should dip downwards in the presence of SOC. In 2D Rashba systems\cite{Agarwal2011,Maiti2015}, this is exactly what happens: SOC drains the spectral weight away from $\mathcal D$ to higher energies, while still satisfying the sum-rule. However, observe from Fig. \ref{fig:DrudeWeight}(a) that $s$ is invariant under changes to the Rashba parameter $\alpha_R$ when $\lambda=0$. This can be understood from the result in Ref. \cite{Maiti2015} which demonstrated that the $c$-axis plasmon is not renormalized by SOC in the absence of $\lambda$. As soon as we introduce $\lambda$ [see Fig. \ref{fig:DrudeWeight}(b)], we observe that $s$ becomes sensitive to the SOC constant indicating that the Drude weight is renormalized (downwards). The lost spectral weight from the Drude weight is recovered in the continuum introduced due to $\lambda$. This is consistent with the observation that our observed $c$-axis plasmon is damped from the continuum, the presence of which would necessarily imply a Drude weight renormalization. While our modelling and interpretation imply renormalization along the $c$-axis of BiTeI, verifying this directly is beyond the scope of an eRS experiment.

\section{Conclusion}\label{Conclusion and outlook}
It has long been known from group theory considerations that the eRS in the fully symmetric channel can couple to the plasmon excitations. But even in resonant eRS, it does so only at finite momentum transfer $q$  and the spectral weight of the plasmon scales as $q^2$. In this work we have demonstrated that in resonant eRS of light in systems with broken inversion, the presence of Rashba-type SOC induces an additional coupling to plasmons at zero momentum transfer. This happens due to two effects: first, as a high energy effect, the fully symmetric channel develops an effective spin vertex which is characterized by a polarization factor $\mathcal F^{\beta\alpha}_i$, that couples to the spin-charge susceptibility of the system; second, as a low energy effect, the spin-charge susceptibility ($\Pi_{i0}$) in the system itself becomes non-zero due to Rashba type SOC. The development of the effective spin vertex happens under resonant conditions and requires Rashba coupling of the spins involved in the resonant interband transitions. This effect is not present if the spins are merely split in the two bands. The resulting coupling to plasmons ($\mathcal F_i\Pi_{i0}$) is particularly significant in giant Rashba systems such as BiTeI.

While the emergence of spin-charge susceptibilities was known, showing the presence of the effective spin vertex in the fully symmetric channel is the main contribution of this work. We developed the general theoretical formalism to capture this effect and explicitly demonstrated the existence of this vertex in a toy Dirac system with SOC where we had access to the wavefunctions of all the relevant states in the system. For BiTeI, although we did not have access to the wavefunction of the intermediate states, we demonstrated all the necessary ingredients to be present to observe this effect. We then verified these ideas by performing resonant eRS on various BiTeI samples demonstrating a charge sector collective mode that scales just like a plasmon with respect to the carrier density. Assuming a simple toy model for BiTeI conduction band we were also able to reproduce the temperature dependence and the laser excitation profile dependence as seen in the experiment. Within the context of BiTeI, we also argued for the presence of an out-of-plane canting of spins that is ultimately responsible for the observation of the plasmon mode in eRS under normal incidence. We additionally predicted a Drude-weight renormalization along the $c$-axis of BiTeI.

This novel high-energy effect of SOC, which primarily arises due to the breaking of SU(2) invariance in the spin-flip interband transitions, not only allows plasmons to couple to optical probes but also opens up a new route to explore the interplay of spin and charge degrees of freedom in semiconductors. In fact, it is quite possible that the observation of many charge collective modes, which are expected to be weighted by $q^2$ but still seen through Raman could be so due to the enhancement from the SOC in such materials. Apart from BiTeI, this can be relevant to other C$_{nv}$ systems like transition-metal-dichalcogenides and multi-layer Graphene. This new coupling channel should be further investigated to study its effect on non-equilibrium ultrafast spectroscopies, an emerging field, as it provides an additional mechanism for energy relaxation for charge and spin excitations alike. This coupling channel could also be exploited to use charge-driven collective modes to manipulate spin. Finally, it is encouraging to note that the spectroscopic landscape has expanded tremendously due to considerations of effects of spin which has been largely underappreciated for a long time. We believe that the physics we outline here will be relevant to studying the interplay of spin-orbit coupled electrons either with chiral phonons which can also couple to eRS\cite{Parlak2023}, with excitations in strongly SOC Floquet driven systems\cite{Arakawa2021}, and to other metallic topologically non-trivial systems that often require strong SOC.\\  

\paragraph*{Acknowledgements.} 
We thank Dmitrii Maslov for the discussions. The work at Concordia University (S.S. and S.M.) was funded by the Natural Sciences and Engineering Research Council of Canada (NSERC) Grant No. RGPIN-2019-05486. 
S.S. was also supported by the Horizon Postdoctoral fellowship from Concordia University. 
The spectroscopic work conducted at Rutgers (A.L. and G.B.) was supported by the NSF Grant No. DMR-2105001. 
The work at NICPB (G.B.) was supported by the European Research Council (ERC) under the European Union’s Horizon 2020 research and innovation programme Grant Agreement No. 885413.

\appendix

\section{Discussion on the Raman processes and their contribution to the eRS spectrum}\label{app:1}
\paragraph{The eRS response from the direct term.} In Eq. (\ref{eq:Raman vertex}) if we only keep the direct term, the final response $R_d(\Omega,\bq)$ can be obtained from the analytic continuation of 
\bea\label{eq:response direct}
\chi_d(Q)=-\int_K{\rm Tr}[\hat \gamma_d\hat G_K\hat\gamma_d\hat G_{K+Q}].
\eea
At this stage, the hat represents matrices in the full Hilbert space of states at the Fermi surface as well as those in the intermediate states. The analytic continuation technique has been detailed in Ref. \cite{Shvaika2005} (although it was done for a one-band system, the technique outlined there still applies to a multiband system, provided the prescription to couple the vector potential is appropriately identified). The frequency sum in $\int_K GG$ leads to the Lindhard form $\frac{n(\ve_{\bk+\bq})-n(\ve_{\bk})}{i\Omega_m+\ve_{\bk+\bq}-\ve_{\bk}}$ which restricts the $\int_\bk$ to states along the Fermi surface (at $\bq\rightarrow 0$). Thus $\chi_d(Q)$ only picks up the contributions from the states at the Fermi surface.

\paragraph{The eRS response from the mixed terms.} A mixed term comprises one vertex depicting the direct process and the other the indirect process. The term of the type $\sum_m\frac{[j^S_\beta]_{nm}[j^I_\alpha]_{mn'}}{\Omega_I-E_g-i\Gamma}$ in the vertex is not conducive to a field theory treatment. However, this term can be seen as an on-shell contribution of the propagator $\hat G$ of the system (defined in the full Hilbert space). This allows the following generalization:
\bea\label{eq:res term}
&&\sum_m [\hat j^S_{\beta}]_{nm}\left[\frac{1}{\ve_n+\Omega_I-\ve_m-i\Gamma}\right]_m[\hat j^I_{\alpha}]_{mn'}\nn\\
&=&\sum_{mm'} [\hat j^S_{\beta}]_{nm}\left[\frac{\delta_{mm'}}{\ve_n+\Omega_I-\ve_m-i\Gamma}\right][\hat j^I_{\alpha}]_{m'n'}\\
&\rightarrow &[\hat j^S_{\beta}\hat G(\ve_n+\Omega_I)\hat j^I_{\alpha}]_{nn'}.
\eea
Here we have used that $\hat G(Q)=\sum_m\frac{|m\rangle\langle m|}{i\omega-\ve_m-i\Gamma_m}$ such that on-shell condition (which would lead to a pole) would be met when $\omega$ (translated to real frequencies) is $\sim \ve_m$. Thus, in the presence of an external laser field from $\Omega_I$, the on-shell contribution selectively comes from the intermediate states. Even if the intermediate states are incoherent or absent, there still remains the off-shell contribution that can produce a finite response from the mixed terms. In common parlance, these are referred to as `virtual' processes. The choice of $\Omega_I$ is what ultimately picks which type (on- or off-shell) virtual process would contribute the most. This promotion of the vertex to the Green's function is what ultimately allows one to draw the diagrams as in Fig. \ref{fig:RamanDiagrams}. Thus, the response $R_m(\Omega,\bq)$ is obtained from analytic continuation of terms such as
\beq\label{eq:mixed1}
\chi^{(1)}_m(Q)=\int_K{\rm Tr}[\hat\gamma_d\hat G_{K+Q}\hat j^I\hat G_{K+Q+Q_I}\hat j^S\hat G_{K}].
\eeq
Here $Q_I\equiv(\Omega^I_m,\bq_I)$, where $\Omega_m^I$ is a Matsubara frequency that would be continued to the incoming laser frequency $\Omega_I$ and $\bq_I$ is the momentum of the incoming photon. The superscript (1) denotes that this is just one of the terms. Other terms arise from interchanging the vertices with direct and indirect processes, and also interchanging the direct process containing $\Omega_I$ with that containing $\Omega_S$ [where we would get $\hat j^I_{\alpha}\hat G(\ve_n-\Omega_S)\hat j^S_{\beta}$]. We don't list all the terms here as we don't need them in this work. It is important to emphasize that the above calculation can be carried out in any basis as it is a trace, but then $\hat{\mathbf j}$ would have to be rotated to that basis from the original basis where $\hat{\mathbf j}=\frac{\partial\hat H}{\partial\mathbf A}$.

\paragraph{The eRS response from indirect terms.} Like in mixed terms, the two vertices corresponding to the indirect processes are again composed of the on-shell and off-shell contributions. Near resonance, picking the on-shell contributions within the same approximations as above, $R_{id}(\Omega,\bq)$ can be computed from terms like
\bea\label{eq:ID}
&&\chi^{(1)}_{id}(Q)=\nn\\
&&\int_K{\rm Tr}[\hat j^S\hat G_{K+Q}\hat j^I\hat G_{K+Q+Q_I}\hat j^S\hat G_K\hat j^I\hat G_{K-Q_I}].\nn\\
\eea

\section{Hilbert space factoring of the Raman vertex due to resonance}\label{app:2}
Although we are only interested in the resonant terms, the Coulomb renormalization happens through the term $\Pi_{\alpha\beta;0}$, as mentioned in the main text, and this mimics the structure of the mixed terms with $\hat\gamma_d\rightarrow1$. It will be instructive to understand the evaluation of one of the mixed terms to see how the condition of resonance allows us to factor the Hilbert space into those involving the Fermi surface states and those involving intermediate states. In particular, consider the evaluation of 
\beq\label{eq:Red1}
L(Q)=\int_K{\rm Tr}[\hat j^I\hat G_{K+Q+Q_I}\hat j^S\hat G_K\hat G_{K+Q}].
\eeq
Let us now specify the system in some orbital basis (with components $a,b$ where the vector potential would be coupled) as 
\beq\label{hamil_LK}
H=\left(\begin{array}{cc}
\mathscr{H}_{aa}&\mathscr{H}_{ab}\\
\mathscr{H}_{ba}&\mathscr{H}_{bb}\end{array}\right),
\eeq
where each of the component above has $(2\times 2)$ internal spin structure. We can now introduce an inversion-breaking SOC (whose energy scale can be denoted by $E_{SOC}$). For convenience, let us switch to a basis (which we can call the $c$ and $v$ basis) that would be diagonal in the absence of SOC and look like
\beq\label{hamil_LK2}
H=\left(\begin{array}{cc}
\mathscr{H}_{cc}&0\\
0&\mathscr{H}_{vv}\end{array}\right).
\eeq
Upon adding SOC, this matrix would acquire off-diagonal entries $\mathscr{H}_{cv},\mathscr{H}_{vc}\sim\mathcal{O}(E_{SOC}/E_g)$, where $E_g$ is the gap between the $c$ states and the $v$ states. We could then split Green's functions (in the $cv$ basis) into
\beq\label{eq:Gree}
\hat G_{K}=\left(\begin{array}{cc}
\mathscr{G}^{cc}&0\\
0&\mathscr{G}^{vv}\end{array}\right)+\mathcal{O}\left(\frac{E_{SOC}}{E_g}\right).
\eeq
The transition elements $j^{S,I}$ can also be written in the $cv$ basis as 
$$
\hat j^{x}=\left(\begin{array}{cc}
\mathscr{J}^{x}_{cc}&\mathscr{J}^{x}_{cv}\\
\mathscr{J}^{x}_{vc}&\mathscr{J}^{x}_{vv}\end{array}\right),~~x\in\{I,S\}.
$$
Plugging these forms into Eq. (\ref{eq:Red1}) and evaluating the trace we get
\bea\label{Eq:Red2}
L(Q)&=&\int_K\Big[\mathscr{J}^{I}_{cc}~\mathscr{G}^{cc}_{K+Q+Q_I}~\mathscr{J}^{S}_{cc}\Big]\mathscr{G}^{cc}_{K}\mathscr{G}^{cc}_{K+Q}\nn\\
&&+\int_K\Big[\mathscr{J}^{I}_{cv}~\mathscr{G}^{vv}_{K+Q+Q_I}~\mathscr{J}^{S}_{vc}\Big]\mathscr{G}^{cc}_{K}\mathscr{G}^{cc}_{K+Q}\nn\\
&&+\int_K\Big[\mathscr{J}^{I}_{vc}~\mathscr{G}^{cc}_{K+Q+Q_I}~\mathscr{J}^{S}_{cv}\Big]\mathscr{G}^{vv}_{K}\mathscr{G}^{vv}_{K+Q}\nn\\
&&+\int_K\Big[\mathscr{J}^{I}_{vv}~\mathscr{G}^{vv}_{K+Q+Q_I}~\mathscr{J}^{S}_{vv}\Big]\mathscr{G}^{vv}_{K}\mathscr{G}^{vv}_{K+Q}.\nn\\
\eea
To proceed further, we can also assume a general structure for the $2\times2$ components of $\mathscr(G)_K^{ii}$ as
\begin{equation}\label{ggf}
\mathscr{G}^{ii}_K=\sum_{s=\pm}\frac{1}{i\omega_m-\varepsilon^{i_s}_{\vec k}}\Big[\mathcal{M}^{i_s}_k\Big]_{2\times2},~~~i\in\{c,v\}. 
\end{equation}
The $s=\pm$ denotes the two flavours resulting from the spin degree of freedom. The exact form of $\mathcal{M}^{i_s}_k$ is explicitly dependent on the system of interest and the type of the perturbation. Performing the fermionic Matsubara sums we get
\begin{widetext}
\begin{eqnarray}\label{eq:Red3}
L(Q)=\sum_{s,s',s''=\pm}\int_{\vec k}\frac{\Big[\mathscr{J}^{I}_{cc}\mathcal{M}_{\vec k'}^{c_{s''}}\mathscr{J}^{S}_{cc}\mathcal{M}_{\vec k}^{c_{s}}\mathcal{M}_{\vec k+\vec q}^{c_{s'}}\Big]} {i\Omega_m+\varepsilon^{c_{s}}_{\vec k}-\varepsilon^{c_{s'}}_{\vec k+\vec q}} \left[\frac{n_F(\varepsilon^{c_{s}}_{\vec k})-n_F(\varepsilon^{c_{s''}}_{\vec k+\vec q_I+\vec q})}{i\Omega_m+i\Omega_I+\varepsilon^{c_{s}}_{\vec k}-\varepsilon^{c_{s''}}_{\vec k+\vec q_I+\vec q}}-\frac{n_F(\varepsilon^{c_{s'}}_{\vec k+\vec q})-n_F(\varepsilon^{c_{s''}}_{\vec k+\vec q_I+\vec q})}{i\Omega_I+\varepsilon^{c_{s'}}_{\vec k+\vec q}-\varepsilon^{c_{s''}}_{\vec k+\vec q_I+\vec q}}\right]\notag\\
+\int_{\vec k}\frac{\Big[\mathscr{J}^{I}_{cv}\mathcal{M}_{\vec k'}^{v_{s''}}\mathscr{J}^{S}_{vc}\mathcal{M}_{\vec k}^{c_{s}}\mathcal{M}_{\vec k+\vec q}^{c_{s'}}\Big]}{i\Omega_m+\varepsilon^{c_{s}}_{\vec k}-\varepsilon^{c_{s'}}_{\vec k+\vec q}} \left[\frac{n_F(\varepsilon^{c_{s}}_{\vec k})-n_F(\varepsilon^{v_{s''}}_{\vec k+\vec q_I+\vec q})}{i\Omega_m+i\Omega_I+\varepsilon^{c_{s}}_{\vec k}-\varepsilon^{v_{s''}}_{\vec k+\vec q_I+\vec q}}-\frac{n_F(\varepsilon^{c_{s'}}_{\vec k+\vec q})-n_F(\varepsilon^{v_{s''}}_{\vec k+\vec q_I+\vec q})}{i\Omega_I+\varepsilon^{c_{s'}}_{\vec k+\vec q}-\varepsilon^{v_{s''}}_{\vec k+\vec q_I+\vec q}}\right]\notag\\
+\int_{\vec k}\frac{\Big[\mathscr{J}^{I}_{vc}\mathcal{M}_{\vec k'}^{c_{s''}}\mathscr{J}^{S}_{cv}\mathcal{M}_{\vec k}^{v_{s}}\mathcal{M}_{\vec k+\vec q}^{v_{s'}}\Big]}{i\Omega_m+\varepsilon^{v_{s}}_{\vec k}-\varepsilon^{v_{s'}}_{\vec k+\vec q}} \left[\frac{n_F(\varepsilon^{v_{s}}_{\vec k})-n_F(\varepsilon^{c_{s''}}_{\vec k+\vec q_I+\vec q})}{i\Omega_m+i\Omega_I+\varepsilon^{v_{s}}_{\vec k}-\varepsilon^{c_{s''}}_{\vec k+\vec q_I+\vec q}}-\frac{n_F(\varepsilon^{v_{s'}}_{\vec k+\vec q})-n_F(\varepsilon^{c_{s''}}_{\vec k+\vec q_I+\vec q})}{i\Omega_I+\varepsilon^{v_{s'}}_{\vec k+\vec q}-\varepsilon^{c_{s''}}_{\vec k+\vec q_I+\vec q}}\right]\notag\\
+\int_{\vec k}\frac{\Big[\mathscr{J}^{I}_{vv}\mathcal{M}_{\vec k'}^{v_{s''}}\mathscr{J}^{S}_{vv}\mathcal{M}_{\vec k}^{v_{s}}\mathcal{M}_{\vec k+\vec q}^{v_{s'}}\Big]}{i\Omega_m+\varepsilon^{v_{s}}_{\vec k}-\varepsilon^{v_{s'}}_{\vec k+\vec q}} \left[\frac{n_F(\varepsilon^{v_{s}}_{\vec k})-n_F(\varepsilon^{v_{s''}}_{\vec k+\vec q_I+\vec q})}{i\Omega_m+i\Omega_I+\varepsilon^{v_{s}}_{\vec k}-\varepsilon^{v_{s''}}_{\vec k+\vec q_I+\vec q}}-\frac{n_F(\varepsilon^{v_{s'}}_{\vec k+\vec q})-n_F(\varepsilon^{v_{s''}}_{\vec k+\vec q_I+\vec q})}{i\Omega_I+\varepsilon^{v_{s'}}_{\vec k+\vec q}-\varepsilon^{v_{s''}}_{\vec k+\vec q_I+\vec q}}\right].
\end{eqnarray} 
Here $n_F(\ve)$ is the Fermi function. For definiteness, let us now choose the chemical potential ($\mu$) to lie in the $v$ band. We will also assume $\ve^c_{\vec k}-\ve^v_{\vec k}\sim E_g>0$. Since SOC is a perturbation, we will have $\varepsilon^{c_{s}}_{\vec k}-\varepsilon^{c_{s'}}_{\vec k+\vec q_I+\vec q}\sim E_{SOC}$, and $\varepsilon^{v_{s}}_{\vec k}-\varepsilon^{v_{s'}}_{\vec k+\vec q_I+\vec q}\sim E_{SOC}$. Resonance condition would imply that $\Omega_I\sim E_g$, and we will be interested in Raman shifts $\Omega\sim E_{SOC}$. With the various variables tuned to this regime, the fourth term in Eq. (\ref{eq:Red3}) gives 
\begin{eqnarray}\label{eq:Red4}
L(Q)\Big|_{4^{th}{\text{term}}}&=&\sum_{s,s',s''=\pm}\int_{\vec k}\frac{\Big[\mathscr{J}^{I}_{vv}\mathcal{M}_{\vec k'}^{v_{s''}}\mathscr{J}^{S}_{vv}\mathcal{M}_{\vec k}^{v_{s}}\mathcal{M}_{\vec k+\vec q}^{v_{s'}}\Big]}{i\Omega_m+\varepsilon^{v_{s}}_{\vec k}-\varepsilon^{v_{s'}}_{\vec k+\vec q}}\left[\frac{n_F(\varepsilon^{v_{s}}_{\vec k})-n_F(\varepsilon^{v_{s''}}_{\vec k+\vec q_I+\vec q})}{i\Omega_I}-\frac{n_F(\varepsilon^{v_{s'}}_{\vec k+\vec q})-n_F(\varepsilon^{v_{s''}}_{\vec k+\vec q_I+\vec q})}{i\Omega_I}\right]\nn\\&&+\mathscr{O}\left(\frac{E_{SOC}}{E_g}\right)\nn\\
&=&\sum_{s,s',s''=\pm}\int_{\vec k}\frac{\Big[\mathscr{J}^{I}_{vv}\mathcal{M}_{\vec k'}^{v_{s''}}\mathscr{J}^{S}_{vv}\mathcal{M}_{\vec k}^{v_{s}}\mathcal{M}_{\vec k+\vec q}^{v_{s'}}\Big]} {i\Omega_I}\left[\frac{n_F(\varepsilon^{v_{s}}_{\vec k})-n_F(\varepsilon^{v_{s'}}_{\vec k+\vec q})}{i\Omega_m+\varepsilon^{v_{s}}_{\vec k}-\varepsilon^{v_{s'}}_{\vec k+\vec q}}\right]+\mathscr{O}\left(\frac{E_{SOC}}{E_g}\right).
\end{eqnarray}
Notice the Linhard term in the square braces in the last line of Eq. (\ref{eq:Red4}). Because $\mu$ is assumed to lie in the $v$ band, this term would contribute to the final result due to finite weight from the ph excitations. The first term for $L(Q)$ in Eq. (\ref{eq:Red3}) would also result in a similar expression but with $v\leftrightarrow c$. But the $n_F\rightarrow 0$ for the $c$ bands as those are empty (since $\ve_c>\ve_v$). These terms of Eq. (\ref{eq:Red4}), as is evident, are not resonantly enhanced: they only have a factor of $\Omega_I$ in the denominator. Following similar steps for the second and third terms of $L(Q)$, we are led to 
\begin{eqnarray}\label{eq:Red5}
L(Q)|_{2+3}&=&\sum_{s,s',s''=\pm}\int_{\vec k}\frac{\Big[\mathscr{J}^{I}_{cv}\mathcal{M}_{\vec k'}^{v_{s''}}\mathscr{J}^{S}_{vc}\mathcal{M}_{\vec k}^{c_{s}}\mathcal{M}_{\vec k+\vec q}^{c_{s'}}\Big]}{i\Omega_m+\varepsilon^{c_{s}}_{\vec k}-\varepsilon^{c_{s'}}_{\vec k+\vec q}} \left[\frac{n_F(\varepsilon^{c_{s}}_{\vec k})-n_F(\varepsilon^{v_{s''}}_{\vec k+\vec q_I+\vec q})}{i\Omega_I+E_g}-\frac{n_F(\varepsilon^{c_{s'}}_{\vec k+\vec q})-n_F(\varepsilon^{v_{s''}}_{\vec k+\vec q_I+\vec q})}{i\Omega_I+E_g}\right]\nn\\
&&+\int_{\vec k}\frac{\Big[\mathscr{J}^{I}_{vc}\mathcal{M}_{\vec k'}^{c_{s''}}\mathscr{J}^{S}_{cv}\mathcal{M}_{\vec k}^{v_{s}}\mathcal{M}_{\vec k+\vec q}^{v_{s'}}\Big]}{i\Omega_m+\varepsilon^{v_{s}}_{\vec k}-\varepsilon^{v_{s'}}_{\vec k+\vec q}} \left[\frac{n_F(\varepsilon^{v_{s}}_{\vec k})-n_F(\varepsilon^{c_{s''}}_{\vec k+\vec q_I+\vec q})}{i\Omega_m+i\Omega_I-E_g}-\frac{n_F(\varepsilon^{v_{s'}}_{\vec k+\vec q})-n_F(\varepsilon^{c_{s''}}_{\vec k+\vec q_I+\vec q})}{i\Omega_I-E_g}\right]+\mathscr{O}\left(\frac{E_{SOC}}{E_g}\right).\nn\\
\end{eqnarray}
Close to the resonance condition, the dominant contribution to the Raman bubble arises from the second term in Eq. (\ref{eq:Red5}). Thus,
\begin{eqnarray}
L(Q)\approx\sum_{s,s',s''=\pm}\int_{\vec k}\frac{\Big[\mathscr{J}^{I}_{vc}\mathcal{M}_{\vec k'}^{c_{s''}}\mathscr{J}^{S}_{cv}\mathcal{M}_{\vec k}^{v_{s}}\mathcal{M}_{\vec k+\vec q}^{v_{s'}}\Big]}{i\Omega_m+\varepsilon^{v_{s}}_{\vec k}-\varepsilon^{v_{s'}}_{\vec k+\vec q}} \left[\frac{n_F(\varepsilon^{v_{s}}_{\vec k})-n_F(\varepsilon^{c_{s''}}_{\vec k+\vec q_I+\vec q})}{i\Omega_m+i\Omega_I-E_g}-\frac{n_F(\varepsilon^{v_{s'}}_{\vec k+\vec q})-n_F(\varepsilon^{c_{s''}}_{\vec k+\vec q_I+\vec q})}{i\Omega_I-E_g}\right]\notag\\
\end{eqnarray}
We expect the intermediate states to be broad and thus have a lifetime $\sim1/\Gamma$. If we are in a regime such that $E_{SOC}\ll \Gamma$, we can further ignore the $\Omega_m$ in the denominator above to get (after continuing $\Omega_I$ to real frequencies)
\begin{eqnarray}\label{do_bu}
L(Q)&\approx&\sum_{s,s',s''=\pm}\int_{\vec k}\frac{\Big[\mathscr{J}^{I}_{vc}\mathcal{M}_{\vec k'}^{c_{s''}}\mathscr{J}^{S}_{cv}\mathcal{M}_{\vec k}^{v_{s}}\mathcal{M}_{\vec k+\vec q}^{v_{s'}}\Big]}{\Omega_I-E_g+i\Gamma} \left[\frac{n_F(\varepsilon^{v_{s}}_{\vec k})-n_F(\varepsilon^{v_{s'}}_{\vec k+\vec q})}{i\Omega_m+\varepsilon^{v_{s}}_{\vec k}-\varepsilon^{v_{s'}}_{\vec k+\vec q}}\right].
\end{eqnarray}
\end{widetext}
The above expression implies that under the resonant condition, the bubble involving a direct and an indirect vertex reduces to a form containing the Linhard factor created out of states near the Fermi level, the resonance enhancement factor, and the precise form associated with the transition from the states near the Fermi level to the intermediate state and back ($\mathscr{J}_{vc}$ and $\mathscr{J}_{cv}$). The main thing to note is that the original Hilbert space involving both the $c$ and $v$ states is factored into a particle-hole contribution from the subspace of states near the Fermi surface and a transition factor associated with $c$-$v$ transitions weighted by the resonance factor. This factorization allows us to decouple the light-excitation processes (of the order of the laser frequency) from the quasiparticle excitations (of the order of the Raman shifts). There are other terms of the order of $\mathcal{O}(E_{SOC}/E_g)$ and $\mathcal{O}(E_{SOC}/\Gamma)$ which are ignored here, but they will only provide corrections to the physical effects captured by the resonant enhancement term. It should be stated that the form is well expected and has been known since the study of Resonant Raman in semiconductors. However, the transition element was only discussed for III-V semiconductors where it was modelled as a constant which is usually referred to as the Kane parameter\cite{Kane1957}. In the above derivation, we have access to the precise form of these transition elements. 

Finally, we note that the above form was obtained for a form of the Greens function as in Eq. (\ref{eq:Gree}). In general, we would have the block-off-diagonal terms $\mathscr{G}_{cv}$ to be non-zero (they are induced by SOC after all). An explicit evaluation of $L(Q)$ in such a case would involve 32 terms. Four of the relevant ones are accounted for above, and the other terms look like 
\bea\label{eq:otherterms}
T_1&:&\int_K\Big[\mathscr{J}^{I}_{vv}~\mathscr{G}^{vc}_{K+Q+Q_I}~\mathscr{J}^{S}_{cv}\Big]\mathscr{G}^{vv}_{K}\mathscr{G}^{vv}_{K+Q},\nn\\
T_2&:&\int_K\Big[\mathscr{J}^{I}_{vv}~\mathscr{G}^{vv}_{K+Q+Q_I}~\mathscr{J}^{S}_{vc}\Big]\mathscr{G}^{cv}_{K}\mathscr{G}^{vv}_{K+Q}.
\eea
In the $T_1$ family of terms, the low energy quasiparticle response comes from well-defined poles $\mathscr{G}^{vv}_{K}\mathscr{G}^{vv}_{K+Q}$. But note that there is no $c\rightarrow v$ transition due to the incoming light as $\mathscr{J}^I$ is associated with the $vv$ index. It is the Greens function that propagates an electron from one band to the other. This adversely affects the resonance enhancement that we were able to get with the other four terms we considered above. Further, a $T_1$ type of term that is allowed in general due to the existence of $\mathscr{G}_{cv}$ term, however, is also suppressed by the gap, $E_g$, between the $c$ and the $v$ bands. In the $T_2$ type of terms the low energy quasiparticle response is already incoherent due to the $\mathscr{G}^{cv}_{K}\mathscr{G}^{vv}_{K+Q}$ structure. Thus, near the resonance, only the four terms that we considered are the relevant processes. Even though not all the four terms contributed in the Eq. (\ref{do_bu}), they would do so if the resonance was considered with bands below the Fermi level. But the $T_1$ and $T_2$ type of terms still would not contribute.

\section{Computation of polarization-charge bubble}\label{app:3}
As is evident from Eq. (\ref{eq:dressed Raman susceptibilty2}), to couple the plasmon pole to the Raman response, we need, what we call the polarization-charge bubble $\Pi_{\alpha\beta;0}$. This bubble is a trace over the structure of $L(Q)$ computed above with $\mathscr{J}$ acquiring the indices $\alpha,\beta$. Let us try to evaluate this expression. First, note that the $s''$ dependence in the Fermi functions drops out under the approximations we made above. Then, by properties of Greens' functions, $\sum_{s''}\mathcal M^{c_s''}_{\vec k}=1$. This leads to the expression
\begin{eqnarray}\label{do_bu2}
L(Q)&\approx&\frac{1}{\Omega_I-E_g+i\Gamma}\sum_{s,s'}\int_{\vec k} \Big[\mathscr{J}^{I}_{vc}\mathscr{J}^{S}_{cv}\Big]\Big[\mathcal{M}_{\vec k}^{v_{s}}\mathcal{M}_{\vec k+\vec q}^{v_{s'}}\Big]\times\nn\\
&&\left[\frac{n_F(\varepsilon^{v_{s}}_{\vec k})-n_F(\varepsilon^{v_{s'}}_{\vec k+\vec q})}{i\Omega_m+\varepsilon^{v_{s}}_{\vec k}-\varepsilon^{v_{s'}}_{\vec k+\vec q}}\right].
\end{eqnarray}
It should be kept in mind that the matrix $\Big[\mathscr{J}^{I}_{vc}\mathscr{J}^{S}_{cv}\Big]$ is $\vec k$-dependent and hence cannot be factored out. However, without the $\mathscr{J}\mathscr{J}$, the expression would simply be exactly what arises in the computation of the charge bubble $\int_K {\rm Tr}[\hat G_K\hat G_{K+Q}]$ prior to the trace operation. In fact, since $\mathscr{J}\mathscr{J}$ is a $2\times2$ matrix, we can write \beq\label{eq:deff}\Big[\mathscr{J}^{I}_{\beta,vc}\mathscr{J}^{S}_{\alpha,cv}\Big]=\mathcal{F}^{\beta\alpha}_i(\vec k)\hat\sigma_i.\eeq This allows us to write 
\begin{eqnarray}\label{do_bu3}
L(Q)&\approx&\frac{1}{\Omega_I-E_g+i\Gamma}\sum_{s,s'}\int_{\vec k} \mathcal{F}^{\beta\alpha}_i(\vec k)\Big[\hat\sigma_i\mathcal{M}_{\vec k}^{v_{s}}\mathcal{M}_{\vec k+\vec q}^{v_{s'}}\Big]\times\nn\\
&&\left[\frac{n_F(\varepsilon^{v_{s}}_{\vec k})-n_F(\varepsilon^{v_{s'}}_{\vec k+\vec q})}{i\Omega_m+\varepsilon^{v_{s}}_{\vec k}-\varepsilon^{v_{s'}}_{\vec k+\vec q}}\right].
\end{eqnarray}
If $\mathcal{F}^{\beta\alpha}(\vec k)$ has an $s$-wave component denoted by $\mathcal{F}^{\beta\alpha}$, then we arrive at
\begin{eqnarray}\label{do_bu4}
L(Q)&\approx&\frac{\mathcal F_i^{\beta\alpha}}{\Omega_I-E_g+i\Gamma}\sum_{s,s'}\int_{\vec k}\Big[\hat\sigma_i\mathcal{M}_{\vec k}^{v_{s}}\mathcal{M}_{\vec k+\vec q}^{v_{s'}}\Big]\times\nn\\
&&\left[\frac{n_F(\varepsilon^{v_{s}}_{\vec k})-n_F(\varepsilon^{v_{s'}}_{\vec k+\vec q})}{i\Omega_m+\varepsilon^{v_{s}}_{\vec k}-\varepsilon^{v_{s'}}_{\vec k+\vec q}}\right]+h.h.\nn\\
&=& \frac{\mathcal F_i^{\beta\alpha}}{\Omega_I-E_g+i\Gamma}\int_{K}[\hat\sigma_i\hat G_K\hat\sigma_0\hat G_{K+Q}]+h.h.\nn\\
\end{eqnarray}
The $h.h.$ stands for higher harmonics which usually integrate to zero for most systems. Upon taking trace we get
\begin{eqnarray}\label{do_bu5}
\Pi_{\alpha\beta;0}(Q)&\approx&m_e{\rm Tr}[L(Q)]\nn\\
&\approx& \frac{\mathcal F_i^{\beta\alpha}\Pi_{i0}(Q)}{\Omega_I-E_g+i\Gamma}.
\end{eqnarray}
This is the form used in the main text. In the case of III-V semiconductors, $\vec p^\alpha\e^\alpha\rightarrow \mathcal P\vec\e$. Thus, following the discussion after Eq. (\ref{eq:f}) in the main text, we can write $\mathcal F^{\beta\alpha}_i\e^\alpha_I\e^\beta_S\rightarrow \mathcal P(\vec \e_I\times\vec \e_S)_i$,  restoring the conventional result.

\section{Calculation details for SOC Dirac system}\label{app:Graphene}
Starting from Eq. (\ref{hamil_gra}), we note that $\hat{\mathbf j}=v_F(\tau_z\hat\sigma_x \hat x+\hat\sigma_y \hat y$. We move to the eigenbasis using the transformation 
\beq\label{eq:trans}
\hat{\mathbf j}^{(cv)}=M^\dag \hat{\mathbf j}^{(ab)}M,
\eeq
where the matrix $M$ is formed by column juxtaposition of the eigenvectors of $H$. In the absence of $\Delta_R$ and $\Delta_Z$, the transformed current becomes
\bea\label{eq:jformx}
\mathscr{J}^{0}_{x,cc}&=&v_F\frac{v_Fk}{\xi_k}\cos\theta~\hat\sigma_0\nn\\
\mathscr{J}^{0}_{x,cv}&=&v_F(i\tau_z\sin\theta+\frac{\Delta}{\xi_k}\cos\theta)\hat\sigma_0\nn\\
\mathscr{J}^{0}_{x,vv}&=&-\mathscr{J}^{0}_{x,cc},\nn\\
\mathscr{J}^{0}_{x,vc}&=&\left(\mathscr{J}^{0}_{x,cv}\right)^\dag.
\eea
Similarly, the various components of $j_y$ current elements are,
\bea\label{eq:jformy}
\mathscr{J}^{0}_{y,cc}&=&v_F\frac{v_Fk}{\xi_k}\sin\theta~\hat s_0\nn\\
\mathscr{J}^{0}_{y,cv}&=&v_F(-i\tau_z\cos\theta+\frac{\Delta}{\xi_k}\sin\theta)\hat s_0\nn\\
\mathscr{J}^{0}_{y,vv}&=&-\mathscr{J}^{0}_{y,cc},\nn\\
\mathscr{J}^{0}_{y,vc}&=&\left(\mathscr{J}^{0}_{y,cv}\right)^\dag.
\eea
Here $\theta$ is the azimuthal angle of $\vec k$. Note the presence of the interband components, $\mathscr{J}_{cv,vc}$, that is responsible for optical interband transitions (as in Graphene). However, there is no spin-splitting in these blocks. When we include $\Delta_R$ in the Hamiltonian, the $M$-transformed current now becomes
\bea\label{eq:jformxR}
\mathscr{J}_{x,cc}&=&\mathscr{J}^{0}_{x,cc}+v_F\frac{\Delta_R}{2\xi_k}\left(\frac{\Delta^2}{\xi_k^2}\cos\theta\hat s_z-\sin\theta\hat s_y\right)\nn\\
\mathscr{J}_{x,cv}&=&\mathscr{J}^{0}_{x,cv}-v_F\frac{\Delta_R}{2\xi_k}\times\nn\\
&&\left(\frac{v_Fk\Delta }{\xi_k^2}\cos\theta\hat s_z+\left[\frac{\Delta}{v_Fk}\sin\theta-i\frac{\tau_z\xi_k}{v_Fk}\cos\theta\right]\hat s_y\right)\nn\\
\mathscr{J}_{x,vv}&=&-\mathscr{J}_{x,cc},\nn\\
\mathscr{J}_{x,vc}&=&\left(\mathscr{J}_{x,cv}\right)^\dagger.
\eea
Similarly, the various components of $j_y$ current elements are,
\bea\label{eq:jformyR}
\mathscr{J}_{y,cc}&=&\mathscr{J}^{0}_{y,cc}+v_F\frac{\Delta_R}{2\xi_k}\left(\frac{\Delta^2}{\xi_k^2}\sin\theta\hat s_z+\cos\theta\hat s_y\right)\nn\\
\mathscr{J}^{0}_{y,cv}&=&\mathscr{J}^{0}_{y,cv}-v_F\frac{\Delta_R}{2\xi_k}\times\nn\\
&&\left(\frac{v_Fk\Delta }{\xi_k^2}\sin\theta\hat s_z-\left[\frac{\Delta}{v_Fk}\cos\theta+i\frac{\tau_z\xi_k}{v_Fk}\sin\theta\right]\hat s_y\right)\nn\\
\mathscr{J}_{y,vv}&=&-\mathscr{J}_{y,cc},\nn\\
\mathscr{J}_{y,vc}&=&\left(\mathscr{J}_{y,cv}\right)^\dagger.
\eea

Note that the $\mathscr{J}_{cv}=\mathscr{J}_{vc}^\dag$ blocks are now spin-dependent. Also, as discussed in the main text, in the prefactors of the interband $\mathscr J_{cv}$'s, the $\Delta_R$'s get promoted to the interband $\delta_R$. Using this form for $\hat{\mathbf{j}}$ and Eq. (\ref{eq:f}), we arrive at the following forms for $\mathcal{F}^{\alpha\beta}_i(\vec k)$.
\bea\label{eq:Fs}
\mathcal{F}^{XX}_0&=&\frac{v_F^2\Delta^2}{2\xi_k^2}+\frac{v_F^4k^2\sin^2\theta}{2\xi_k^2},\nn\\
\mathcal{F}^{YY}_0&=&\frac{v_F^2\Delta^2}{2\xi_k^2}+\frac{v_F^4k^2\cos^2\theta}{2\xi_k^2},\nn\\
\mathcal{F}^{XY}_0&=&\frac{i\tau_zv_F^2\Delta}{\xi_k}-\frac{v_F^4k^2\sin2\theta}{2\xi_k^2},\nn\\
\mathcal{F}^{\beta\alpha}_1&=&0,\nn\\
\mathcal{F}^{XX}_2&=&\frac{\delta_Rv_F^3k\sin2\theta}{2\xi_k^2},\nn\\
\mathcal{F}^{YY}_2&=&-\frac{\delta_Rv_F^3k\sin2\theta}{2\xi_k^2},\nn\\
\mathcal{F}^{XY}_2&=&-\frac{\delta_Rv_F^3k\cos2\theta}{2\xi_k^2},\nn\\
\mathcal{F}^{XX}_3&=&-\frac{\delta_Rv_F^3k\Delta^2\cos^2\theta}{\xi_k^4},\nn\\
\mathcal{F}^{YY}_3&=&-\frac{\delta_Rv_F^3k\Delta^2\sin^2\theta}{\xi_k^4},\nn\\
\mathcal{F}^{XY}_3&=&-\frac{i\tau_z\delta_Rv_F^3k\Delta}{2\xi_k^3}-\frac{\delta_Rv_F^3k\Delta^2\sin2\theta}{2\xi_k^4}.
\eea

In the main text, we only keep the $s$-wave component of the above terms. The other angle-dependent terms evaluate to zero in the bubble of interest and are expected to be small even when non-resonant terms are accounted for.

When we consider only the Valley Zeeman term, the blocks of the Hamiltonian takes the form
\bea
H^{cc}=\hat{s}_0\xi_k+\frac{\Delta_Z}{2}\hat{s}_z,&&H^{vv}=-\hat{s}_0\xi_k+\frac{\Delta_Z}{2}\hat{s}_z,\nn\\
H^{cv}=0,&&H^{vc}=0.
\eea
Since there is no $\vec k$-dependent correction, the expression for the current remains the same as Eqs. (\ref{eq:jformx}) and (\ref{eq:jformy}) where $\mathscr{J}_{cv}=0=\mathscr{J}_{vc}$. Likewise $\mathcal F^{\alpha\beta}_0$ remains the same as in Eq. (\ref{eq:Fs}) and $\mathcal F^{\alpha\beta}_i=0$ for $i=1,2,3$.\\

\section{Calculation details for BiTeI}\label{app:BiTeI}
In this section we present calculation details of $\Pi_{i0}(Q)$ for $i\in\{0,1,2,3\}$. First, we observe that all the integrals involved in the computation of $\Pi_{i0}$ are convergent in $\bk$ integration. We shall perform a change of variable of $\vec k\rightarrow\vec k-\vec q/2$. This makes the calculation a lot simpler. For brevity, let us define 
\beq\label{eq:brevity}
L_{s,s'}(\vec k,\vec q)\equiv\frac{n_F(\varepsilon^s_{\vec k-\vec q/2})-n_F(\varepsilon^{s'}_{\vec k+\vec q/2})}{i\Omega_m+\varepsilon^s_{\vec k-\vec q/2}-\varepsilon^{s'}_{\vec k+\vec q/2}}.
\eeq
Then we have
\beq\label{eq:defi}
\Pi_{i0}(Q)=\sum_{ss'}\int_{\vec k} \mathcal{N}^i_{s,s'}(\vec k,\vec q)L_{s,s'}(\vec k,\vec q).
\eeq

\subsection{Out-of-plane $\Pi_{i0}(Q)$:}
Since we are interested in small $q_3$, to leading orders in it we have
\bea\label{eq:Lxp}
L_{s,s}(\vec k,\vec q)&=&-\partial_{\varepsilon^s_{\vec k}} n_F(\varepsilon^s_{\vec k})\left[\frac{\partial_{k_3}\varepsilon^s_{\vec k}~q_3}{i\Omega_m}-\frac{(\partial_{k_3}\varepsilon^s_{\vec k})^2~q_3^2}{\Omega_m^2}\right]\nn\\
&&~~~~~~~~~~~~~~~~~~~~~~~~~~~~~~~~~~~~~~~+\mathcal{O}(q_3^3)\nn\\
L_{s,s'}(\vec k,\vec q)&=&\left[\frac{n_F(\varepsilon^s_{\vec k})-n_F(\varepsilon^{s'}_{\vec k})}{i\Omega_m+(s-s')D_k}\right]+\mathcal{O}(q_3).
\eea
We have retained different orders in different terms so as to ensure that we get the appropriate leading order terms in the full calculation of $\Pi_{i0}$. Next we calculate the various $\mathcal N^i_{ss'}$ factors to leading order in $q_3$:
\bea\label{eq:N}
\mathcal{N}^0_{++}&=&1+\mathcal{O}(q_3^2),\nn\\
\mathcal{N}^0_{--}&=&\mathcal{N}^0_{++},\nn\\
\mathcal{N}^0_{+-}&=&\frac{\alpha^2k_{||}^2\lambda^2q_3^2}{4D_k^4}+\mathcal{O}(q_3^3),\nn\\
\mathcal{N}^0_{-+}&=&\mathcal{N}^0_{+-},\nn\\
\mathcal{N}^1_{++}&=&\frac{\alpha k_2}{D_k}-i\frac{\alpha k_1\lambda q_3}{2(D_k)^2} +\mathcal{O}(q_3^2),\nn\\
\mathcal{N}^1_{--}&=&-(\mathcal{N}^1_{+,+})^*,\nn\\
\mathcal{N}^1_{+-}&=&\frac{i\alpha k_1(D_k)+\alpha\lambda k_2k_3}{2(D_k)^{3}}\lambda q_3+\mathcal{O}(q_3^3),\nn\\
\mathcal{N}^1_{-+}&=&-(\mathcal{N}^1_{+-})^*,\nn\\
\mathcal{N}^2_{ss'}&=&\mathcal{N}^1_{ss'}(k_1\leftrightarrow k_2),\nn\\
\mathcal{N}^3_{++}&=&\frac{\lambda k_3}{(D_k)}+\mathcal{O}(q_3^2),\nn\\
\mathcal{N}^3_{--}&=&-\mathcal{N}^3_{++},\nn\\
\mathcal{N}^3_{+-}&=&-\frac{\alpha^2k_{||}^2}{2(D_k)^{3}}\lambda q_3+\mathcal{O}(q_3^3),\nn\\
\mathcal{N}^3_{-+}&=&-\mathcal{N}^3_{+-}.
\eea
To calculate $\Pi_{00}(Q)$, the leading order contribution is $\mathcal{O}(q_3^2)$ and it is given by (below we have assumed that our chemical potential is in the lower band $\ve^-_{\vec k}$)
\bea\label{eq:Pi00}
\Pi_{00}(Q)&=&q_3^2\int_{\vec k}[-\partial_{\ve^-} n_F(\ve^-)]\frac{(\partial_{k_3}\ve^-)^2}{-\Omega_m^2}\nn\\
&&+\lambda^2q_3^2\int_{\vec k}\frac{[n_F(\ve^+)-n_F(\ve^-)]}{D_k^3}\frac{\alpha^2k^2_\parallel}{\Omega_m^2+4D_k^2}.\nn\\
\eea
Notice that this is like the single-band contribution plus a SOC-induced interband correction. Notice also that this correction would be absent if $\lambda=0$. To calculate $\Pi_{10}(Q)$ observe that the inter and intra band terms are odd functions of $k_1$ and $k_2$. The $L_{ss}$ function only has a term that is odd in $k_3$. Thus, $\Pi_{10}(Q)$ vanishes for $\vec q= (0,0,q_3)$. Since $\Pi_{20}(Q)$ simply interchanges $k_1\leftrightarrow k_2$, it also vanishes. Finally, to calculate $\Pi_{30}(Q)$ we only need contribution to $\mathcal O(q_3)$. Using the $\mathcal N_{ss'}$ and the $L_{ss'}$ functions above, we arrive at 
\bea\label{eq:Pi30}
\Pi_{30}(Q)&=&-\lambda q_3\int_{\vec k}\frac{[-\partial_{\ve^-} n_F(\ve^-)]}{D_k}\frac{k_3\partial_{k_3}\ve^-}{i\Omega_m}\nn\\
&&+\alpha^2\lambda q_3\int_{\vec k}\frac{[n_F(\ve^+)-n_F(\ve^-)]}{D_k^3}\frac{i\Omega_m k^2_\parallel}{\Omega_m^2+4D_k^2}.\nn\\
\eea



\subsection{In-plane $\Pi_{i0}(Q)$:}
For completeness, we can also look at small in-plane momentum transfer $q_1$. To leading orders in $q_1$ we have
\bea\label{eq:Lxp2}
L_{s,s}(\vec k,\vec q)&=&-\partial_{\varepsilon^s_{\vec k}} n_F(\varepsilon^s_{\vec k})\left[\frac{\partial_{k_1}\varepsilon^s_{\vec k}~q_1}{i\Omega_m}-\frac{(\partial_{k_1}\varepsilon^s_{\vec k})^2~q_1^2}{\Omega_m^2}\right]\nn\\
&&~~~~~~~~~~~~~~~~~~~~~~~~~~~~~~~~~~~~~~~+\mathcal{O}(q_1^3)\nn\\
L_{s,s'}(\vec k,\vec q)&=&\left[\frac{n_F(\varepsilon^s_{\vec k})-n_F(\varepsilon^{s'}_{\vec k})}{i\Omega_m+(s-s')D_k}\right]+\mathcal{O}(q_1).
\eea

Next we calculate the various $\mathcal N^i_{ss'}$ factors to leading order in $q_1$:
\bea\label{eq:N2}
\mathcal{N}^0_{++}&=&1+\mathcal{O}(q_1^2),\nn\\
\mathcal{N}^0_{--}&=&\mathcal{N}^0_{++},\nn\\
\mathcal{N}^0_{+-}&=&\frac{\alpha^2(\alpha^2k_2^2+\lambda^2k_3^2)q_1^2}{4D_k^4}+\mathcal{O}(q_1^3),\nn\\
\mathcal{N}^0_{-+}&=&\mathcal{N}^0_{+-},\nn\\
\mathcal{N}^1_{++}&=&\frac{\alpha k_2}{D_k}+i\frac{\alpha k_3\lambda q_1}{2(D_k)^2} +\mathcal{O}(q_1^2),\nn\\
\mathcal{N}^1_{--}&=&-(\mathcal{N}^1_{+,+})^*,\nn\\
\mathcal{N}^1_{+-}&=&\frac{-i\alpha\lambda k_3(D_k)+\alpha^3 k_1k_2}{2(D_k)^{3}}q_1+\mathcal{O}(q_1^3),\nn\\
\mathcal{N}^1_{-+}&=&-(\mathcal{N}^1_{+-})^*,\nn\\
\mathcal{N}^2_{++}&=&-\frac{\alpha k_1}{D_k} +\mathcal{O}(q_1^2),\nn\\
\mathcal{N}^2_{--}&=&-\mathcal{N}^1_{+,+},\nn\\
\mathcal{N}^2_{+-}&=&\frac{\alpha(\alpha^2k_2^2+\lambda^2k_3^2)q_1}{2D_k^3}+\mathcal{O}(q_1^3),\nn\\
\mathcal{N}^2_{-+}&=&-\mathcal{N}^1_{+-},\nn\\
\mathcal{N}^3_{++}&=&\frac{\lambda k_3}{(D_k)}-i\frac{\alpha^2k_2q_1}{2(D_k)^2}+\mathcal{O}(q_1^2),\nn\\
\mathcal{N}^3_{--}&=&-(\mathcal{N}^3_{++})^*,\nn\\
\mathcal{N}^3_{+-}&=&\frac{\alpha^2}{2}\left(\frac{\lambda k_1k_3}{2(D_k)^{3}} + \frac{ik_2}{2(D_k)^{2}}\right)q_1+\mathcal{O}(q_3^3),\nn\\
\mathcal{N}^3_{-+}&=&-(\mathcal{N}^3_{+-})^*.
\eea

These can be used to calculate $\Pi_{00}(Q)$ to the leading order contribution to $\mathcal{O}(q_1^2)$ and it is given by
\bea\label{}
\Pi_{00}(Q)&=&q_1^2\int_{\vec k}[-\partial_{\ve^-} n_F(\ve^-)]\frac{(\partial_{k_1}\ve^-)^2}{-\Omega_m^2}\nn\\
&&+\alpha^2q_1^2\int_{\vec k}\frac{[n_F(\ve^+)-n_F(\ve^-)]}{D_k^3}\frac{\alpha^2k^2_2+\lambda^2k^2_3}{\Omega_m^2+4D_k^2}.\nn\\
\eea

For $\Pi_{10}(Q)$ and $\Pi_{30}(Q)$ observe that the inter and intra band terms are odd functions of $k_2$ and $k_3$. The $L_{ss}$ function only has a term that is odd in $k_1$. Thus, they vanish for $\vec q= (q_1,0,0)$. Finally, to calculate $\Pi_{20}(Q)$ we only need contribution to $\mathcal O(q_1)$. Using the $\mathcal N_{ss'}$ and the $L_{ss'}$ functions above, we arrive at 
\bea\label{}
\Pi_{20}(Q)&=&-\alpha q_1\int_{\vec k}\frac{[-\partial_{\ve^-} n_F(\ve^-)]}{D_k}\frac{k_1\partial_{k_1}\ve^-}{i\Omega_m}\nn\\
&&+\alpha q_1\int_{\vec k}\frac{[n_F(\ve^+)-n_F(\ve^-)]}{D_k^3}\frac{i\Omega_m (\alpha^2k_2^2+\lambda^2k_3^2)}{\Omega_m^2+4D_k^2}.\nn\\
\eea
As expected, the charge-charge susceptibility has a correction $\propto \alpha^2$ and the spin-charge susceptibility is $\propto \alpha$. In this case we don't need any canting effect (from the $\lambda$-term).

\section{Derivation of Plasma frequency}\label{app:omega_pl}
We need to evaluate $\Pi_{00}(\bq,\Omega+i0^+)$. It is instructive to rescale the integration variables and integrands to dimensionless variables as such: $\tilde \varepsilon\equiv\varepsilon/|\mu|$, $\tilde k_i\equiv k_i/\sqrt{2m_i|\mu|}$ ($i\in\{1,2,3\}$), $\tilde T=T/|\mu|$, $\tilde \alpha\equiv\alpha_R \sqrt{2m_1/|\mu|}$ and $\tilde \lambda\equiv\lambda \sqrt{2m_3/|\mu|}$, where $\mu$ is the chemical potential relative to the Dirac point. 
This leads to    
\begin{align*}V_{\bq}\Pi_{00}(\bq,\Omega)&=\frac{e^2}{\epsilon_\infty}\frac{\sqrt{(2m_1|\mu|)(2m_1|\mu|)(2m_3|\mu|)}}{(2m_q|\mu|)}\frac1\mu\left[\frac1{\tilde q^2}\tilde\Pi_{00}\right].
\end{align*}
where $m_q$ is the effective mass along the direction of $\bq$, and $\tilde\Pi_{00}$ is the same as $\Pi_{00}$ [Eq. (\ref{eq:Pi00})], but expressed in terms of dimensionless variable $\tilde\e$, $\tilde T$, $\tilde\alpha$ and $\tilde \lambda$. We can rewrite the above expression as
\begin{align*}V_{\bq}\Pi_{00}&=\frac{\Omega_0^2}{\Omega^2}\frac{\sqrt{m_1^*m_1^*m_3^*}}{m_q^*}\underbrace{\left[3\pi^2\frac{\tilde\Omega^2}{\tilde q^2}\tilde\Pi_{00}\right]}_{f_{pl}(\tilde \alpha, \tilde\lambda)},
\end{align*}
where $m^*_i=m_i/m_0$ is the effective mass. Since in our problem, $\bq$ is along the $3$-axis, we have $m^*_q=m^*_3$. Solving for $V_{\bq}\Pi_{00}=1$ we arrive at the Eq. (\ref{eq:plasmafreq}).

\bibliography{ResRaman}

\begin{thebibliography}{72}%
\makeatletter
\providecommand \@ifxundefined [1]{%
 \@ifx{#1\undefined}
}%
\providecommand \@ifnum [1]{%
 \ifnum #1\expandafter \@firstoftwo
 \else \expandafter \@secondoftwo
 \fi
}%
\providecommand \@ifx [1]{%
 \ifx #1\expandafter \@firstoftwo
 \else \expandafter \@secondoftwo
 \fi
}%
\providecommand \natexlab [1]{#1}%
\providecommand \enquote  [1]{``#1''}%
\providecommand \bibnamefont  [1]{#1}%
\providecommand \bibfnamefont [1]{#1}%
\providecommand \citenamefont [1]{#1}%
\providecommand \href@noop [0]{\@secondoftwo}%
\providecommand \href [0]{\begingroup \@sanitize@url \@href}%
\providecommand \@href[1]{\@@startlink{#1}\@@href}%
\providecommand \@@href[1]{\endgroup#1\@@endlink}%
\providecommand \@sanitize@url [0]{\catcode `\\12\catcode `\$12\catcode
  `\&12\catcode `\#12\catcode `\^12\catcode `\_12\catcode `\%12\relax}%
\providecommand \@@startlink[1]{}%
\providecommand \@@endlink[0]{}%
\providecommand \url  [0]{\begingroup\@sanitize@url \@url }%
\providecommand \@url [1]{\endgroup\@href {#1}{\urlprefix }}%
\providecommand \urlprefix  [0]{URL }%
\providecommand \Eprint [0]{\href }%
\providecommand \doibase [0]{https://doi.org/}%
\providecommand \selectlanguage [0]{\@gobble}%
\providecommand \bibinfo  [0]{\@secondoftwo}%
\providecommand \bibfield  [0]{\@secondoftwo}%
\providecommand \translation [1]{[#1]}%
\providecommand \BibitemOpen [0]{}%
\providecommand \bibitemStop [0]{}%
\providecommand \bibitemNoStop [0]{.\EOS\space}%
\providecommand \EOS [0]{\spacefactor3000\relax}%
\providecommand \BibitemShut  [1]{\csname bibitem#1\endcsname}%
\let\auto@bib@innerbib\@empty
\bibitem [{\citenamefont {Abstreiter}\ \emph {et~al.}(1984)\citenamefont
  {Abstreiter}, \citenamefont {Cardona},\ and\ \citenamefont
  {Pinczuk}}]{Abstreiter1984}%
  \BibitemOpen
  \bibfield  {author} {\bibinfo {author} {\bibfnamefont {G.}~\bibnamefont
  {Abstreiter}}, \bibinfo {author} {\bibfnamefont {M.}~\bibnamefont
  {Cardona}},\ and\ \bibinfo {author} {\bibfnamefont {A.}~\bibnamefont
  {Pinczuk}},\ }\bibinfo {title} {Light scattering by free carrier excitations
  in semiconductors},\ in\ \href {https://doi.org/10.1007/3-540-11942-6_20}
  {\emph {\bibinfo {booktitle} {Light Scattering in Solids IV: Electronics
  Scattering, Spin Effects, SERS, and Morphic Effects}}},\ \bibinfo {editor}
  {edited by\ \bibinfo {editor} {\bibfnamefont {M.}~\bibnamefont {Cardona}}\
  and\ \bibinfo {editor} {\bibfnamefont {G.}~\bibnamefont {G{\"u}ntherodt}}}\
  (\bibinfo  {publisher} {Springer Berlin Heidelberg},\ \bibinfo {address}
  {Berlin, Heidelberg},\ \bibinfo {year} {1984})\ pp.\ \bibinfo {pages}
  {5--150}\BibitemShut {NoStop}%
\bibitem [{\citenamefont {Hayes}\ and\ \citenamefont
  {Loudon}(2012)}]{BookHayes2012}%
  \BibitemOpen
  \bibfield  {author} {\bibinfo {author} {\bibfnamefont {W.}~\bibnamefont
  {Hayes}}\ and\ \bibinfo {author} {\bibfnamefont {R.}~\bibnamefont {Loudon}},\
  }\href {https://books.google.ca/books?id=8N4rU\_gtHgAC} {\emph {\bibinfo
  {title} {Scattering of Light by Crystals}}},\ Dover Books on Physics\
  (\bibinfo  {publisher} {Courier Corporation},\ \bibinfo {year}
  {2012})\BibitemShut {NoStop}%
\bibitem [{\citenamefont {Placzek}(1959)}]{placzek1959rayleigh}%
  \BibitemOpen
  \bibfield  {author} {\bibinfo {author} {\bibfnamefont {G.}~\bibnamefont
  {Placzek}},\ }\href@noop {} {\emph {\bibinfo {title} {{The Rayleigh and Raman
  scattering}}}},\ Vol.\ \bibinfo {volume} {526}\ (\bibinfo  {publisher}
  {Lawrence Radiation Laboratory},\ \bibinfo {year} {1959})\BibitemShut
  {NoStop}%
\bibitem [{\citenamefont {Bairamova}\ \emph {et~al.}(1993)\citenamefont
  {Bairamova}, \citenamefont {Ipatovaa},\ and\ \citenamefont
  {Voitenko}}]{Bairamova1993}%
  \BibitemOpen
  \bibfield  {author} {\bibinfo {author} {\bibfnamefont {B.}~\bibnamefont
  {Bairamova}}, \bibinfo {author} {\bibfnamefont {I.}~\bibnamefont
  {Ipatovaa}},\ and\ \bibinfo {author} {\bibfnamefont {V.}~\bibnamefont
  {Voitenko}},\ }\href@noop {} {\bibfield  {journal} {\bibinfo  {journal}
  {Physics Reports}\ }\textbf {\bibinfo {volume} {229}},\ \bibinfo {pages}
  {221} (\bibinfo {year} {1993})}\BibitemShut {NoStop}%
\bibitem [{\citenamefont {Ivchenko}(2004)}]{Ivchenko2004}%
  \BibitemOpen
  \bibfield  {author} {\bibinfo {author} {\bibfnamefont {E.}~\bibnamefont
  {Ivchenko}},\ }\href@noop {} {\emph {\bibinfo {title} {Optical Spectroscopy
  of semiconductor nanostructures}}}\ (\bibinfo  {publisher} {Springer},\
  \bibinfo {year} {2004})\BibitemShut {NoStop}%
\bibitem [{\citenamefont {Pinczuk}\ \emph {et~al.}(1971)\citenamefont
  {Pinczuk}, \citenamefont {Brillson}, \citenamefont {Burstein},\ and\
  \citenamefont {Anastassakis}}]{pinczuk1971resonant}%
  \BibitemOpen
  \bibfield  {author} {\bibinfo {author} {\bibfnamefont {A.}~\bibnamefont
  {Pinczuk}}, \bibinfo {author} {\bibfnamefont {L.}~\bibnamefont {Brillson}},
  \bibinfo {author} {\bibfnamefont {E.}~\bibnamefont {Burstein}},\ and\
  \bibinfo {author} {\bibfnamefont {E.}~\bibnamefont {Anastassakis}},\ }\href
  {https://doi.org/10.1103/PhysRevLett.27.317} {\bibfield  {journal} {\bibinfo
  {journal} {Phys. Rev. Lett.}\ }\textbf {\bibinfo {volume} {27}},\ \bibinfo
  {pages} {317} (\bibinfo {year} {1971})}\BibitemShut {NoStop}%
\bibitem [{\citenamefont {Pinczuk}\ \emph {et~al.}(1989)\citenamefont
  {Pinczuk}, \citenamefont {Schmitt-Rink}, \citenamefont {Danan}, \citenamefont
  {Valladares}, \citenamefont {Pfeiffer},\ and\ \citenamefont
  {West}}]{pinczuk1989large}%
  \BibitemOpen
  \bibfield  {author} {\bibinfo {author} {\bibfnamefont {A.}~\bibnamefont
  {Pinczuk}}, \bibinfo {author} {\bibfnamefont {S.}~\bibnamefont
  {Schmitt-Rink}}, \bibinfo {author} {\bibfnamefont {G.}~\bibnamefont {Danan}},
  \bibinfo {author} {\bibfnamefont {J.~P.}\ \bibnamefont {Valladares}},
  \bibinfo {author} {\bibfnamefont {L.~N.}\ \bibnamefont {Pfeiffer}},\ and\
  \bibinfo {author} {\bibfnamefont {K.~W.}\ \bibnamefont {West}},\ }\href
  {https://doi.org/10.1103/PhysRevLett.63.1633} {\bibfield  {journal} {\bibinfo
   {journal} {Phys. Rev. Lett.}\ }\textbf {\bibinfo {volume} {63}},\ \bibinfo
  {pages} {1633} (\bibinfo {year} {1989})}\BibitemShut {NoStop}%
\bibitem [{\citenamefont {Go\~ni}\ \emph
  {et~al.}(1991{\natexlab{a}})\citenamefont {Go\~ni}, \citenamefont {Pinczuk},
  \citenamefont {Weiner}, \citenamefont {Calleja}, \citenamefont {Dennis},
  \citenamefont {Pfeiffer},\ and\ \citenamefont
  {West}}]{goni1991onedimensional}%
  \BibitemOpen
  \bibfield  {author} {\bibinfo {author} {\bibfnamefont {A.~R.}\ \bibnamefont
  {Go\~ni}}, \bibinfo {author} {\bibfnamefont {A.}~\bibnamefont {Pinczuk}},
  \bibinfo {author} {\bibfnamefont {J.~S.}\ \bibnamefont {Weiner}}, \bibinfo
  {author} {\bibfnamefont {J.~M.}\ \bibnamefont {Calleja}}, \bibinfo {author}
  {\bibfnamefont {B.~S.}\ \bibnamefont {Dennis}}, \bibinfo {author}
  {\bibfnamefont {L.~N.}\ \bibnamefont {Pfeiffer}},\ and\ \bibinfo {author}
  {\bibfnamefont {K.~W.}\ \bibnamefont {West}},\ }\href
  {https://doi.org/10.1103/PhysRevLett.67.3298} {\bibfield  {journal} {\bibinfo
   {journal} {Phys. Rev. Lett.}\ }\textbf {\bibinfo {volume} {67}},\ \bibinfo
  {pages} {3298} (\bibinfo {year} {1991}{\natexlab{a}})}\BibitemShut {NoStop}%
\bibitem [{\citenamefont {Lee}\ \emph {et~al.}(2022)\citenamefont {Lee},
  \citenamefont {Peng}, \citenamefont {Du}, \citenamefont {Kung}, \citenamefont
  {Monserrat}, \citenamefont {Cheong}, \citenamefont {Won},\ and\ \citenamefont
  {Blumberg}}]{lee2022chiral}%
  \BibitemOpen
  \bibfield  {author} {\bibinfo {author} {\bibfnamefont {A.~C.}\ \bibnamefont
  {Lee}}, \bibinfo {author} {\bibfnamefont {B.}~\bibnamefont {Peng}}, \bibinfo
  {author} {\bibfnamefont {K.}~\bibnamefont {Du}}, \bibinfo {author}
  {\bibfnamefont {H.-H.}\ \bibnamefont {Kung}}, \bibinfo {author}
  {\bibfnamefont {B.}~\bibnamefont {Monserrat}}, \bibinfo {author}
  {\bibfnamefont {S.~W.}\ \bibnamefont {Cheong}}, \bibinfo {author}
  {\bibfnamefont {C.~J.}\ \bibnamefont {Won}},\ and\ \bibinfo {author}
  {\bibfnamefont {G.}~\bibnamefont {Blumberg}},\ }\href
  {https://doi.org/10.1103/PhysRevB.105.L161105} {\bibfield  {journal}
  {\bibinfo  {journal} {Phys. Rev. B}\ }\textbf {\bibinfo {volume} {105}},\
  \bibinfo {pages} {L161105} (\bibinfo {year} {2022})}\BibitemShut {NoStop}%
\bibitem [{\citenamefont {Perez}(2009)}]{Perez2009}%
  \BibitemOpen
  \bibfield  {author} {\bibinfo {author} {\bibfnamefont {F.}~\bibnamefont
  {Perez}},\ }\href {https://doi.org/10.1103/PhysRevB.79.045306} {\bibfield
  {journal} {\bibinfo  {journal} {Phys. Rev. B}\ }\textbf {\bibinfo {volume}
  {79}},\ \bibinfo {pages} {045306} (\bibinfo {year} {2009})}\BibitemShut
  {NoStop}%
\bibitem [{\citenamefont {Baboux}\ \emph {et~al.}(2013)\citenamefont {Baboux},
  \citenamefont {Perez}, \citenamefont {Ullrich}, \citenamefont {D'Amico},
  \citenamefont {Karczewski},\ and\ \citenamefont {Wojtowicz}}]{Baboux2013}%
  \BibitemOpen
  \bibfield  {author} {\bibinfo {author} {\bibfnamefont {F.}~\bibnamefont
  {Baboux}}, \bibinfo {author} {\bibfnamefont {F.}~\bibnamefont {Perez}},
  \bibinfo {author} {\bibfnamefont {C.~A.}\ \bibnamefont {Ullrich}}, \bibinfo
  {author} {\bibfnamefont {I.}~\bibnamefont {D'Amico}}, \bibinfo {author}
  {\bibfnamefont {G.}~\bibnamefont {Karczewski}},\ and\ \bibinfo {author}
  {\bibfnamefont {T.}~\bibnamefont {Wojtowicz}},\ }\href
  {https://doi.org/10.1103/PhysRevB.87.121303} {\bibfield  {journal} {\bibinfo
  {journal} {Phys. Rev. B}\ }\textbf {\bibinfo {volume} {87}},\ \bibinfo
  {pages} {121303} (\bibinfo {year} {2013})}\BibitemShut {NoStop}%
\bibitem [{\citenamefont {Baboux}\ \emph {et~al.}(2015)\citenamefont {Baboux},
  \citenamefont {Perez}, \citenamefont {Ullrich}, \citenamefont {Karczewski},\
  and\ \citenamefont {Wojtowicz}}]{Baboux2015}%
  \BibitemOpen
  \bibfield  {author} {\bibinfo {author} {\bibfnamefont {F.}~\bibnamefont
  {Baboux}}, \bibinfo {author} {\bibfnamefont {F.}~\bibnamefont {Perez}},
  \bibinfo {author} {\bibfnamefont {C.~A.}\ \bibnamefont {Ullrich}}, \bibinfo
  {author} {\bibfnamefont {G.}~\bibnamefont {Karczewski}},\ and\ \bibinfo
  {author} {\bibfnamefont {T.}~\bibnamefont {Wojtowicz}},\ }\href
  {https://doi.org/10.1103/PhysRevB.92.125307} {\bibfield  {journal} {\bibinfo
  {journal} {Phys. Rev. B}\ }\textbf {\bibinfo {volume} {92}},\ \bibinfo
  {pages} {125307} (\bibinfo {year} {2015})}\BibitemShut {NoStop}%
\bibitem [{\citenamefont {Perez}\ \emph {et~al.}(2016)\citenamefont {Perez},
  \citenamefont {Baboux}, \citenamefont {Ullrich}, \citenamefont {D'Amico},
  \citenamefont {Vignale}, \citenamefont {Karczewski},\ and\ \citenamefont
  {Wojtowicz}}]{Perez2016}%
  \BibitemOpen
  \bibfield  {author} {\bibinfo {author} {\bibfnamefont {F.}~\bibnamefont
  {Perez}}, \bibinfo {author} {\bibfnamefont {F.}~\bibnamefont {Baboux}},
  \bibinfo {author} {\bibfnamefont {C.~A.}\ \bibnamefont {Ullrich}}, \bibinfo
  {author} {\bibfnamefont {I.}~\bibnamefont {D'Amico}}, \bibinfo {author}
  {\bibfnamefont {G.}~\bibnamefont {Vignale}}, \bibinfo {author} {\bibfnamefont
  {G.}~\bibnamefont {Karczewski}},\ and\ \bibinfo {author} {\bibfnamefont
  {T.}~\bibnamefont {Wojtowicz}},\ }\href
  {https://doi.org/10.1103/PhysRevLett.117.137204} {\bibfield  {journal}
  {\bibinfo  {journal} {Phys. Rev. Lett.}\ }\textbf {\bibinfo {volume} {117}},\
  \bibinfo {pages} {137204} (\bibinfo {year} {2016})}\BibitemShut {NoStop}%
\bibitem [{\citenamefont {Kung}\ \emph {et~al.}(2017)\citenamefont {Kung},
  \citenamefont {Maiti}, \citenamefont {Wang}, \citenamefont {Cheong},
  \citenamefont {Maslov},\ and\ \citenamefont {Blumberg}}]{Kung2017}%
  \BibitemOpen
  \bibfield  {author} {\bibinfo {author} {\bibfnamefont {H.-H.}\ \bibnamefont
  {Kung}}, \bibinfo {author} {\bibfnamefont {S.}~\bibnamefont {Maiti}},
  \bibinfo {author} {\bibfnamefont {X.}~\bibnamefont {Wang}}, \bibinfo {author}
  {\bibfnamefont {S.-W.}\ \bibnamefont {Cheong}}, \bibinfo {author}
  {\bibfnamefont {D.~L.}\ \bibnamefont {Maslov}},\ and\ \bibinfo {author}
  {\bibfnamefont {G.}~\bibnamefont {Blumberg}},\ }\href
  {https://doi.org/10.1103/PhysRevLett.119.136802} {\bibfield  {journal}
  {\bibinfo  {journal} {Phys. Rev. Lett.}\ }\textbf {\bibinfo {volume} {119}},\
  \bibinfo {pages} {136802} (\bibinfo {year} {2017})}\BibitemShut {NoStop}%
\bibitem [{\citenamefont {Ishizaka}\ \emph {et~al.}(2011)\citenamefont
  {Ishizaka}, \citenamefont {Bahramy}, \citenamefont {Murakawa}, \citenamefont
  {Sakano}, \citenamefont {Shimojima}, \citenamefont {Sonobe}, \citenamefont
  {Koizumi}, \citenamefont {Shin}, \citenamefont {Miyahara}, \citenamefont
  {Kimura}, \citenamefont {Miyamoto}, \citenamefont {Okuda}, \citenamefont
  {Namatame}, \citenamefont {Taniguchi}, \citenamefont {Arita}, \citenamefont
  {Nagaosa}, \citenamefont {Kobayashi}, \citenamefont {Murakami}, \citenamefont
  {Kumai}, \citenamefont {Kaneko}, \citenamefont {Onose},\ and\ \citenamefont
  {Tokura}}]{ishizaka2011giant}%
  \BibitemOpen
  \bibfield  {author} {\bibinfo {author} {\bibfnamefont {K.}~\bibnamefont
  {Ishizaka}}, \bibinfo {author} {\bibfnamefont {M.~S.}\ \bibnamefont
  {Bahramy}}, \bibinfo {author} {\bibfnamefont {H.}~\bibnamefont {Murakawa}},
  \bibinfo {author} {\bibfnamefont {M.}~\bibnamefont {Sakano}}, \bibinfo
  {author} {\bibfnamefont {T.}~\bibnamefont {Shimojima}}, \bibinfo {author}
  {\bibfnamefont {T.}~\bibnamefont {Sonobe}}, \bibinfo {author} {\bibfnamefont
  {K.}~\bibnamefont {Koizumi}}, \bibinfo {author} {\bibfnamefont
  {S.}~\bibnamefont {Shin}}, \bibinfo {author} {\bibfnamefont {H.}~\bibnamefont
  {Miyahara}}, \bibinfo {author} {\bibfnamefont {A.}~\bibnamefont {Kimura}},
  \bibinfo {author} {\bibfnamefont {K.}~\bibnamefont {Miyamoto}}, \bibinfo
  {author} {\bibfnamefont {T.}~\bibnamefont {Okuda}}, \bibinfo {author}
  {\bibfnamefont {H.}~\bibnamefont {Namatame}}, \bibinfo {author}
  {\bibfnamefont {M.}~\bibnamefont {Taniguchi}}, \bibinfo {author}
  {\bibfnamefont {R.}~\bibnamefont {Arita}}, \bibinfo {author} {\bibfnamefont
  {N.}~\bibnamefont {Nagaosa}}, \bibinfo {author} {\bibfnamefont
  {K.}~\bibnamefont {Kobayashi}}, \bibinfo {author} {\bibfnamefont
  {Y.}~\bibnamefont {Murakami}}, \bibinfo {author} {\bibfnamefont
  {R.}~\bibnamefont {Kumai}}, \bibinfo {author} {\bibfnamefont
  {Y.}~\bibnamefont {Kaneko}}, \bibinfo {author} {\bibfnamefont
  {Y.}~\bibnamefont {Onose}},\ and\ \bibinfo {author} {\bibfnamefont
  {Y.}~\bibnamefont {Tokura}},\ }\href {https://doi.org/10.1038/nmat3051}
  {\bibfield  {journal} {\bibinfo  {journal} {Nature Materials}\ }\textbf
  {\bibinfo {volume} {10}},\ \bibinfo {pages} {521} (\bibinfo {year}
  {2011})}\BibitemShut {NoStop}%
\bibitem [{\citenamefont {Chen}\ \emph {et~al.}(2013)\citenamefont {Chen},
  \citenamefont {Kanou}, \citenamefont {Liu}, \citenamefont {Zhang},
  \citenamefont {Sobota}, \citenamefont {Leuenberger}, \citenamefont {Mo},
  \citenamefont {Zhou}, \citenamefont {Yang}, \citenamefont {Kirchmann},
  \citenamefont {Lu}, \citenamefont {Moore}, \citenamefont {Hussain},
  \citenamefont {Shen}, \citenamefont {Qi},\ and\ \citenamefont
  {Sasagawa}}]{chen2013discovery}%
  \BibitemOpen
  \bibfield  {author} {\bibinfo {author} {\bibfnamefont {Y.~L.}\ \bibnamefont
  {Chen}}, \bibinfo {author} {\bibfnamefont {M.}~\bibnamefont {Kanou}},
  \bibinfo {author} {\bibfnamefont {Z.~K.}\ \bibnamefont {Liu}}, \bibinfo
  {author} {\bibfnamefont {H.~J.}\ \bibnamefont {Zhang}}, \bibinfo {author}
  {\bibfnamefont {J.~A.}\ \bibnamefont {Sobota}}, \bibinfo {author}
  {\bibfnamefont {D.}~\bibnamefont {Leuenberger}}, \bibinfo {author}
  {\bibfnamefont {S.~K.}\ \bibnamefont {Mo}}, \bibinfo {author} {\bibfnamefont
  {B.}~\bibnamefont {Zhou}}, \bibinfo {author} {\bibfnamefont {S.-L.}\
  \bibnamefont {Yang}}, \bibinfo {author} {\bibfnamefont {P.~S.}\ \bibnamefont
  {Kirchmann}}, \bibinfo {author} {\bibfnamefont {D.~H.}\ \bibnamefont {Lu}},
  \bibinfo {author} {\bibfnamefont {R.~G.}\ \bibnamefont {Moore}}, \bibinfo
  {author} {\bibfnamefont {Z.}~\bibnamefont {Hussain}}, \bibinfo {author}
  {\bibfnamefont {Z.~X.}\ \bibnamefont {Shen}}, \bibinfo {author}
  {\bibfnamefont {X.~L.}\ \bibnamefont {Qi}},\ and\ \bibinfo {author}
  {\bibfnamefont {T.}~\bibnamefont {Sasagawa}},\ }\href
  {https://doi.org/10.1038/nphys2768} {\bibfield  {journal} {\bibinfo
  {journal} {Nature Physics}\ }\textbf {\bibinfo {volume} {9}},\ \bibinfo
  {pages} {704} (\bibinfo {year} {2013})}\BibitemShut {NoStop}%
\bibitem [{\citenamefont {Sakano}\ \emph {et~al.}(2013)\citenamefont {Sakano},
  \citenamefont {Bahramy}, \citenamefont {Katayama}, \citenamefont {Shimojima},
  \citenamefont {Murakawa}, \citenamefont {Kaneko}, \citenamefont {Malaeb},
  \citenamefont {Shin}, \citenamefont {Ono}, \citenamefont {Kumigashira},
  \citenamefont {Arita}, \citenamefont {Nagaosa}, \citenamefont {Hwang},
  \citenamefont {Tokura},\ and\ \citenamefont {Ishizaka}}]{sakano2013strongly}%
  \BibitemOpen
  \bibfield  {author} {\bibinfo {author} {\bibfnamefont {M.}~\bibnamefont
  {Sakano}}, \bibinfo {author} {\bibfnamefont {M.~S.}\ \bibnamefont {Bahramy}},
  \bibinfo {author} {\bibfnamefont {A.}~\bibnamefont {Katayama}}, \bibinfo
  {author} {\bibfnamefont {T.}~\bibnamefont {Shimojima}}, \bibinfo {author}
  {\bibfnamefont {H.}~\bibnamefont {Murakawa}}, \bibinfo {author}
  {\bibfnamefont {Y.}~\bibnamefont {Kaneko}}, \bibinfo {author} {\bibfnamefont
  {W.}~\bibnamefont {Malaeb}}, \bibinfo {author} {\bibfnamefont
  {S.}~\bibnamefont {Shin}}, \bibinfo {author} {\bibfnamefont {K.}~\bibnamefont
  {Ono}}, \bibinfo {author} {\bibfnamefont {H.}~\bibnamefont {Kumigashira}},
  \bibinfo {author} {\bibfnamefont {R.}~\bibnamefont {Arita}}, \bibinfo
  {author} {\bibfnamefont {N.}~\bibnamefont {Nagaosa}}, \bibinfo {author}
  {\bibfnamefont {H.~Y.}\ \bibnamefont {Hwang}}, \bibinfo {author}
  {\bibfnamefont {Y.}~\bibnamefont {Tokura}},\ and\ \bibinfo {author}
  {\bibfnamefont {K.}~\bibnamefont {Ishizaka}},\ }\href
  {https://doi.org/10.1103/PhysRevLett.110.107204} {\bibfield  {journal}
  {\bibinfo  {journal} {Phys. Rev. Lett.}\ }\textbf {\bibinfo {volume} {110}},\
  \bibinfo {pages} {107204} (\bibinfo {year} {2013})}\BibitemShut {NoStop}%
\bibitem [{\citenamefont {Maiti}\ \emph {et~al.}(2015)\citenamefont {Maiti},
  \citenamefont {Zyuzin},\ and\ \citenamefont {Maslov}}]{Maiti2015}%
  \BibitemOpen
  \bibfield  {author} {\bibinfo {author} {\bibfnamefont {S.}~\bibnamefont
  {Maiti}}, \bibinfo {author} {\bibfnamefont {V.}~\bibnamefont {Zyuzin}},\ and\
  \bibinfo {author} {\bibfnamefont {D.~L.}\ \bibnamefont {Maslov}},\ }\href
  {https://doi.org/10.1103/PhysRevB.91.035106} {\bibfield  {journal} {\bibinfo
  {journal} {Phys. Rev. B}\ }\textbf {\bibinfo {volume} {91}},\ \bibinfo
  {pages} {035106} (\bibinfo {year} {2015})}\BibitemShut {NoStop}%
\bibitem [{\citenamefont {Maiti}\ \emph {et~al.}(2016)\citenamefont {Maiti},
  \citenamefont {Imran},\ and\ \citenamefont {Maslov}}]{Maiti2016}%
  \BibitemOpen
  \bibfield  {author} {\bibinfo {author} {\bibfnamefont {S.}~\bibnamefont
  {Maiti}}, \bibinfo {author} {\bibfnamefont {M.}~\bibnamefont {Imran}},\ and\
  \bibinfo {author} {\bibfnamefont {D.~L.}\ \bibnamefont {Maslov}},\ }\href
  {https://doi.org/10.1103/PhysRevB.93.045134} {\bibfield  {journal} {\bibinfo
  {journal} {Phys. Rev. B}\ }\textbf {\bibinfo {volume} {93}},\ \bibinfo
  {pages} {045134} (\bibinfo {year} {2016})}\BibitemShut {NoStop}%
\bibitem [{\citenamefont {Devereaux}\ and\ \citenamefont
  {Hackl}(2007)}]{Devereaux2007}%
  \BibitemOpen
  \bibfield  {author} {\bibinfo {author} {\bibfnamefont {T.~P.}\ \bibnamefont
  {Devereaux}}\ and\ \bibinfo {author} {\bibfnamefont {R.}~\bibnamefont
  {Hackl}},\ }\href {https://doi.org/10.1103/RevModPhys.79.175} {\bibfield
  {journal} {\bibinfo  {journal} {Rev. Mod. Phys.}\ }\textbf {\bibinfo {volume}
  {79}},\ \bibinfo {pages} {175} (\bibinfo {year} {2007})}\BibitemShut
  {NoStop}%
\bibitem [{\citenamefont {Landau}\ and\ \citenamefont
  {Lifshitz}(1980)}]{LL_StatPhys2}%
  \BibitemOpen
  \bibfield  {author} {\bibinfo {author} {\bibfnamefont {L.}~\bibnamefont
  {Landau}}\ and\ \bibinfo {author} {\bibfnamefont {E.}~\bibnamefont
  {Lifshitz}},\ }\href@noop {} {\emph {\bibinfo {title} {Statistical
  Physics,Ppart 2}}},\ Vol.~\bibinfo {volume} {9}\ (\bibinfo  {publisher}
  {Pergamon Press},\ \bibinfo {year} {1980})\BibitemShut {NoStop}%
\bibitem [{\citenamefont {Aldrich}\ \emph {et~al.}(1976)\citenamefont
  {Aldrich}, \citenamefont {Pethick},\ and\ \citenamefont
  {Pines}}]{Aldrich1976}%
  \BibitemOpen
  \bibfield  {author} {\bibinfo {author} {\bibfnamefont {C.~H.}\ \bibnamefont
  {Aldrich}}, \bibinfo {author} {\bibfnamefont {C.~J.}\ \bibnamefont
  {Pethick}},\ and\ \bibinfo {author} {\bibfnamefont {D.}~\bibnamefont
  {Pines}},\ }\href {https://doi.org/10.1103/PhysRevLett.37.845} {\bibfield
  {journal} {\bibinfo  {journal} {Phys. Rev. Lett.}\ }\textbf {\bibinfo
  {volume} {37}},\ \bibinfo {pages} {845} (\bibinfo {year} {1976})}\BibitemShut
  {NoStop}%
\bibitem [{\citenamefont {Bohm}\ and\ \citenamefont
  {Pines}(1953)}]{Bohm1951_3}%
  \BibitemOpen
  \bibfield  {author} {\bibinfo {author} {\bibfnamefont {D.}~\bibnamefont
  {Bohm}}\ and\ \bibinfo {author} {\bibfnamefont {D.}~\bibnamefont {Pines}},\
  }\href {https://doi.org/10.1103/PhysRev.92.609} {\bibfield  {journal}
  {\bibinfo  {journal} {Phys. Rev.}\ }\textbf {\bibinfo {volume} {92}},\
  \bibinfo {pages} {609} (\bibinfo {year} {1953})}\BibitemShut {NoStop}%
\bibitem [{\citenamefont {Anderson}(1963)}]{Anderson1963}%
  \BibitemOpen
  \bibfield  {author} {\bibinfo {author} {\bibfnamefont {P.~W.}\ \bibnamefont
  {Anderson}},\ }\href {https://doi.org/10.1103/PhysRev.130.439} {\bibfield
  {journal} {\bibinfo  {journal} {Phys. Rev.}\ }\textbf {\bibinfo {volume}
  {130}},\ \bibinfo {pages} {439} (\bibinfo {year} {1963})}\BibitemShut
  {NoStop}%
\bibitem [{\citenamefont {Pinczuk}\ \emph {et~al.}(1992)\citenamefont
  {Pinczuk}, \citenamefont {Dennis}, \citenamefont {Heiman}, \citenamefont
  {Kallin}, \citenamefont {Brey}, \citenamefont {Tejedor}, \citenamefont
  {Schmitt-Rink}, \citenamefont {Pfeiffer},\ and\ \citenamefont
  {West}}]{Pinczuk1992}%
  \BibitemOpen
  \bibfield  {author} {\bibinfo {author} {\bibfnamefont {A.}~\bibnamefont
  {Pinczuk}}, \bibinfo {author} {\bibfnamefont {B.~S.}\ \bibnamefont {Dennis}},
  \bibinfo {author} {\bibfnamefont {D.}~\bibnamefont {Heiman}}, \bibinfo
  {author} {\bibfnamefont {C.}~\bibnamefont {Kallin}}, \bibinfo {author}
  {\bibfnamefont {L.}~\bibnamefont {Brey}}, \bibinfo {author} {\bibfnamefont
  {C.}~\bibnamefont {Tejedor}}, \bibinfo {author} {\bibfnamefont
  {S.}~\bibnamefont {Schmitt-Rink}}, \bibinfo {author} {\bibfnamefont {L.~N.}\
  \bibnamefont {Pfeiffer}},\ and\ \bibinfo {author} {\bibfnamefont {K.~W.}\
  \bibnamefont {West}},\ }\href {https://doi.org/10.1103/PhysRevLett.68.3623}
  {\bibfield  {journal} {\bibinfo  {journal} {Phys. Rev. Lett.}\ }\textbf
  {\bibinfo {volume} {68}},\ \bibinfo {pages} {3623} (\bibinfo {year}
  {1992})}\BibitemShut {NoStop}%
\bibitem [{\citenamefont {Klein}(1981)}]{Klein1981}%
  \BibitemOpen
  \bibfield  {author} {\bibinfo {author} {\bibfnamefont {M.~V.}\ \bibnamefont
  {Klein}},\ }in\ \href@noop {} {\emph {\bibinfo {booktitle} {Lasers and
  Applications}}},\ \bibinfo {editor} {edited by\ \bibinfo {editor}
  {\bibfnamefont {W.~O.~N.}\ \bibnamefont {Guimaraes}}, \bibinfo {editor}
  {\bibfnamefont {C.-T.}\ \bibnamefont {Lin}},\ and\ \bibinfo {editor}
  {\bibfnamefont {A.}~\bibnamefont {Mooradian}}}\ (\bibinfo  {publisher}
  {Springer Berlin Heidelberg},\ \bibinfo {address} {Berlin, Heidelberg},\
  \bibinfo {year} {1981})\ pp.\ \bibinfo {pages} {45--54}\BibitemShut {NoStop}%
\bibitem [{\citenamefont {Klein}(2010)}]{Klein2010}%
  \BibitemOpen
  \bibfield  {author} {\bibinfo {author} {\bibfnamefont {M.~V.}\ \bibnamefont
  {Klein}},\ }\href {https://doi.org/10.1103/PhysRevB.82.014507} {\bibfield
  {journal} {\bibinfo  {journal} {Phys. Rev. B}\ }\textbf {\bibinfo {volume}
  {82}},\ \bibinfo {pages} {014507} (\bibinfo {year} {2010})}\BibitemShut
  {NoStop}%
\bibitem [{\citenamefont {Blumberg}\ \emph {et~al.}(2007)\citenamefont
  {Blumberg}, \citenamefont {Mialitsin}, \citenamefont {Dennis}, \citenamefont
  {Klein}, \citenamefont {Zhigadlo},\ and\ \citenamefont
  {Karpinski}}]{Blumberg2007}%
  \BibitemOpen
  \bibfield  {author} {\bibinfo {author} {\bibfnamefont {G.}~\bibnamefont
  {Blumberg}}, \bibinfo {author} {\bibfnamefont {A.}~\bibnamefont {Mialitsin}},
  \bibinfo {author} {\bibfnamefont {B.~S.}\ \bibnamefont {Dennis}}, \bibinfo
  {author} {\bibfnamefont {M.~V.}\ \bibnamefont {Klein}}, \bibinfo {author}
  {\bibfnamefont {N.~D.}\ \bibnamefont {Zhigadlo}},\ and\ \bibinfo {author}
  {\bibfnamefont {J.}~\bibnamefont {Karpinski}},\ }\href
  {https://doi.org/10.1103/PhysRevLett.99.227002} {\bibfield  {journal}
  {\bibinfo  {journal} {Phys. Rev. Lett.}\ }\textbf {\bibinfo {volume} {99}},\
  \bibinfo {pages} {227002} (\bibinfo {year} {2007})}\BibitemShut {NoStop}%
\bibitem [{\citenamefont {Cea}\ and\ \citenamefont {Benfatto}(2016)}]{Cea2016}%
  \BibitemOpen
  \bibfield  {author} {\bibinfo {author} {\bibfnamefont {T.}~\bibnamefont
  {Cea}}\ and\ \bibinfo {author} {\bibfnamefont {L.}~\bibnamefont {Benfatto}},\
  }\href {https://doi.org/10.1103/PhysRevB.94.064512} {\bibfield  {journal}
  {\bibinfo  {journal} {Phys. Rev. B}\ }\textbf {\bibinfo {volume} {94}},\
  \bibinfo {pages} {064512} (\bibinfo {year} {2016})}\BibitemShut {NoStop}%
\bibitem [{\citenamefont {Maiti}\ \emph {et~al.}(2017)\citenamefont {Maiti},
  \citenamefont {Chubukov},\ and\ \citenamefont {Hirschfeld}}]{Maiti2017b}%
  \BibitemOpen
  \bibfield  {author} {\bibinfo {author} {\bibfnamefont {S.}~\bibnamefont
  {Maiti}}, \bibinfo {author} {\bibfnamefont {A.~V.}\ \bibnamefont
  {Chubukov}},\ and\ \bibinfo {author} {\bibfnamefont {P.~J.}\ \bibnamefont
  {Hirschfeld}},\ }\href {https://doi.org/10.1103/PhysRevB.96.014503}
  {\bibfield  {journal} {\bibinfo  {journal} {Phys. Rev. B}\ }\textbf {\bibinfo
  {volume} {96}},\ \bibinfo {pages} {014503} (\bibinfo {year}
  {2017})}\BibitemShut {NoStop}%
\bibitem [{\citenamefont {Das~Sarma}\ and\ \citenamefont
  {Wang}(1999)}]{Sarma1999}%
  \BibitemOpen
  \bibfield  {author} {\bibinfo {author} {\bibfnamefont {S.}~\bibnamefont
  {Das~Sarma}}\ and\ \bibinfo {author} {\bibfnamefont {D.-W.}\ \bibnamefont
  {Wang}},\ }\href {https://doi.org/10.1103/PhysRevLett.83.816} {\bibfield
  {journal} {\bibinfo  {journal} {Phys. Rev. Lett.}\ }\textbf {\bibinfo
  {volume} {83}},\ \bibinfo {pages} {816} (\bibinfo {year} {1999})}\BibitemShut
  {NoStop}%
\bibitem [{\citenamefont {Go\~ni}\ \emph
  {et~al.}(1991{\natexlab{b}})\citenamefont {Go\~ni}, \citenamefont {Pinczuk},
  \citenamefont {Weiner}, \citenamefont {Calleja}, \citenamefont {Dennis},
  \citenamefont {Pfeiffer},\ and\ \citenamefont {West}}]{Goni1991}%
  \BibitemOpen
  \bibfield  {author} {\bibinfo {author} {\bibfnamefont {A.~R.}\ \bibnamefont
  {Go\~ni}}, \bibinfo {author} {\bibfnamefont {A.}~\bibnamefont {Pinczuk}},
  \bibinfo {author} {\bibfnamefont {J.~S.}\ \bibnamefont {Weiner}}, \bibinfo
  {author} {\bibfnamefont {J.~M.}\ \bibnamefont {Calleja}}, \bibinfo {author}
  {\bibfnamefont {B.~S.}\ \bibnamefont {Dennis}}, \bibinfo {author}
  {\bibfnamefont {L.~N.}\ \bibnamefont {Pfeiffer}},\ and\ \bibinfo {author}
  {\bibfnamefont {K.~W.}\ \bibnamefont {West}},\ }\href
  {https://doi.org/10.1103/PhysRevLett.67.3298} {\bibfield  {journal} {\bibinfo
   {journal} {Phys. Rev. Lett.}\ }\textbf {\bibinfo {volume} {67}},\ \bibinfo
  {pages} {3298} (\bibinfo {year} {1991}{\natexlab{b}})}\BibitemShut {NoStop}%
\bibitem [{\citenamefont {Rodina}\ and\ \citenamefont
  {Ivchenko}(2022)}]{Rodina2022}%
  \BibitemOpen
  \bibfield  {author} {\bibinfo {author} {\bibfnamefont {A.~V.}\ \bibnamefont
  {Rodina}}\ and\ \bibinfo {author} {\bibfnamefont {E.~L.}\ \bibnamefont
  {Ivchenko}},\ }\href {https://doi.org/10.1103/PhysRevB.106.245202} {\bibfield
   {journal} {\bibinfo  {journal} {Phys. Rev. B}\ }\textbf {\bibinfo {volume}
  {106}},\ \bibinfo {pages} {245202} (\bibinfo {year} {2022})}\BibitemShut
  {NoStop}%
\bibitem [{\citenamefont {Shastry}\ and\ \citenamefont
  {Shraiman}(1990)}]{Shastry1990}%
  \BibitemOpen
  \bibfield  {author} {\bibinfo {author} {\bibfnamefont {B.~S.}\ \bibnamefont
  {Shastry}}\ and\ \bibinfo {author} {\bibfnamefont {B.~I.}\ \bibnamefont
  {Shraiman}},\ }\href {https://doi.org/10.1103/PhysRevLett.65.1068} {\bibfield
   {journal} {\bibinfo  {journal} {Phys. Rev. Lett.}\ }\textbf {\bibinfo
  {volume} {65}},\ \bibinfo {pages} {1068} (\bibinfo {year}
  {1990})}\BibitemShut {NoStop}%
\bibitem [{\citenamefont {Shastry}\ and\ \citenamefont
  {Shraiman}(1991)}]{Shastry1991}%
  \BibitemOpen
  \bibfield  {author} {\bibinfo {author} {\bibfnamefont {B.~S.}\ \bibnamefont
  {Shastry}}\ and\ \bibinfo {author} {\bibfnamefont {B.~I.}\ \bibnamefont
  {Shraiman}},\ }\href {https://doi.org/10.1142/S0217979291000237} {\bibfield
  {journal} {\bibinfo  {journal} {International Journal of Modern Physics B}\
  }\textbf {\bibinfo {volume} {05}},\ \bibinfo {pages} {365} (\bibinfo {year}
  {1991})}\BibitemShut {NoStop}%
\bibitem [{\citenamefont {Buhot}\ \emph {et~al.}(2014)\citenamefont {Buhot},
  \citenamefont {M\'easson}, \citenamefont {Gallais}, \citenamefont {Cazayous},
  \citenamefont {Sacuto}, \citenamefont {Lapertot},\ and\ \citenamefont
  {Aoki}}]{Buhot2014}%
  \BibitemOpen
  \bibfield  {author} {\bibinfo {author} {\bibfnamefont {J.}~\bibnamefont
  {Buhot}}, \bibinfo {author} {\bibfnamefont {M.-A.}\ \bibnamefont
  {M\'easson}}, \bibinfo {author} {\bibfnamefont {Y.}~\bibnamefont {Gallais}},
  \bibinfo {author} {\bibfnamefont {M.}~\bibnamefont {Cazayous}}, \bibinfo
  {author} {\bibfnamefont {A.}~\bibnamefont {Sacuto}}, \bibinfo {author}
  {\bibfnamefont {G.}~\bibnamefont {Lapertot}},\ and\ \bibinfo {author}
  {\bibfnamefont {D.}~\bibnamefont {Aoki}},\ }\href
  {https://doi.org/10.1103/PhysRevLett.113.266405} {\bibfield  {journal}
  {\bibinfo  {journal} {Phys. Rev. Lett.}\ }\textbf {\bibinfo {volume} {113}},\
  \bibinfo {pages} {266405} (\bibinfo {year} {2014})}\BibitemShut {NoStop}%
\bibitem [{\citenamefont {Kung}\ \emph {et~al.}(2015)\citenamefont {Kung},
  \citenamefont {Baumbach}, \citenamefont {Bauer}, \citenamefont
  {Thorsm{\o}lle}, \citenamefont {Zhang}, \citenamefont {Haule}, \citenamefont
  {Mydosh},\ and\ \citenamefont {Blumberg}}]{kung2015chirality}%
  \BibitemOpen
  \bibfield  {author} {\bibinfo {author} {\bibfnamefont {H.-H.}\ \bibnamefont
  {Kung}}, \bibinfo {author} {\bibfnamefont {R.~E.}\ \bibnamefont {Baumbach}},
  \bibinfo {author} {\bibfnamefont {E.~D.}\ \bibnamefont {Bauer}}, \bibinfo
  {author} {\bibfnamefont {V.~K.}\ \bibnamefont {Thorsm{\o}lle}}, \bibinfo
  {author} {\bibfnamefont {W.-L.}\ \bibnamefont {Zhang}}, \bibinfo {author}
  {\bibfnamefont {K.}~\bibnamefont {Haule}}, \bibinfo {author} {\bibfnamefont
  {J.~A.}\ \bibnamefont {Mydosh}},\ and\ \bibinfo {author} {\bibfnamefont
  {G.}~\bibnamefont {Blumberg}},\ }\href
  {https://doi.org/10.1126/science.1259729} {\bibfield  {journal} {\bibinfo
  {journal} {Science}\ }\textbf {\bibinfo {volume} {347}},\ \bibinfo {pages}
  {1339} (\bibinfo {year} {2015})}\BibitemShut {NoStop}%
\bibitem [{\citenamefont {Kung}\ \emph {et~al.}(2016)\citenamefont {Kung},
  \citenamefont {Ran}, \citenamefont {Kanchanavatee}, \citenamefont {Krapivin},
  \citenamefont {Lee}, \citenamefont {Mydosh}, \citenamefont {Haule},
  \citenamefont {Maple},\ and\ \citenamefont {Blumberg}}]{Kung2016}%
  \BibitemOpen
  \bibfield  {author} {\bibinfo {author} {\bibfnamefont {H.-H.}\ \bibnamefont
  {Kung}}, \bibinfo {author} {\bibfnamefont {S.}~\bibnamefont {Ran}}, \bibinfo
  {author} {\bibfnamefont {N.}~\bibnamefont {Kanchanavatee}}, \bibinfo {author}
  {\bibfnamefont {V.}~\bibnamefont {Krapivin}}, \bibinfo {author}
  {\bibfnamefont {A.}~\bibnamefont {Lee}}, \bibinfo {author} {\bibfnamefont
  {J.~A.}\ \bibnamefont {Mydosh}}, \bibinfo {author} {\bibfnamefont
  {K.}~\bibnamefont {Haule}}, \bibinfo {author} {\bibfnamefont {M.~B.}\
  \bibnamefont {Maple}},\ and\ \bibinfo {author} {\bibfnamefont
  {G.}~\bibnamefont {Blumberg}},\ }\href
  {https://doi.org/10.1103/PhysRevLett.117.227601} {\bibfield  {journal}
  {\bibinfo  {journal} {Phys. Rev. Lett.}\ }\textbf {\bibinfo {volume} {117}},\
  \bibinfo {pages} {227601} (\bibinfo {year} {2016})}\BibitemShut {NoStop}%
\bibitem [{\citenamefont {Shekhter}\ \emph {et~al.}(2005)\citenamefont
  {Shekhter}, \citenamefont {Khodas},\ and\ \citenamefont
  {Finkel'stein}}]{Shekhter2005}%
  \BibitemOpen
  \bibfield  {author} {\bibinfo {author} {\bibfnamefont {A.}~\bibnamefont
  {Shekhter}}, \bibinfo {author} {\bibfnamefont {M.}~\bibnamefont {Khodas}},\
  and\ \bibinfo {author} {\bibfnamefont {A.~M.}\ \bibnamefont {Finkel'stein}},\
  }\href {https://doi.org/10.1103/PhysRevB.71.165329} {\bibfield  {journal}
  {\bibinfo  {journal} {Phys. Rev. B}\ }\textbf {\bibinfo {volume} {71}},\
  \bibinfo {pages} {165329} (\bibinfo {year} {2005})}\BibitemShut {NoStop}%
\bibitem [{\citenamefont {Ashrafi}\ and\ \citenamefont
  {Maslov}(2012)}]{Ashrafi2012}%
  \BibitemOpen
  \bibfield  {author} {\bibinfo {author} {\bibfnamefont {A.}~\bibnamefont
  {Ashrafi}}\ and\ \bibinfo {author} {\bibfnamefont {D.~L.}\ \bibnamefont
  {Maslov}},\ }\href {https://doi.org/10.1103/PhysRevLett.109.227201}
  {\bibfield  {journal} {\bibinfo  {journal} {Phys. Rev. Lett.}\ }\textbf
  {\bibinfo {volume} {109}},\ \bibinfo {pages} {227201} (\bibinfo {year}
  {2012})}\BibitemShut {NoStop}%
\bibitem [{\citenamefont {Raghu}\ \emph {et~al.}(2010)\citenamefont {Raghu},
  \citenamefont {Chung}, \citenamefont {Qi},\ and\ \citenamefont
  {Zhang}}]{raghu2010collective}%
  \BibitemOpen
  \bibfield  {author} {\bibinfo {author} {\bibfnamefont {S.}~\bibnamefont
  {Raghu}}, \bibinfo {author} {\bibfnamefont {S.~B.}\ \bibnamefont {Chung}},
  \bibinfo {author} {\bibfnamefont {X.-L.}\ \bibnamefont {Qi}},\ and\ \bibinfo
  {author} {\bibfnamefont {S.-C.}\ \bibnamefont {Zhang}},\ }\href
  {https://doi.org/10.1103/PhysRevLett.104.116401} {\bibfield  {journal}
  {\bibinfo  {journal} {Phys. Rev. Lett.}\ }\textbf {\bibinfo {volume} {104}},\
  \bibinfo {pages} {116401} (\bibinfo {year} {2010})}\BibitemShut {NoStop}%
\bibitem [{\citenamefont {Baboux}\ \emph {et~al.}(2012)\citenamefont {Baboux},
  \citenamefont {Perez}, \citenamefont {Ullrich}, \citenamefont {D'Amico},
  \citenamefont {G\'omez},\ and\ \citenamefont {Bernard}}]{baboux2012giant}%
  \BibitemOpen
  \bibfield  {author} {\bibinfo {author} {\bibfnamefont {F.}~\bibnamefont
  {Baboux}}, \bibinfo {author} {\bibfnamefont {F.}~\bibnamefont {Perez}},
  \bibinfo {author} {\bibfnamefont {C.~A.}\ \bibnamefont {Ullrich}}, \bibinfo
  {author} {\bibfnamefont {I.}~\bibnamefont {D'Amico}}, \bibinfo {author}
  {\bibfnamefont {J.}~\bibnamefont {G\'omez}},\ and\ \bibinfo {author}
  {\bibfnamefont {M.}~\bibnamefont {Bernard}},\ }\href
  {https://doi.org/10.1103/PhysRevLett.109.166401} {\bibfield  {journal}
  {\bibinfo  {journal} {Phys. Rev. Lett.}\ }\textbf {\bibinfo {volume} {109}},\
  \bibinfo {pages} {166401} (\bibinfo {year} {2012})}\BibitemShut {NoStop}%
\bibitem [{\citenamefont {Vitlina}\ \emph {et~al.}(2012)\citenamefont
  {Vitlina}, \citenamefont {Magarill},\ and\ \citenamefont
  {Chaplik}}]{Vitlina2012}%
  \BibitemOpen
  \bibfield  {author} {\bibinfo {author} {\bibfnamefont {R.~Z.}\ \bibnamefont
  {Vitlina}}, \bibinfo {author} {\bibfnamefont {L.~I.}\ \bibnamefont
  {Magarill}},\ and\ \bibinfo {author} {\bibfnamefont {A.~V.}\ \bibnamefont
  {Chaplik}},\ }\href {https://doi.org/10.1134/S0021364012050098} {\bibfield
  {journal} {\bibinfo  {journal} {JETP Letters}\ }\textbf {\bibinfo {volume}
  {95}},\ \bibinfo {pages} {253} (\bibinfo {year} {2012})}\BibitemShut
  {NoStop}%
\bibitem [{\citenamefont {Maiti}\ and\ \citenamefont
  {Maslov}(2017)}]{Maiti2017}%
  \BibitemOpen
  \bibfield  {author} {\bibinfo {author} {\bibfnamefont {S.}~\bibnamefont
  {Maiti}}\ and\ \bibinfo {author} {\bibfnamefont {D.~L.}\ \bibnamefont
  {Maslov}},\ }\href {https://doi.org/10.1103/PhysRevB.95.134425} {\bibfield
  {journal} {\bibinfo  {journal} {Phys. Rev. B}\ }\textbf {\bibinfo {volume}
  {95}},\ \bibinfo {pages} {134425} (\bibinfo {year} {2017})}\BibitemShut
  {NoStop}%
\bibitem [{\citenamefont {Kane}(1957)}]{Kane1957}%
  \BibitemOpen
  \bibfield  {author} {\bibinfo {author} {\bibfnamefont {E.~O.}\ \bibnamefont
  {Kane}},\ }\href
  {https://doi.org/https://doi.org/10.1016/0022-3697(57)90013-6} {\bibfield
  {journal} {\bibinfo  {journal} {Journal of Physics and Chemistry of Solids}\
  }\textbf {\bibinfo {volume} {1}},\ \bibinfo {pages} {249} (\bibinfo {year}
  {1957})}\BibitemShut {NoStop}%
\bibitem [{\citenamefont {Kashuba}\ and\ \citenamefont
  {Fal'ko}(2009)}]{Kashuba2009}%
  \BibitemOpen
  \bibfield  {author} {\bibinfo {author} {\bibfnamefont {O.}~\bibnamefont
  {Kashuba}}\ and\ \bibinfo {author} {\bibfnamefont {V.~I.}\ \bibnamefont
  {Fal'ko}},\ }\href {https://doi.org/10.1103/PhysRevB.80.241404} {\bibfield
  {journal} {\bibinfo  {journal} {Phys. Rev. B}\ }\textbf {\bibinfo {volume}
  {80}},\ \bibinfo {pages} {241404} (\bibinfo {year} {2009})}\BibitemShut
  {NoStop}%
\bibitem [{\citenamefont {Heller}\ \emph {et~al.}(2016)\citenamefont {Heller},
  \citenamefont {Yang}, \citenamefont {Kocia}, \citenamefont {Chen},
  \citenamefont {Fang}, \citenamefont {Borunda},\ and\ \citenamefont
  {Kaxiras}}]{Heller2016}%
  \BibitemOpen
  \bibfield  {author} {\bibinfo {author} {\bibfnamefont {E.}~\bibnamefont
  {Heller}}, \bibinfo {author} {\bibfnamefont {Y.}~\bibnamefont {Yang}},
  \bibinfo {author} {\bibfnamefont {L.}~\bibnamefont {Kocia}}, \bibinfo
  {author} {\bibfnamefont {W.}~\bibnamefont {Chen}}, \bibinfo {author}
  {\bibfnamefont {S.}~\bibnamefont {Fang}}, \bibinfo {author} {\bibfnamefont
  {M.}~\bibnamefont {Borunda}},\ and\ \bibinfo {author} {\bibfnamefont
  {E.}~\bibnamefont {Kaxiras}},\ }\href
  {https://doi.org/10.1021/acsnano.5b07676} {\bibfield  {journal} {\bibinfo
  {journal} {ACS Nano}\ }\textbf {\bibinfo {volume} {10}},\ \bibinfo {pages}
  {2803{\textendash}2818} (\bibinfo {year} {2016})},\ \bibinfo {note} {pMID:
  26799915}\BibitemShut {NoStop}%
\bibitem [{\citenamefont {Riccardi}\ \emph {et~al.}(2016)\citenamefont
  {Riccardi}, \citenamefont {M\'easson}, \citenamefont {Cazayous},
  \citenamefont {Sacuto},\ and\ \citenamefont {Gallais}}]{Riccardi2016}%
  \BibitemOpen
  \bibfield  {author} {\bibinfo {author} {\bibfnamefont {E.}~\bibnamefont
  {Riccardi}}, \bibinfo {author} {\bibfnamefont {M.-A.}\ \bibnamefont
  {M\'easson}}, \bibinfo {author} {\bibfnamefont {M.}~\bibnamefont {Cazayous}},
  \bibinfo {author} {\bibfnamefont {A.}~\bibnamefont {Sacuto}},\ and\ \bibinfo
  {author} {\bibfnamefont {Y.}~\bibnamefont {Gallais}},\ }\href
  {https://doi.org/10.1103/PhysRevLett.116.066805} {\bibfield  {journal}
  {\bibinfo  {journal} {Phys. Rev. Lett.}\ }\textbf {\bibinfo {volume} {116}},\
  \bibinfo {pages} {066805} (\bibinfo {year} {2016})}\BibitemShut {NoStop}%
\bibitem [{\citenamefont {Cardona}(1983)}]{cardona1983lightscattering}%
  \BibitemOpen
  \bibfield  {author} {\bibinfo {author} {\bibfnamefont {M.}~\bibnamefont
  {Cardona}},\ }\href@noop {} {\emph {\bibinfo {title} {{Light Scattering in
  Solids I}}}}\ (\bibinfo  {publisher} {Springer-Verlag, Berlin},\ \bibinfo
  {year} {1983})\BibitemShut {NoStop}%
\bibitem [{\citenamefont {Shvaika}\ \emph {et~al.}(2005)\citenamefont
  {Shvaika}, \citenamefont {Vorobyov}, \citenamefont {Freericks},\ and\
  \citenamefont {Devereaux}}]{Shvaika2005}%
  \BibitemOpen
  \bibfield  {author} {\bibinfo {author} {\bibfnamefont {A.~M.}\ \bibnamefont
  {Shvaika}}, \bibinfo {author} {\bibfnamefont {O.}~\bibnamefont {Vorobyov}},
  \bibinfo {author} {\bibfnamefont {J.~K.}\ \bibnamefont {Freericks}},\ and\
  \bibinfo {author} {\bibfnamefont {T.~P.}\ \bibnamefont {Devereaux}},\ }\href
  {https://doi.org/10.1103/PhysRevB.71.045120} {\bibfield  {journal} {\bibinfo
  {journal} {Phys. Rev. B}\ }\textbf {\bibinfo {volume} {71}},\ \bibinfo
  {pages} {045120} (\bibinfo {year} {2005})}\BibitemShut {NoStop}%
\bibitem [{\citenamefont {Bercioux}\ and\ \citenamefont
  {Lucignano}(2015)}]{Bercioux2015}%
  \BibitemOpen
  \bibfield  {author} {\bibinfo {author} {\bibfnamefont {D.}~\bibnamefont
  {Bercioux}}\ and\ \bibinfo {author} {\bibfnamefont {P.}~\bibnamefont
  {Lucignano}},\ }\href {https://doi.org/10.1088/0034-4885/78/10/106001}
  {\bibfield  {journal} {\bibinfo  {journal} {Reports on Progress in Physics}\
  }\textbf {\bibinfo {volume} {78}},\ \bibinfo {pages} {106001} (\bibinfo
  {year} {2015})}\BibitemShut {NoStop}%
\bibitem [{\citenamefont {Wang}\ \emph {et~al.}(2015)\citenamefont {Wang},
  \citenamefont {Ki}, \citenamefont {Chen}, \citenamefont {Berger},
  \citenamefont {MacDonald},\ and\ \citenamefont {Morpurgo}}]{Wang2015}%
  \BibitemOpen
  \bibfield  {author} {\bibinfo {author} {\bibfnamefont {Z.}~\bibnamefont
  {Wang}}, \bibinfo {author} {\bibfnamefont {D.-K.}\ \bibnamefont {Ki}},
  \bibinfo {author} {\bibfnamefont {H.}~\bibnamefont {Chen}}, \bibinfo {author}
  {\bibfnamefont {H.}~\bibnamefont {Berger}}, \bibinfo {author} {\bibfnamefont
  {A.~H.}\ \bibnamefont {MacDonald}},\ and\ \bibinfo {author} {\bibfnamefont
  {A.~F.}\ \bibnamefont {Morpurgo}},\ }\href
  {http://dx.doi.org/10.1038/ncomms9339} {\bibfield  {journal} {\bibinfo
  {journal} {Nature Communications}\ }\textbf {\bibinfo {volume} {6}},\
  \bibinfo {pages} {8339 EP } (\bibinfo {year} {2015})}\BibitemShut {NoStop}%
\bibitem [{\citenamefont {Cummings}\ \emph {et~al.}(2017)\citenamefont
  {Cummings}, \citenamefont {Garcia}, \citenamefont {Fabian},\ and\
  \citenamefont {Roche}}]{Cummings2017}%
  \BibitemOpen
  \bibfield  {author} {\bibinfo {author} {\bibfnamefont {A.~W.}\ \bibnamefont
  {Cummings}}, \bibinfo {author} {\bibfnamefont {J.~H.}\ \bibnamefont
  {Garcia}}, \bibinfo {author} {\bibfnamefont {J.}~\bibnamefont {Fabian}},\
  and\ \bibinfo {author} {\bibfnamefont {S.}~\bibnamefont {Roche}},\ }\href
  {https://doi.org/10.1103/PhysRevLett.119.206601} {\bibfield  {journal}
  {\bibinfo  {journal} {Phys. Rev. Lett.}\ }\textbf {\bibinfo {volume} {119}},\
  \bibinfo {pages} {206601} (\bibinfo {year} {2017})}\BibitemShut {NoStop}%
\bibitem [{\citenamefont {Garcia}\ \emph {et~al.}(2018)\citenamefont {Garcia},
  \citenamefont {Vila}, \citenamefont {Cummings},\ and\ \citenamefont
  {Roche}}]{Garcia2018}%
  \BibitemOpen
  \bibfield  {author} {\bibinfo {author} {\bibfnamefont {J.~H.}\ \bibnamefont
  {Garcia}}, \bibinfo {author} {\bibfnamefont {M.}~\bibnamefont {Vila}},
  \bibinfo {author} {\bibfnamefont {A.~W.}\ \bibnamefont {Cummings}},\ and\
  \bibinfo {author} {\bibfnamefont {S.}~\bibnamefont {Roche}},\ }\href
  {https://doi.org/10.1039/C7CS00864C} {\bibfield  {journal} {\bibinfo
  {journal} {Chem. Soc. Rev.}\ }\textbf {\bibinfo {volume} {47}},\ \bibinfo
  {pages} {3359} (\bibinfo {year} {2018})}\BibitemShut {NoStop}%
\bibitem [{\citenamefont {Kumar}\ \emph {et~al.}(2021)\citenamefont {Kumar},
  \citenamefont {Maiti},\ and\ \citenamefont {Maslov}}]{Kumar2021}%
  \BibitemOpen
  \bibfield  {author} {\bibinfo {author} {\bibfnamefont {A.}~\bibnamefont
  {Kumar}}, \bibinfo {author} {\bibfnamefont {S.}~\bibnamefont {Maiti}},\ and\
  \bibinfo {author} {\bibfnamefont {D.~L.}\ \bibnamefont {Maslov}},\ }\href
  {https://doi.org/10.1103/PhysRevB.104.155138} {\bibfield  {journal} {\bibinfo
   {journal} {Phys. Rev. B}\ }\textbf {\bibinfo {volume} {104}},\ \bibinfo
  {pages} {155138} (\bibinfo {year} {2021})}\BibitemShut {NoStop}%
\bibitem [{\citenamefont {Sakano}\ \emph {et~al.}(2012)\citenamefont {Sakano},
  \citenamefont {Miyawaki}, \citenamefont {Chainani}, \citenamefont {Takata},
  \citenamefont {Sonobe}, \citenamefont {Shimojima}, \citenamefont {Oura},
  \citenamefont {Shin}, \citenamefont {Bahramy}, \citenamefont {Arita},
  \citenamefont {Nagaosa}, \citenamefont {Murakawa}, \citenamefont {Kaneko},
  \citenamefont {Tokura},\ and\ \citenamefont {Ishizaka}}]{Ishizakaprb12}%
  \BibitemOpen
  \bibfield  {author} {\bibinfo {author} {\bibfnamefont {M.}~\bibnamefont
  {Sakano}}, \bibinfo {author} {\bibfnamefont {J.}~\bibnamefont {Miyawaki}},
  \bibinfo {author} {\bibfnamefont {A.}~\bibnamefont {Chainani}}, \bibinfo
  {author} {\bibfnamefont {Y.}~\bibnamefont {Takata}}, \bibinfo {author}
  {\bibfnamefont {T.}~\bibnamefont {Sonobe}}, \bibinfo {author} {\bibfnamefont
  {T.}~\bibnamefont {Shimojima}}, \bibinfo {author} {\bibfnamefont
  {M.}~\bibnamefont {Oura}}, \bibinfo {author} {\bibfnamefont {S.}~\bibnamefont
  {Shin}}, \bibinfo {author} {\bibfnamefont {M.~S.}\ \bibnamefont {Bahramy}},
  \bibinfo {author} {\bibfnamefont {R.}~\bibnamefont {Arita}}, \bibinfo
  {author} {\bibfnamefont {N.}~\bibnamefont {Nagaosa}}, \bibinfo {author}
  {\bibfnamefont {H.}~\bibnamefont {Murakawa}}, \bibinfo {author}
  {\bibfnamefont {Y.}~\bibnamefont {Kaneko}}, \bibinfo {author} {\bibfnamefont
  {Y.}~\bibnamefont {Tokura}},\ and\ \bibinfo {author} {\bibfnamefont
  {K.}~\bibnamefont {Ishizaka}},\ }\href
  {https://doi.org/10.1103/PhysRevB.86.085204} {\bibfield  {journal} {\bibinfo
  {journal} {Phys. Rev. B}\ }\textbf {\bibinfo {volume} {86}},\ \bibinfo
  {pages} {085204} (\bibinfo {year} {2012})}\BibitemShut {NoStop}%
\bibitem [{\citenamefont {VanGennep}\ \emph {et~al.}(2014)\citenamefont
  {VanGennep}, \citenamefont {Maiti}, \citenamefont {Graf}, \citenamefont
  {Tozer}, \citenamefont {Martin}, \citenamefont {Berger}, \citenamefont
  {Maslov},\ and\ \citenamefont {Hamlin}}]{VanGennep14}%
  \BibitemOpen
  \bibfield  {author} {\bibinfo {author} {\bibfnamefont {D.}~\bibnamefont
  {VanGennep}}, \bibinfo {author} {\bibfnamefont {S.}~\bibnamefont {Maiti}},
  \bibinfo {author} {\bibfnamefont {D.}~\bibnamefont {Graf}}, \bibinfo {author}
  {\bibfnamefont {S.~W.}\ \bibnamefont {Tozer}}, \bibinfo {author}
  {\bibfnamefont {C.}~\bibnamefont {Martin}}, \bibinfo {author} {\bibfnamefont
  {H.}~\bibnamefont {Berger}}, \bibinfo {author} {\bibfnamefont {D.~L.}\
  \bibnamefont {Maslov}},\ and\ \bibinfo {author} {\bibfnamefont {J.~J.}\
  \bibnamefont {Hamlin}},\ }\href
  {https://doi.org/10.1088/0953-8984/26/34/342202} {\bibfield  {journal}
  {\bibinfo  {journal} {Journal of Physics: Condensed Matter}\ }\textbf
  {\bibinfo {volume} {26}},\ \bibinfo {pages} {342202} (\bibinfo {year}
  {2014})}\BibitemShut {NoStop}%
\bibitem [{\citenamefont {Cai}\ \emph {et~al.}(2022)\citenamefont {Cai},
  \citenamefont {Yu}, \citenamefont {Zhao}, \citenamefont {Li}, \citenamefont
  {Feng}, \citenamefont {Zhou}, \citenamefont {Wang}, \citenamefont {Zhang},
  \citenamefont {Zhang}, \citenamefont {Shi}, \citenamefont {Zhang},
  \citenamefont {Yang},\ and\ \citenamefont {Jiang}}]{cai_22}%
  \BibitemOpen
  \bibfield  {author} {\bibinfo {author} {\bibfnamefont {L.}~\bibnamefont
  {Cai}}, \bibinfo {author} {\bibfnamefont {C.}~\bibnamefont {Yu}}, \bibinfo
  {author} {\bibfnamefont {W.}~\bibnamefont {Zhao}}, \bibinfo {author}
  {\bibfnamefont {Y.}~\bibnamefont {Li}}, \bibinfo {author} {\bibfnamefont
  {H.}~\bibnamefont {Feng}}, \bibinfo {author} {\bibfnamefont {H.-A.}\
  \bibnamefont {Zhou}}, \bibinfo {author} {\bibfnamefont {L.}~\bibnamefont
  {Wang}}, \bibinfo {author} {\bibfnamefont {X.}~\bibnamefont {Zhang}},
  \bibinfo {author} {\bibfnamefont {Y.}~\bibnamefont {Zhang}}, \bibinfo
  {author} {\bibfnamefont {Y.}~\bibnamefont {Shi}}, \bibinfo {author}
  {\bibfnamefont {J.}~\bibnamefont {Zhang}}, \bibinfo {author} {\bibfnamefont
  {L.}~\bibnamefont {Yang}},\ and\ \bibinfo {author} {\bibfnamefont
  {W.}~\bibnamefont {Jiang}},\ }\href
  {https://doi.org/10.1021/acs.nanolett.2c02354} {\bibfield  {journal}
  {\bibinfo  {journal} {Nano Letters}\ }\textbf {\bibinfo {volume} {22}},\
  \bibinfo {pages} {7441} (\bibinfo {year} {2022})},\ \bibinfo {note} {pMID:
  36099337},\ \Eprint
  {https://arxiv.org/abs/https://doi.org/10.1021/acs.nanolett.2c02354}
  {https://doi.org/10.1021/acs.nanolett.2c02354} \BibitemShut {NoStop}%
\bibitem [{\citenamefont {Maa{\ss}}\ \emph {et~al.}(2016)\citenamefont
  {Maa{\ss}}, \citenamefont {Bentmann}, \citenamefont {Seibel}, \citenamefont
  {Tusche}, \citenamefont {Eremeev}, \citenamefont {Peixoto}, \citenamefont
  {Tereshchenko}, \citenamefont {Kokh}, \citenamefont {Chulkov}, \citenamefont
  {Kirschner},\ and\ \citenamefont {Reinert}}]{Maas2016}%
  \BibitemOpen
  \bibfield  {author} {\bibinfo {author} {\bibfnamefont {H.}~\bibnamefont
  {Maa{\ss}}}, \bibinfo {author} {\bibfnamefont {H.}~\bibnamefont {Bentmann}},
  \bibinfo {author} {\bibfnamefont {C.}~\bibnamefont {Seibel}}, \bibinfo
  {author} {\bibfnamefont {C.}~\bibnamefont {Tusche}}, \bibinfo {author}
  {\bibfnamefont {S.~V.}\ \bibnamefont {Eremeev}}, \bibinfo {author}
  {\bibfnamefont {T.~R.~F.}\ \bibnamefont {Peixoto}}, \bibinfo {author}
  {\bibfnamefont {O.~E.}\ \bibnamefont {Tereshchenko}}, \bibinfo {author}
  {\bibfnamefont {K.~A.}\ \bibnamefont {Kokh}}, \bibinfo {author}
  {\bibfnamefont {E.~V.}\ \bibnamefont {Chulkov}}, \bibinfo {author}
  {\bibfnamefont {J.}~\bibnamefont {Kirschner}},\ and\ \bibinfo {author}
  {\bibfnamefont {F.}~\bibnamefont {Reinert}},\ }\href
  {https://doi.org/10.1038/ncomms11621} {\bibfield  {journal} {\bibinfo
  {journal} {Nature Communications}\ }\textbf {\bibinfo {volume} {7}},\
  \bibinfo {pages} {11621} (\bibinfo {year} {2016})}\BibitemShut {NoStop}%
\bibitem [{\citenamefont {Lee}\ \emph {et~al.}(2023)\citenamefont {Lee},
  \citenamefont {Sarkar}, \citenamefont {Maiti}, \citenamefont {Du},
  \citenamefont {Kung}, \citenamefont {Cheong}, \citenamefont {Won},
  \citenamefont {Kang},\ and\ \citenamefont {Blumberg}}]{lee2023resonant}%
  \BibitemOpen
  \bibfield  {author} {\bibinfo {author} {\bibfnamefont {A.}~\bibnamefont
  {Lee}}, \bibinfo {author} {\bibfnamefont {S.}~\bibnamefont {Sarkar}},
  \bibinfo {author} {\bibfnamefont {S.}~\bibnamefont {Maiti}}, \bibinfo
  {author} {\bibfnamefont {K.}~\bibnamefont {Du}}, \bibinfo {author}
  {\bibfnamefont {H.-H.}\ \bibnamefont {Kung}}, \bibinfo {author}
  {\bibfnamefont {S.}~\bibnamefont {Cheong}}, \bibinfo {author} {\bibfnamefont
  {C.}~\bibnamefont {Won}}, \bibinfo {author} {\bibfnamefont {K.}~\bibnamefont
  {Kang}},\ and\ \bibinfo {author} {\bibfnamefont {G.}~\bibnamefont
  {Blumberg}},\ }\href@noop {} {\bibfield  {journal} {\bibinfo  {journal}
  {arXiv}\ } (\bibinfo {year} {2023})}\BibitemShut {NoStop}%
\bibitem [{\citenamefont {Pletyukhov}\ and\ \citenamefont
  {Gritsev}(2006)}]{Pletyukhov2006}%
  \BibitemOpen
  \bibfield  {author} {\bibinfo {author} {\bibfnamefont {M.}~\bibnamefont
  {Pletyukhov}}\ and\ \bibinfo {author} {\bibfnamefont {V.}~\bibnamefont
  {Gritsev}},\ }\href {https://doi.org/10.1103/PhysRevB.74.045307} {\bibfield
  {journal} {\bibinfo  {journal} {Phys. Rev. B}\ }\textbf {\bibinfo {volume}
  {74}},\ \bibinfo {pages} {045307} (\bibinfo {year} {2006})}\BibitemShut
  {NoStop}%
\bibitem [{\citenamefont {Lee}\ \emph {et~al.}(2011)\citenamefont {Lee},
  \citenamefont {Schober}, \citenamefont {Bahramy}, \citenamefont {Murakawa},
  \citenamefont {Onose}, \citenamefont {Arita}, \citenamefont {Nagaosa},\ and\
  \citenamefont {Tokura}}]{lee_prl11}%
  \BibitemOpen
  \bibfield  {author} {\bibinfo {author} {\bibfnamefont {J.~S.}\ \bibnamefont
  {Lee}}, \bibinfo {author} {\bibfnamefont {G.~A.~H.}\ \bibnamefont {Schober}},
  \bibinfo {author} {\bibfnamefont {M.~S.}\ \bibnamefont {Bahramy}}, \bibinfo
  {author} {\bibfnamefont {H.}~\bibnamefont {Murakawa}}, \bibinfo {author}
  {\bibfnamefont {Y.}~\bibnamefont {Onose}}, \bibinfo {author} {\bibfnamefont
  {R.}~\bibnamefont {Arita}}, \bibinfo {author} {\bibfnamefont
  {N.}~\bibnamefont {Nagaosa}},\ and\ \bibinfo {author} {\bibfnamefont
  {Y.}~\bibnamefont {Tokura}},\ }\href
  {https://doi.org/10.1103/PhysRevLett.107.117401} {\bibfield  {journal}
  {\bibinfo  {journal} {Phys. Rev. Lett.}\ }\textbf {\bibinfo {volume} {107}},\
  \bibinfo {pages} {117401} (\bibinfo {year} {2011})}\BibitemShut {NoStop}%
\bibitem [{\citenamefont {Schwalbe}\ \emph {et~al.}(2016)\citenamefont
  {Schwalbe}, \citenamefont {Wirnata}, \citenamefont {Starke}, \citenamefont
  {Schober},\ and\ \citenamefont {Kortus}}]{schwalbe_prb16}%
  \BibitemOpen
  \bibfield  {author} {\bibinfo {author} {\bibfnamefont {S.}~\bibnamefont
  {Schwalbe}}, \bibinfo {author} {\bibfnamefont {R.}~\bibnamefont {Wirnata}},
  \bibinfo {author} {\bibfnamefont {R.}~\bibnamefont {Starke}}, \bibinfo
  {author} {\bibfnamefont {G.~A.~H.}\ \bibnamefont {Schober}},\ and\ \bibinfo
  {author} {\bibfnamefont {J.}~\bibnamefont {Kortus}},\ }\href
  {https://doi.org/10.1103/PhysRevB.94.205130} {\bibfield  {journal} {\bibinfo
  {journal} {Phys. Rev. B}\ }\textbf {\bibinfo {volume} {94}},\ \bibinfo
  {pages} {205130} (\bibinfo {year} {2016})}\BibitemShut {NoStop}%
\bibitem [{\citenamefont {Gavoret}\ \emph {et~al.}(1969)\citenamefont
  {Gavoret}, \citenamefont {Nozières}, \citenamefont {Roulet},\ and\
  \citenamefont {Combescot}}]{Gavoret}%
  \BibitemOpen
  \bibfield  {author} {\bibinfo {author} {\bibfnamefont {J.}~\bibnamefont
  {Gavoret}}, \bibinfo {author} {\bibfnamefont {P.}~\bibnamefont {Nozières}},
  \bibinfo {author} {\bibfnamefont {B.}~\bibnamefont {Roulet}},\ and\ \bibinfo
  {author} {\bibfnamefont {M.}~\bibnamefont {Combescot}},\ }\href
  {https://doi.org/10.1051/jphys:019690030011-12098700} {\bibfield  {journal}
  {\bibinfo  {journal} {J. Phys. France}\ }\textbf {\bibinfo {volume} {30}},\
  \bibinfo {pages} {987} (\bibinfo {year} {1969})}\BibitemShut {NoStop}%
\bibitem [{\citenamefont {Sharma}\ \emph {et~al.}(2021)\citenamefont {Sharma},
  \citenamefont {Principi},\ and\ \citenamefont {Maslov}}]{Sharma2021}%
  \BibitemOpen
  \bibfield  {author} {\bibinfo {author} {\bibfnamefont {P.}~\bibnamefont
  {Sharma}}, \bibinfo {author} {\bibfnamefont {A.}~\bibnamefont {Principi}},\
  and\ \bibinfo {author} {\bibfnamefont {D.~L.}\ \bibnamefont {Maslov}},\
  }\href {https://doi.org/10.1103/PhysRevB.104.045142} {\bibfield  {journal}
  {\bibinfo  {journal} {Phys. Rev. B}\ }\textbf {\bibinfo {volume} {104}},\
  \bibinfo {pages} {045142} (\bibinfo {year} {2021})}\BibitemShut {NoStop}%
\bibitem [{\citenamefont {Goyal}\ \emph {et~al.}(2023)\citenamefont {Goyal},
  \citenamefont {Sharma},\ and\ \citenamefont {Maslov}}]{Goyal2023}%
  \BibitemOpen
  \bibfield  {author} {\bibinfo {author} {\bibfnamefont {A.}~\bibnamefont
  {Goyal}}, \bibinfo {author} {\bibfnamefont {P.}~\bibnamefont {Sharma}},\ and\
  \bibinfo {author} {\bibfnamefont {D.~L.}\ \bibnamefont {Maslov}},\ }\href
  {https://doi.org/10.48550/arXiv.2303.08705} {\bibfield  {journal} {\bibinfo
  {journal} {arXiv:2303.08705}\ } (\bibinfo {year} {2023})}\BibitemShut
  {NoStop}%
\bibitem [{\citenamefont {Chen}(1993)}]{Chen1993}%
  \BibitemOpen
  \bibfield  {author} {\bibinfo {author} {\bibfnamefont {H.}~\bibnamefont
  {Chen}},\ }\href {https://doi.org/10.1103/PhysRevLett.71.2304} {\bibfield
  {journal} {\bibinfo  {journal} {Phys. Rev. Lett.}\ }\textbf {\bibinfo
  {volume} {71}},\ \bibinfo {pages} {2304} (\bibinfo {year}
  {1993})}\BibitemShut {NoStop}%
\bibitem [{\citenamefont {Klein}\ and\ \citenamefont
  {Blumberg}(1999)}]{Klein1999}%
  \BibitemOpen
  \bibfield  {author} {\bibinfo {author} {\bibfnamefont {M.~V.}\ \bibnamefont
  {Klein}}\ and\ \bibinfo {author} {\bibfnamefont {G.}~\bibnamefont
  {Blumberg}},\ }\href {https://doi.org/10.1126/science.283.5398.42} {\bibfield
   {journal} {\bibinfo  {journal} {Science}\ }\textbf {\bibinfo {volume}
  {283}},\ \bibinfo {pages} {42} (\bibinfo {year} {1999})},\ \Eprint
  {https://arxiv.org/abs/https://www.science.org/doi/pdf/10.1126/science.283.5398.42}
  {https://www.science.org/doi/pdf/10.1126/science.283.5398.42} \BibitemShut
  {NoStop}%
\bibitem [{\citenamefont {Basov}\ \emph {et~al.}(1999)\citenamefont {Basov},
  \citenamefont {Woods}, \citenamefont {Katz}, \citenamefont {Singley},
  \citenamefont {Dynes}, \citenamefont {Xu}, \citenamefont {Hinks},
  \citenamefont {Homes},\ and\ \citenamefont {Strongin}}]{basov1999}%
  \BibitemOpen
  \bibfield  {author} {\bibinfo {author} {\bibfnamefont {D.}~\bibnamefont
  {Basov}}, \bibinfo {author} {\bibfnamefont {S.}~\bibnamefont {Woods}},
  \bibinfo {author} {\bibfnamefont {A.}~\bibnamefont {Katz}}, \bibinfo {author}
  {\bibfnamefont {E.}~\bibnamefont {Singley}}, \bibinfo {author} {\bibfnamefont
  {R.}~\bibnamefont {Dynes}}, \bibinfo {author} {\bibfnamefont
  {M.}~\bibnamefont {Xu}}, \bibinfo {author} {\bibfnamefont {D.}~\bibnamefont
  {Hinks}}, \bibinfo {author} {\bibfnamefont {C.}~\bibnamefont {Homes}},\ and\
  \bibinfo {author} {\bibfnamefont {M.}~\bibnamefont {Strongin}},\ }\href@noop
  {} {\bibfield  {journal} {\bibinfo  {journal} {Science}\ }\textbf {\bibinfo
  {volume} {283}},\ \bibinfo {pages} {49} (\bibinfo {year} {1999})}\BibitemShut
  {NoStop}%
\bibitem [{\citenamefont {Agarwal}\ \emph {et~al.}(2011)\citenamefont
  {Agarwal}, \citenamefont {Chesi}, \citenamefont {Jungwirth}, \citenamefont
  {Sinova}, \citenamefont {Vignale},\ and\ \citenamefont
  {Polini}}]{Agarwal2011}%
  \BibitemOpen
  \bibfield  {author} {\bibinfo {author} {\bibfnamefont {A.}~\bibnamefont
  {Agarwal}}, \bibinfo {author} {\bibfnamefont {S.}~\bibnamefont {Chesi}},
  \bibinfo {author} {\bibfnamefont {T.}~\bibnamefont {Jungwirth}}, \bibinfo
  {author} {\bibfnamefont {J.}~\bibnamefont {Sinova}}, \bibinfo {author}
  {\bibfnamefont {G.}~\bibnamefont {Vignale}},\ and\ \bibinfo {author}
  {\bibfnamefont {M.}~\bibnamefont {Polini}},\ }\href
  {https://doi.org/10.1103/PhysRevB.83.115135} {\bibfield  {journal} {\bibinfo
  {journal} {Phys. Rev. B}\ }\textbf {\bibinfo {volume} {83}},\ \bibinfo
  {pages} {115135} (\bibinfo {year} {2011})}\BibitemShut {NoStop}%
\bibitem [{\citenamefont {Parlak}\ \emph {et~al.}(2023)\citenamefont {Parlak},
  \citenamefont {Ghosh},\ and\ \citenamefont {Garate}}]{Parlak2023}%
  \BibitemOpen
  \bibfield  {author} {\bibinfo {author} {\bibfnamefont {S.~m.~c.}\
  \bibnamefont {Parlak}}, \bibinfo {author} {\bibfnamefont {S.}~\bibnamefont
  {Ghosh}},\ and\ \bibinfo {author} {\bibfnamefont {I.}~\bibnamefont
  {Garate}},\ }\href {https://doi.org/10.1103/PhysRevB.107.104308} {\bibfield
  {journal} {\bibinfo  {journal} {Phys. Rev. B}\ }\textbf {\bibinfo {volume}
  {107}},\ \bibinfo {pages} {104308} (\bibinfo {year} {2023})}\BibitemShut
  {NoStop}%
\bibitem [{\citenamefont {Arakawa}\ and\ \citenamefont
  {Yonemitsu}(2021)}]{Arakawa2021}%
  \BibitemOpen
  \bibfield  {author} {\bibinfo {author} {\bibfnamefont {N.}~\bibnamefont
  {Arakawa}}\ and\ \bibinfo {author} {\bibfnamefont {K.}~\bibnamefont
  {Yonemitsu}},\ }\href {https://doi.org/10.1103/PhysRevB.103.L100408}
  {\bibfield  {journal} {\bibinfo  {journal} {Phys. Rev. B}\ }\textbf {\bibinfo
  {volume} {103}},\ \bibinfo {pages} {L100408} (\bibinfo {year}
  {2021})}\BibitemShut {NoStop}%
\end{thebibliography}%

\end{document}